\documentclass[superscriptaddress,aps,preprintnumbers,amsmath,amssymb,prd,nofootinbib,preprint,longbibliography]{revtex4-1}
\pdfoutput=1

\usepackage{graphicx,array}
\usepackage{color}
\usepackage{latexsym}
\usepackage{amsthm}
\usepackage{amsmath}
\usepackage{amssymb}
\usepackage{hyperref} 
\usepackage{bbold}
\usepackage{mathtools}
\usepackage{enumitem}
\usepackage{makecell}
\usepackage[table]{xcolor}
\usepackage{multirow}
\usepackage{booktabs}
\usepackage{setspace}

\numberwithin{equation}{section}

\makeatletter
\renewcommand{\p@subsection}{}
\renewcommand{\p@subsubsection}{}
\makeatother

\def\simgt{\mathrel{\lower2.5pt\vbox{\lineskip=0pt\baselineskip=0pt
           \hbox{$>$}\hbox{$\sim$}}}}
\def\simlt{\mathrel{\lower2.5pt\vbox{\lineskip=0pt\baselineskip=0pt
           \hbox{$<$}\hbox{$\sim$}}}}

\newcommand{\be}{\begin{equation}}
\newcommand{\ee}{\end{equation}}
\newcommand{\bea}{\begin{eqnarray}}
\newcommand{\eea}{\end{eqnarray}}

\newcommand{\GeV}{\textrm{ GeV}}
\newcommand{\TeV}{\textrm{ TeV}}
\newcommand{\gsim}{\lower.7ex\hbox{$\;\stackrel{\textstyle>}{\sim}\;$}}
\newcommand{\lsim}{\lower.7ex\hbox{$\;\stackrel{\textstyle<}{\sim}\;$}}

\newcommand{\MPl}{M_{\rm Pl}}

\definecolor{nicered}{rgb}{0.7,0.1,0.1}
\definecolor{nicegreen}{rgb}{0.1,0.5,0.1}
\definecolor{PatColor}{rgb}{0,.8,0}
\hypersetup{colorlinks,citecolor= blue,linkcolor= black}

\begin{document}

\title{Lepto-axiogenesis and the scale of supersymmetry}

\author{Patrick Barnes}
\affiliation{Leinweber Center for Theoretical Physics, Department of Physics, University of Michigan, Ann Arbor, MI 48109, USA}
\author{Raymond T. Co}
\affiliation{William I. Fine Theoretical Physics Institute, School of Physics and Astronomy, University of Minnesota, Minneapolis, MN 55455, USA}
\author{Keisuke Harigaya}
\affiliation{Theoretical Physics Department, CERN, Geneva, Switzerland}
\author{Aaron Pierce}
\affiliation{Leinweber Center for Theoretical Physics, Department of Physics, University of Michigan, Ann Arbor, MI 48109, USA}

\date{\today}

\begin{abstract}
If the Peccei-Quinn field containing the QCD axion undergoes rotations in the early universe, the dimension-five operator responsible for neutrino masses can generate a lepton asymmetry that ultimately gives rise to the observed baryon asymmetry of the Universe.  This lepto-axiogenesis scenario requires a flat potential for the radial direction of the Peccei-Quinn field, naturally realized in supersymmetric models. We carefully compute the efficiency of this mechanism for the Dine-Fischler-Srednicki-Zhitnitsky (DFSZ) and Kim-Shifman-Vainshtein-Zakharov (KSVZ) axion models and place lower bounds on the masses of scalar superpartners required to reproduce the observed baryon asymmetry. For the KSVZ model, we find an efficiency for generation of the asymmetry six times larger than the previously extant computation  
after including scattering channels involving superpartners.  In this case, the superpartner scale should be above $\sim$ 30 TeV for a domain wall number of one; the lower bound weakens for larger domain wall numbers.  We find that the superpartner mass scale may also be as low as 30 TeV for the DFSZ model.  In all cases, the lower bound on the superpartner masses is inversely proportional to the sum of the squares of the neutrino masses and so can strengthen as the upper bound on the neutrino mass improves. We identify the parameter space where the axion rotation can simultaneously produce axion dark matter via kinetic misalignment; in this case it is possible to put an upper bound of order PeV on the masses of scalar superpartners. 
\end{abstract}

\preprint{LCTP-22-12, UMN-TH-4130/22, FTPI-MINN-22-21, CERN-TH-2022-136}

\maketitle

\vspace{1cm}

\maketitle

\begingroup
\hypersetup{linkcolor=black}
\renewcommand{\baselinestretch}{1.3}\normalsize
\tableofcontents
\renewcommand{\baselinestretch}{2}\normalsize
\endgroup

\newpage

\section{Introduction}
\label{sec:intro}
The Peccei-Quinn (PQ) symmetry~\cite{Peccei:1977hh,Peccei:1977ur} provides an attractive solution to the strong CP problem.  The pseudo Nambu-Goldstone boson associated with this symmetry, the axion~\cite{Weinberg:1977ma, Wilczek:1977pj}, can have important implications for cosmology.  It is a cold dark matter candidate, and it can also play a central role in the generation of the matter-antimatter asymmetry.

One possibility is that axion dark matter can be generated by the misalignment mechanism~\cite{Preskill:1982cy,Abbott:1982af,Dine:1982ah}, wherein the axion field is displaced from the zero-temperature minimum of its potential in the early universe.  In this case, the axion begins its motion from rest when the mass generated by the QCD anomaly becomes comparable to the Hubble expansion rate.  However, similar to fields in models of Affleck-Dine baryogenesis, the complex PQ field that contains the axion may receive a kick at early times and rotate in field space.  This has ramifications for cosmology.  First, axion dark matter may be produced not from the misalignment mechanism, but rather the so-called ``kinetic misalignment mechanism"~\cite{Co:2019jts,Eroncel:2022vjg}, wherein the energy contained in the motion in field space is converted to axions. The observed abundance of dark matter points to heavier, less weakly-coupled axions than in the conventional misalignment case. Second, there is a PQ charge associated with the angular momentum in field space.   This is analogous to the baryon/lepton number carried by Affleck-Dine fields. In the presence of chirality- and baryon/lepton number-violating interactions, the PQ charge is converted to baryon number, a mechanism known as axiogenesis~\cite{Co:2019wyp}.  

In its minimal form, axiogenesis does not simultaneously explain the dark matter and baryon abundances; once the dark matter abundance is fixed, too little baryon asymmetry is produced.  A successful simultaneous prediction requires additional physics beyond the Standard Model~\cite{Co:2019wyp,Co:2020xlh,Harigaya:2021txz,Chakraborty:2021fkp,Co:2021qgl,Co:2022aav} to increase the efficiency of the transfer of PQ charge to baryon number.  A particularly simple solution takes advantage of lepton-number violation present when neutrino masses are explained by a Majorana mass, a scenario known as lepto-axiogenesis~\cite{Co:2020jtv,Kawamura:2021xpu}. The Majorana mass
allows transfer of the PQ charge to baryon minus lepton number $B-L$, which can eventually be converted to baryon number by weak sphalerons.

In this paper, we revisit lepto-axiogenesis, considering both Dine-Fischler-Srednicki-Zhitnitsky (DFSZ)~\cite{Zhitnitsky:1980tq,Dine:1981rt} and Kim-Shifman-Vainshtein-Zakharov (KSVZ)~\cite{Kim:1979if,Shifman:1979if}  axion models.  We focus on the case where lepto-axiogenesis is embedded in a supersymmetric model.  As we will discuss below, supersymmetric scenarios provide the most natural setting for axiogenesis.  As in the original lepto-axiogenesis proposal,  lepton-number violation is provided by the supersymmetric generalization of the $\Delta L =2$ Weinberg operator~\cite{Weinberg:1979sa} that is responsible for neutrino masses, $(L H_{u})(L H_{u})$.

In the DFSZ case, the PQ field couples directly to the Higgs fields.  Then, the non-trivial dynamics of the PQ field can impact the masses of the Higgs fields present in the Weinberg operator and therefore the transfer of the lepton asymmetry.  On the other hand, in the KSVZ case the PQ field couples to heavy quarks and not directly to the fields of the Standard Model, so the above effect is absent.
 
The precise baryon asymmetry depends on the details of the cosmological history, including the reheat temperature $T_R$ of the universe following inflation.  In our discussion, we pay attention to constraints placed on $T_R$ from, for example, avoiding disruption of Big Bang Nucleosynthesis (BBN) by superpartner decays~\cite{Kawasaki:2008qe,Kawasaki:2017bqm}. We also carefully account for whether various Yukawa interactions are in equilibrium throughout the thermal history. This can affect the efficiency of the asymmetry transfer.

In Sec.~\ref{sec:rotation}, we  review the dynamics of the rotating field and how dark matter is produced in the kinetic misalignment mechanism. We then discuss the computation of the baryon asymmetry in Sec.~\ref{sec:asymmetry}.  In comparison to Ref.~\cite{Co:2020jtv}, we take special care to account for the presence of superpartners, which impacts the rate at which the lepton asymmetry is generated.   We then present detailed results for the DFSZ model including the thermalization of the PQ field in Sec.~\ref{sec:details_DFSZ}. The outcome of our analysis is a prediction for the minimum scale of supersymmetry-breaking scalar masses.
We also find parameter space where dark matter and the baryon asymmetry may be simultaneously explained. The scalar superpartner masses are bounded from above ($\lesssim$ 300 TeV), and the axion decay constant is predicted to be approximately $10^9$ GeV.
We also discuss the possible production of a non-topological soliton, which in principle could disrupt the prediction of the baryon asymmetry. In Sec.~\ref{sec:discussion}, we summarize the results. The scale of supersymmetry breaking required by this mechanism is consistent with that indicated by the observed Higgs boson mass.

\section{Dynamics of the rotating field}
\label{sec:rotation}
We define our complex PQ field $P$ containing the axion as
\begin{equation}
    P = \frac{(f_a N_{\rm DW} +S)}{\sqrt{2}} e^{i \theta/N_{\rm DW}}.
\end{equation}
Here $f_a$ is the decay constant, $N_{\rm DW}$ is the domain wall number, $S$ is the radial direction which we call the saxion, and $\theta = a /f_a$ is the angular direction. We assume that the potential of $S$ is nearly quadratic.  This assumption allows large field values for $S$ in the early universe.  This is necessary for initiating the rotation in field space, as we will discuss below. 
A nearly quadratic potential can be naturally realized in supersymmetric theories, where the potential can be flat up to supersymmetry-breaking corrections.
This is the case for a two-field model, with superpotential and soft masses given by
\begin{align}
\label{eq:two-field}
    W = \lambda X (P \bar{P} - v_{\rm PQ}^2),~~V_{\rm soft} = m_P^2 |P|^2 + m_{\bar{P}}^2|\bar{P}|^2. 
\end{align}
Here, $X$ is a chiral multiplet whose $F$-term potential fixes the PQ-charged fields $P$ and $\bar{P}$ to $P\bar{P}= v_{\rm PQ}^2$. 
Without loss of generality, we take $|P| \gg v_{\rm PQ} \gg |\bar{P}|$ in the early universe.
We may then consider effective single-field dynamics for $P$ with a nearly quadratic potential $m_P^2|P|^2$, while $\bar{P}$ is fixed to a small field value by $\bar{P} =v_{\rm PQ}^2 / P$ and is irrelevant. $X$ will be fixed near the origin because of the large mass $\simeq \lambda P$. A nearly quadratic potential is also achieved by a one-field model with logarithmic corrections~\cite{Moxhay:1984am}
\begin{align}
\label{eq:one-field}
V(P) = \frac{1}{2} m_S^2 |P|^2 \left( {\rm ln} \frac{2 |P|^2}{f_a^2 N_{\rm DW}^2} -1 \right) ,
\end{align}
with $m_S$ the mass of the saxion. The logarithmic corrections arise from the quantum corrections due to a Yukawa coupling of $P$, which can be that with the KSVZ quark in the KSVZ model, while extra fields are required in the DFSZ model.

In the one-field model, the mass of the fermionic superpartner of the axion, the axino, is generated by one-loop corrections and is suppressed relative to the typical scale of scalar soft masses. This tends to make the axino the lightest supersymmetric particle (LSP). In the two-field case, $R$- or supersymmetry-breaking effects will induce an axino mass of order the gravitino mass, and an axino LSP is less likely. If stable, an axino LSP has the potential to be problematic because it will typically overclose the universe.%
\footnote{In general, we expect that the axino will thermalize via the supersymmetric analog of the couplings that thermalize the saxion (see  Sec.~\ref{sec:thermalization}), which would overproduce axinos. It is conceivable that saxion thermalization might not occur until temperatures near the EW scale, in which case supersymmetry-breaking masses would be non-negligible, and the axino might not thermalize even if the saxion does. However, we have checked that even in this case the suppressed freeze-in abundance of an axino LSP would be problematically large.}
The axino may decay if $R$-parity violation is introduced, or
an axino LSP could be avoided if a bino and/or Higgsino were sufficiently light. See Appendix~\ref{app:relics} for details, where we also discuss potential constraints from BBN.

In both the one- and two-field models, assuming the simplest mediation scheme of supersymmetry breaking by Planck-suppressed interactions, the saxion mass is expected to be of the same order as the soft scalar masses of the Minimal Supersymmetric Standard Model (MSSM). We will see that this curvature impacts the rotation of the axion in field space and the generation of the baryon asymmetry, and so the scalar mass may be constrained or predicted. In the one-field model, the curvature of the potential depends logarithmically on the field value of $S$. When we present results, we neglect this logarithmic dependence. So, they apply directly to the two-field case, but a small correction should be applied when interpreting results in the context of the one-field model, see Sec.~\ref{sec:one-field}.

\subsection{Initiation and evolution of rotation}
During inflation, the presence of a Hubble-induced mass term can induce a large field value for $P$~\cite{Dine:1995kz}.  Then, at these early times, operators that explicitly break the PQ symmetry of the form
\begin{equation}
\label{eqn:higherD}
W=\frac{1}{q}\frac{P^{q}}{M^{q-3}}
\end{equation}
can be enhanced, where $q$ is an integer.  Even if these operators are suppressed today so as to not spoil the solution to the strong CP problem, they can have important implications in the early universe.

The potential of $P$ is, for $S\gg f_a$,
\begin{align}
\label{eq:VP}
    V(P) = (m_S^2- c_H H^2)|P|^2 + \frac{|P|^{2q-2}}{M^{2q-6}} + \left( A\frac{P^q}{M^{q-3}} + {\rm h.c.}\right),
\end{align}
where $H$ is the Hubble expansion rate, $A$ is a constant coming from $R$-symmetry breaking, and $c_{H}$, the coefficient of the Hubble-induced mass term, is a constant expected to be ${\mathcal O}(1)$~\cite{Dine:1995kz}. Here, $m_{S}$ is a soft supersymmetry-breaking mass, which in the two-field case would be identified with $m_{P}$. We focus on gravity-mediated scenarios, where $A$ is of the same order as $m_S$. The superpotential in Eq.~(\ref{eqn:higherD}) preserves a linear combination of PQ-symmetry and $R$-symmetry, so the explicit breaking of the $U(1)$ symmetry of $P$ requires $R$-symmetry breaking.
Assuming $c_H>0$ and that the Hubble scale during inflation is larger than $m_S$, $P$ is driven to a large field value where the Hubble-induced mass term and the $|P|^{2q-2}$ term balance with each other. After inflation, $P$ follows the minimum where two terms balance with each other~\cite{Dine:1995kz,Harigaya:2015hha}. When $3H \simeq m_S$, $P$ begins oscillations. At the same time, the $A$-term provides a kick for $P$ in the angular direction, and $P$ begins to rotate. This occurs at a temperature $T_{\rm osc}$,
\begin{align}
\label{eq:Tosc}
T_{\rm osc} \simeq 
4 \times 10^9 \GeV 
\left( \frac{m_S}{\rm TeV} \right)^{ \scalebox{1.01}{$\frac{1}{4}$} }
\left( \frac{T_R}{10^9 \GeV} \right)^{ \scalebox{1.01}{$\frac{1}{2}$} }
\left( \frac{g_{\rm MSSM}}{g_*(T_{\rm osc})} \right)^{ \scalebox{1.01}{$\frac{1}{8}$} } \ \ {\rm for} \ \ T_R < T_{\rm osc} ,
\end{align}
where $T_R$ is the reheat temperature after inflation and $g_*$ denotes the number of  relativistic degrees of freedom in the bath with a full MSSM value of $g_{\rm MSSM} = 228.75$. We assume that inflationary reheating proceeds via perturbative inflaton decay, and thus the scale factor $R$ obeys $R^3 \propto T^{-8}$ during reheating~\cite{Kolb:1990vq}. The PQ charge density associated with the rotation is 
\begin{equation}
\label{eq:chargedensity}
    n_{\theta} =  \frac{i}{N_{\rm DW}} \left(\dot{P}P^*-  \dot{P}^* P \right) 
    = - \dot{\theta} \left(f_a + \frac{S}{N_{\rm DW}}\right)^2.
\end{equation}
We normalize the charge density so that it coincides with $-\dot{\theta} f_a^2$ for $S=0$. The charge density normalized by the entropy density for $T_R < T_{\rm osc}$ can be computed as follows. The inflaton energy density $\rho_{\rm inf}$ scales  in the same way as $n_\theta$ after the initiation of the rotation (as $R^{-3}$), so $n_\theta / \rho_{\rm inf}$ remains constant until $T = T_R$. The result is 
\begin{align}
    Y_{\theta} \equiv \frac{n_\theta}{s} = \left.\frac{n_\theta}{\rho_{\rm inf}} \right|_{T_{\rm osc}} \hspace{-0.5cm} \times \left. \frac{\rho_{\rm inf}}{s} \right|_{T_R} \simeq 
    10 
    \left( \frac{3}{N_{\rm DW}} \right)
    \left( \frac{A}{m_{S}} \right)
    \left( \frac{\rm TeV}{m_{S}} \right)
    \left(\frac{T_R}{10^{9} \GeV} \right) 
    \left(\frac{S(T_{\rm osc})}{10^{16} \GeV} \right)^2 .
\end{align}
The $A/m_S$ is the ratio of the potential gradient in the angular and radial directions at $T_{\rm osc}$.

There is also energy density $\rho_{S}$ stored in the oscillations of the radial mode $S$. Whether $\rho_{S}$ is of importance depends on the cosmological history and at what temperature $T_{\rm th}$ this mode is thermalized. The saxion energy density may come to dominate the energy of the universe if this thermalization is late. We comment on this scenario further at the end of this section. Following thermalization, the motion of the PQ field becomes circular due to PQ charge conservation: the radial mode dissipates, but much of the axial motion remains---for while part of the charge can be transferred into a charge asymmetry of particles in the thermal bath, it is free-energetically favored to keep almost all of the charge in the rotation~\cite{Co:2019wyp,Domcke:2022wpb}. 
 The field will rotate around the body of the potential, with the radial direction eventually settling down to its minimum $N_{\rm DW} f_a$. The energy in rotation $\rho_\theta$, accounting for both the potential and kinetic energy, is given as $-\dot{\theta} n_{\theta}$.  Before the radial direction $S$ reaches its  minimum, which occurs at a temperature denoted by $T_S$, $\dot{\theta}$ is a constant, and conservation of the PQ charge implies the energy density of the rotation scales as matter, $\rho_\theta \propto R^{-3}$. 
For $T < T_S$, the scaling of the rotational energy density resembles that of kination, $\rho_\theta \propto R^{-6}$. This scaling can be derived by noting that conservation of charge $n_\theta R^3$ at constant radial field value implies $\dot{\theta} \propto R^{-3}$. 
When the saxion settles to its minimum  at $T = T_S$, we know both $\left| \dot\theta \right| \simeq N_{\rm DW} m_S$ and the PQ yield $Y_\theta = - \dot\theta f_a^2 / s$, so we can derive
\begin{equation}
\label{eq:TS_general}
T_S \simeq 1.4 \times 10^6 \GeV 
    \left( \frac{100}{Y_\theta} \right)^{ \scalebox{1.01}{$\frac{1}{3}$} }
    \left( \frac{f_a}{10^9 \GeV} \right)^{ \scalebox{1.01}{$\frac{2}{3}$} }
    \left( \frac{m_S}{10 \TeV} \right)^{ \scalebox{1.01}{$\frac{1}{3}$} }
    \left( \frac{N_{\rm DW}}{3} \right)^{ \scalebox{1.01}{$\frac{1}{3}$} }
    \left( \frac{g_{\rm MSSM}}{g_*(T_S)} \right)^{ \scalebox{1.01}{$\frac{1}{3}$} } .
\end{equation}

\begin{figure}
\includegraphics[width=\linewidth]{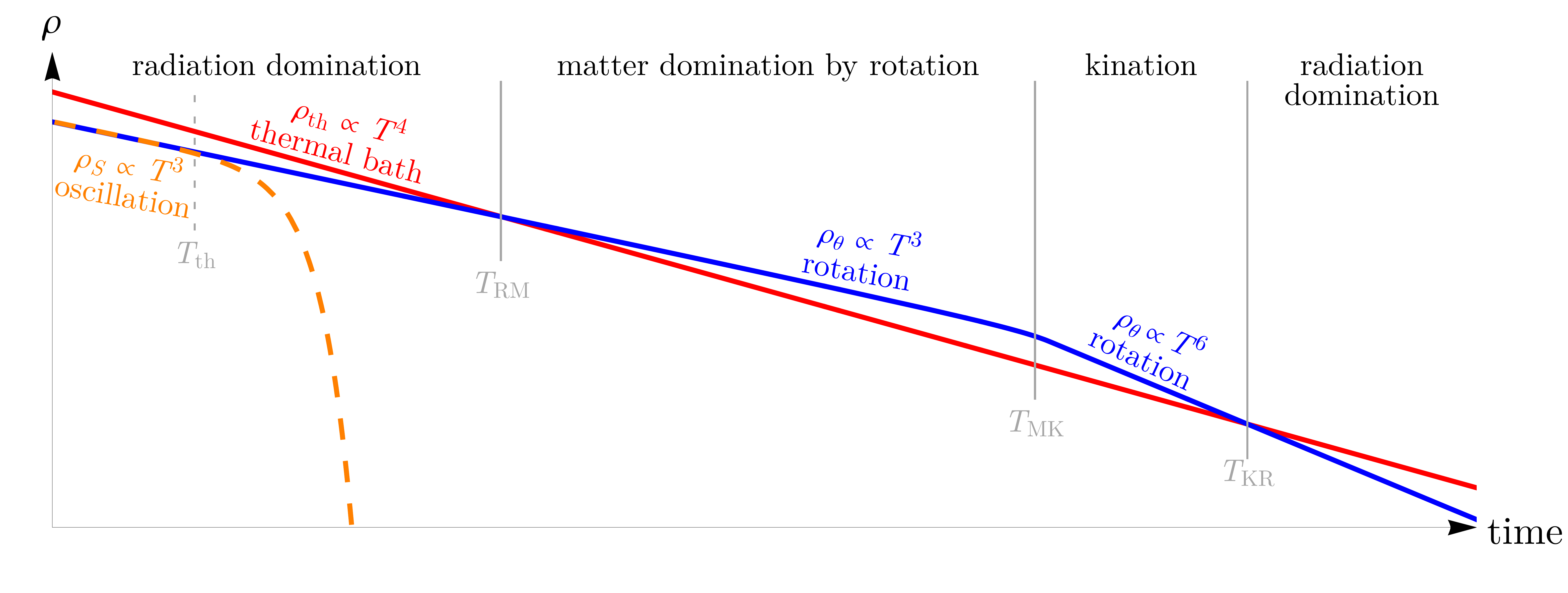} 
\caption{An example evolution of energy densities as a function time for radiation (red), oscillations in the radial direction (orange), and rotations of the PQ-breaking field (blue). Relevant temperatures are labeled in gray and corresponding cosmological eras are labeled in black.}
\label{fig:schematic_rho}
\end{figure}

If the energy of the rotation dominates the energy of radiation at this time and if the saxion has already undergone thermalization  (i.e. $T_{\rm th} > T_S)$, then this $T_S$ is also the temperature $T_{\rm MK}$ at which the universe transitions from a matter-dominated to a kination-dominated one. This history is illustrated in Fig.~\ref{fig:schematic_rho}. We denote the temperature at which the universe transitions from radiation domination to matter domination as $T_{\rm RM}$.  We emphasize that the matter domination we refer to here is domination by an energy density of rotation that scales as matter, not ordinary matter. This occurs at temperature
\begin{equation}
\label{eq:TRM_general}
T_{\rm RM} = \frac{4}{3} N_{\rm DW} m_S Y_{\theta} 
    =  4 \times10^6 \GeV
    \left(\frac{Y_\theta}{100}\right)
    \left(\frac{m_S}{10 \TeV}\right)
    \left( \frac{N_{\rm DW}}{3} \right) .
\end{equation}
This expression is general as long as no entropy is produced after $T_{\rm RM}$, and $Y_\theta$ refers to the charge yield evaluated at $T_{\rm RM}$. In particular, this result applies whether or not there was an era where the saxion came to dominate the energy density of the universe prior to $T_{\rm RM}$. 
The kination-dominated era ends by the redshift of $\rho_\theta = \dot\theta^2 f_a^2 / 2 = n_\theta^2/(2f_a^2)$ at temperature
\begin{equation}
\label{eq:TKR_general}
T_{\rm KR} 
    = \left(\frac{135}{4 \pi^2 g_*}\right)^{ \scalebox{1.01}{$\frac{1}{2}$} }
    \frac{f_a}{Y_\theta}
    \simeq 1.2 \times 10^6 \GeV 
    \left( \frac{100}{Y_\theta} \right)
    \left( \frac{f_a}{10^9 \GeV} \right)
    \left( \frac{g_{\rm MSSM}}{g_*(T_{\rm KR})} \right)^{ \scalebox{1.01}{$\frac{1}{2}$} } .
\end{equation}
A matter-dominated era followed by a kination-dominated one would modify the primordial gravitational wave spectrum in a way that potentially provides a unique signal~\cite{Co:2021lkc, Gouttenoire:2021wzu, Gouttenoire:2021jhk}. 

It is also possible that the energy density due to rotation remains subdominant to the thermal bath.  As we will see, in this case the temperature $T_S$ where the saxion reaches its minimum is still of significance for determination of the baryon asymmetry, as it marks the time where $\dot{\theta}$ changes its scaling.  However, this temperature would no longer mark the onset of a kination era because radiation remains dominant. 

We now comment on the possibility that saxion thermalization is late so that the saxion comes to dominate the energy density of the universe. We define the ratio $r$ of the axion rotation to the saxion oscillation energy densities, which is in turn determined by the ratio of the  potential gradients between the angular and radial modes,
\begin{equation}
\label{eq:r}
    r \equiv \frac{\rho_\theta}{\rho_S} \simeq \frac{A}{m_S}.
\end{equation}
This $r$ is inversely related to the ellipticity of the initial motion and $r = 1$ corresponds to nearly circular rotations. Here it is assumed that the angular direction is not accidentally close to the minimum of the potential in Eq.~(\ref{eq:VP}); otherwise $r$ becomes smaller than $A/m_S$.  The thermal bath created from the saxion is at a temperature $T_{\rm th}$ upon completion of thermalization. 
This fact allows us to predict 
$T_{\rm RM}$ because $\rho_S \times r = \rho_\theta$ should hold at $T_{\rm th}$. This gives $\frac{\pi^2}{30} g_* T_{\rm th}^4 \times r =  \frac{\pi^2}{30} g_* T_{\rm RM}^4 (T_{\rm th}/T_{\rm RM})^3$, or equivalently,
\begin{equation}
\label{eq:TRM_r_Tth}
T_{\rm RM} = r T_{\rm th} \quad \quad \textrm{(for saxion domination)}.  \end{equation}

\subsection{Kinetic misalignment and production of axion dark matter}
In the conventional misalignment mechanism, the value of the axion field is initially frozen by Hubble friction. But once $3H < m_a(T)$, the axion field begins to oscillate around the minimum of its potential. In the axiogenesis framework, on the other hand, the axion is not frozen, rather the PQ field is already rotating with high velocity.  This qualitatively changes the dark matter production story.  The kinetic misalignment mechanism (KMM) occurs when the kinetic energy of the axion field is greater than the potential energy.  
The KMM delays the oscillations around the minimum of the potential. In fact, via parametric resonance~\cite{Dolgov:1989us, Traschen:1990sw, Kofman:1994rk, Shtanov:1994ce, Kofman:1997yn}, the axion rotation fragments into fluctuations around the QCD confinement scale in this scenario; this effect was noted in the context of axion monodromy in Refs.~\cite{Jaeckel:2016qjp, Berges:2019dgr, Fonseca:2019ypl,Morgante:2021bks}. The axions generated in this way are relativistic upon production with momenta determined by the resonance peak $k_a \simeq \dot\theta/2$. With this, we can estimate the yield of the axion as~\cite{Co:2021rhi,Eroncel:2022vjg}
\begin{equation}
Y_{a} \simeq \frac{\rho_\theta}{s \dot{\theta}/2} =\frac{\dot{\theta} f_a^2}{s}.
\end{equation}
The axion yield is equal to the charge yield associated with rotation $Y_\theta$. 
The present day axion energy density is given by $m_{a} Y_{a}$.  Setting this equal to the observed dark matter abundance $\rho_{\rm DM}/s \simeq 0.44$ eV, allows us to determine the required charge yield, 
\begin{equation}
\label{eq:Ytheta_KMM}
    Y_{\theta,\,{\rm KMM}} \simeq 70 \left( \frac{f_a}{10^9 \GeV} \right),
\end{equation}
the required temperature $T_S$ at which the axion settles to its minimum using Eq.~(\ref{eq:TS_general}),
\begin{equation}
\label{eq:TS_KMM}
T_{S,\,{\rm KMM}} \simeq 1.6 \times 10^6 \GeV 
\left( \frac{N_{\rm DW}}{3} \right)^{ \scalebox{1.01}{$\frac{1}{3}$} }
\left( \frac{m_S}{10 \TeV} \right)^{ \scalebox{1.01}{$\frac{1}{3}$} } 
\left( \frac{f_a}{10^9 \GeV} \right)^{ \scalebox{1.01}{$\frac{1}{3}$} }
\left(\frac{g_{\rm MSSM}}{g_*(T_S)}\right)^{ \scalebox{1.01}{$\frac{1}{3}$} },
\end{equation}
and the temperature at the transition from radiation to matter domination using Eq.~(\ref{eq:TRM_general})
\begin{equation}
\label{eq:TRM_KMM}
T_{\rm RM,\,KMM} \simeq 2.9 \times 10^6 \GeV 
    \left(\frac{m_S}{10 \TeV}\right)
    \left(\frac{f_a}{10^9 \GeV}\right)
    \left( \frac{N_{\rm DW}}{3} \right).
\end{equation}

\section{Computation of the baryon asymmetry}
\label{sec:asymmetry}

In this section, we describe the computation of the baryon asymmetry in lepto-axiogenesis.

\subsection{Basics of lepto-axiogenesis}

The axion rotation couples to the thermal bath via the gluon in the KSVZ theory and via the Higgs fields in the DFSZ theory. The PQ-charge is transferred to a particle-antiparticle asymmetry of particles in the thermal bath, and in equilibrium the charge asymmetry in the bath is of the order of $\dot{\theta} T^2$~\cite{Co:2020jtv}.  The total $B-L$ charge vanishes in the absence of $B-L$ violation, and the baryon asymmetry is fixed at the electroweak phase transition~\cite{Co:2019wyp}.

The $B-L$ symmetry is broken if the observed non-zero neutrino mass is explained by a Majorana mass term.
We consider the Majorana mass given by the Weinberg operator~\cite{Weinberg:1979sa}, whose supersymmetrization is given by the superpotential
\begin{equation}
\label{eq:Wnu}
    W_{\nu}=c_{ij} \frac{(L_{i}  H_u) (L_{j} H_u) }{2\Lambda},
\end{equation}
which can be UV-completed by the seesaw mechanism~\cite{Yanagida:1979as,GellMann:1980vs,Minkowski:1977sc,Mohapatra:1979ia}.
This operator gives rise to neutrino mass terms, in terms of the vacuum expectation value $v_{H_u}$ of the up-type Higgs field with $v_{H_u}^2 + v_{H_{d}}^2 \simeq (174$ GeV$)^2$,
\begin{equation}
\label{eq:mnu}
    m_{\nu}^{ij}= \frac{c_{ij} v_{H_u}^2}{\Lambda}, 
\end{equation}
related to eigenvalues through the Pontecorvo–Maki–Nakagawa–Sakata (PMNS) matrix 
\begin{equation}
\label{eq:PMNS_relation}
    U_{\rm PMNS}^T m_{\nu} U_{\rm PMNS} = {\rm diag} (m_1,m_2,m_3).
\end{equation}

The Weinberg operator will transfer the particle-antiparticle asymmetry of $L_i$ and $H_u$ to $B-L$ through  scattering between the lepton and Higgs fields (and their superpartners) in the bath. The scattering is not in equilibrium for temperatures $T\lsim 10^{12} \GeV \times (0.03 {\rm\; eV}/ m_{\nu})^2$, so the transfer of the PQ charge to $B-L$ is suppressed by a factor of $\Gamma_L/H$ with $\Gamma_L$ the lepton-number violating rate. That is, $B-L$ asymmetry is produced by a  ``freeze-in" process~\cite{Co:2019wyp}. This $B-L$ asymmetry is ultimately further processed by electroweak sphalerons to give a baryon asymmetry $n_{B} \simeq (28/79)n_{B-L}$~\cite{Harvey:1990qw}. 

\begin{figure}
\includegraphics[width=0.49\linewidth]{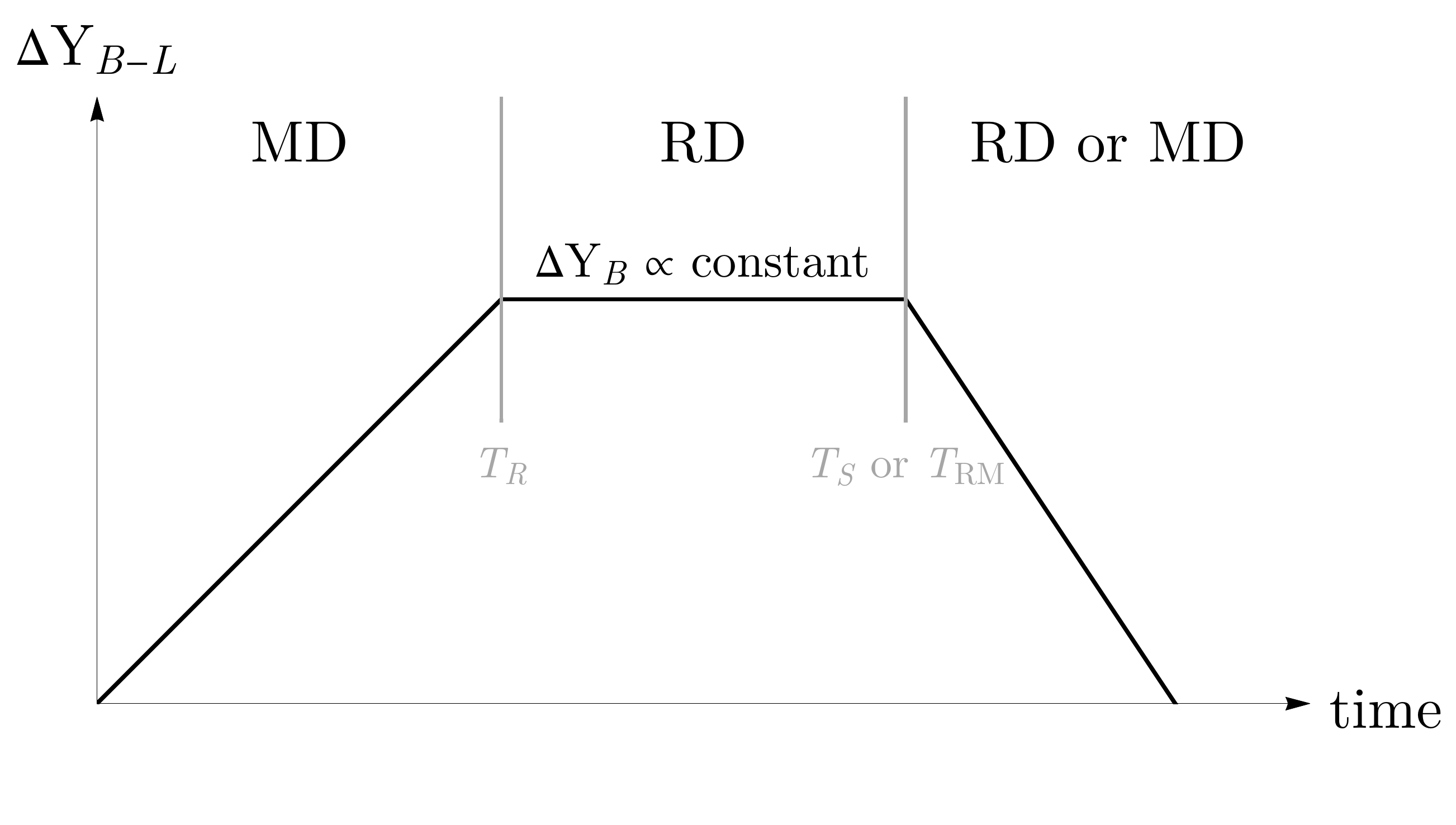}
\includegraphics[width=0.49\linewidth]{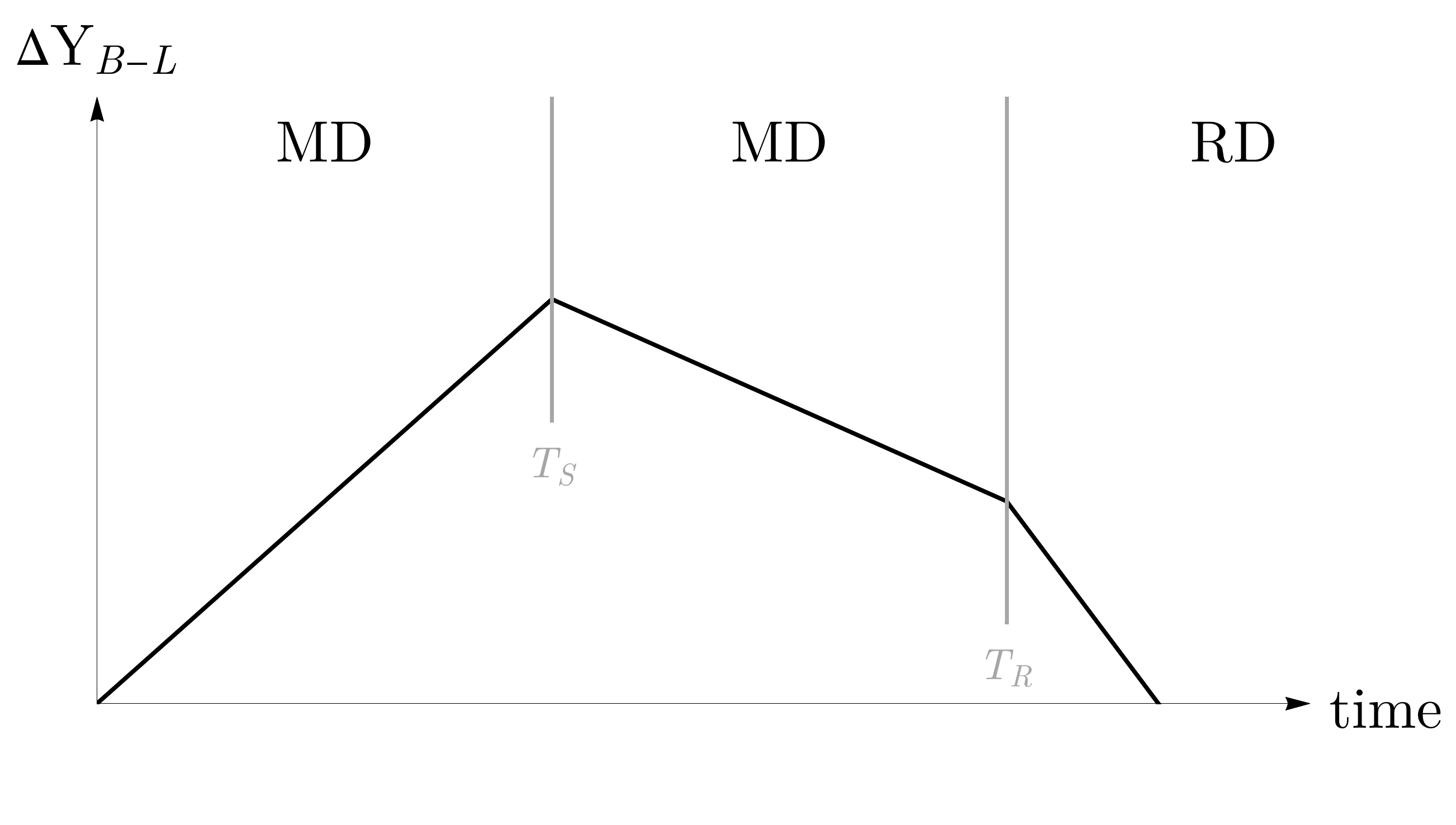}
\caption{The baryon minus lepton asymmetry produced per Hubble time $\Delta Y_{B-L}$ as a function of time in log-log scales during radiation-dominated and matter-dominated eras. Relevant temperatures are labeled in gray and corresponding cosmological eras are labeled in black.}
\label{fig:schematic_YB}
\end{figure}

To calculate the rate of $B-L$ asymmetry production, we must account for all scattering processes due to the operator in Eq.~(\ref{eq:Wnu}). Each contribution takes the form 
\begin{align}
    \dot{n}_{B-L} \supset  2 \int d\Pi_a d\Pi_b d\Pi_c d\Pi_{d}   e^{- \frac{E_a + E_b}{T} } \left(e^{\frac{\mu_a}{T} + \frac{\mu_b}{T}}- e^{\frac{\mu_c}{T} +\frac{\mu_d}{T}} \right)     (2\pi)^4 \delta^{(4)}(p_{a}+p_{b}-p_{c}-p_{d}) 
    |\mathcal{M}|^2,
\end{align}
where $a,b,c,$ and $d$ are field labels, momenta $\{p_{a}$, $p_{b}\}$ are incoming and $\{p_{c}$, $p_{d}\}$ are outgoing, and $d\Pi_X \equiv \frac{1}{(2\pi)^3}\frac{d^3p_{X}}{2E_X}$ with $E_X$  the energy of field $X$. The initial factor of two is because the processes have  $\Delta L = 2$.  Here, $\mu_{X}$ is the chemical potential of field $X$. See the Appendix of Ref.~\cite{Co:2021qgl} for  discussion in a similar context. For chemical potentials much smaller than temperature, the sum over all scattering processes gives
\begin{equation}
    \label{eq:nDotB-L_General}
    \dot{n}_{B-L} = \frac{72}{\pi^5}T^5\sum_i\sum_j\left | \frac{c_{ij}}{2\Lambda}\right |^2\left [\frac{1}{2}(\mu_{\ell_i}+\mu_{\ell_j})+\mu_{\widetilde{H}_u}+\mu_{\lambda}\right ],
\end{equation}
with $i$ and $j$ running over the three generations. We have assumed that processes involving scattering between Higgsinos $\widetilde{H}_{u}$, gauginos $\lambda$, and Higgs bosons are in equilibrium (and similarly for sleptons), which is typically the case.
We have included scattering processes involving superpartners, which were neglected in Ref.~\cite{Co:2020jtv}. We also go beyond the one-generation approximation used there; this has a smaller effect. We have also assumed that the masses of $\ell$ and $H_u$ are smaller than $T$.
As we will discuss in Sec.~\ref{sec:DFSZ}, this is not true for the DFSZ model for sufficiently high temperatures, 
since a large field value of $S$ may impart a mass $>T$ to $H_u$ and $H_d$.

The coefficient of the Weinberg operator can be related to the neutrino masses and mixings as in Eqs.~\eqref{eq:mnu} and \eqref{eq:PMNS_relation}, so the production rate of the $B-L$ asymmetry may be recast as 
\begin{align}
\label{eq:nB_dot}
    \dot{n}_{B-L} = \sum C_i(T) m_{\nu_{i}}^2 \frac{\dot\theta T^5}{v_{H_u}^4},
\end{align}
where $m_{\nu_{i}}$ is the $i^{\rm th}$ neutrino mass eigenvalue. 
The coefficients $C_i(T)$ (which are in general a function of PMNS mixing) are determined by calculating the relevant chemical potentials.  Their values depend on what interactions are in equilibrium at a given temperature as well as the choice of axion model.
Results for $C_{i}(T)$,  generally of order $10^{-2}$--$10^{-3}$, and the details of their computation are given in Appendix~\ref{app:cB}.
The yield of the $B-L$ asymmetry produced per Hubble time may then be estimated as
\begin{align}
\label{eq:DeltaY_B-L}
\Delta Y_{B-L} \simeq \frac{\dot{n}_{B-L}}{s H} = 
\left(\frac{45}{2 \pi^2 g_*}\right) \sum C_i(T) m_{\nu_{i}}^2 \frac{\langle \dot\theta \rangle T^2}{H v_{H_u}^4} ,
\end{align}
with $\langle \dot\theta \rangle$ the time average of $\dot\theta$.

During radiation domination, $H \propto T^2$, so for $T > T_S$, where $\langle \dot\theta \rangle \simeq N_{\rm DW} m_S $ is a constant~\cite{Co:2020jtv},  the temperature dependence of $\Delta Y_{B-L}$ in Eq.~(\ref{eq:DeltaY_B-L}) is especially simple.  It is independent of the temperature, except for a small implicit dependence through the determination of $C_{i}$. On the other hand, $\Delta Y_{B-L}$ decreases with temperature after $T < T_S$ because then $\dot\theta \propto T^3$. The scaling of $\Delta Y_{B-L}$ during different epochs is summarized in Table~\ref{tab:YB} in Appendix~\ref{app:YB_scaling}, and is illustrated in the left panel of Fig.~\ref{fig:schematic_YB}. 

An era of constant $\Delta Y_{B-L}$ indicates a logarithmic enhancement in the integrated production of $Y_{B-L}$. For the case of a long radiation-dominated era, we derive the expression of the final asymmetry $Y_{B-L}$ by integrating $\dot{n}_{B-L}/s$ over time from $T_i$ to $T_f$ using Eq.~(\ref{eq:nB_dot}), 
\begin{align}
    Y_{B-L} = \int \frac{\dot{n}_{B-L}}{s} dt = - \int_{T_i}^{T_f} \frac{\dot{n}_{B-L}}{s H T} dT, \hspace{1 cm} H = \sqrt{\frac{\pi^2 g_*}{90}} \frac{T^2}{\MPl}.
\end{align}
We obtain an analytic result
\begin{align}
\label{eq:YB_RD}
    Y_{B-L} & = \left(\frac{90}{\pi^2 g_*}\right)^{\frac{1}{2}} \left(\frac{45}{2 \pi^2 g_*}\right)
    \sum C_j m_{\nu_{j}}^2 
    \frac{\MPl N_{\rm DW} m_S}{v_{H_u}^4} \ln \left( \frac{T_i}{T_f} \right) \ \ \ {\rm for} \ \ \ T_i \gg T_f ,
\end{align}
where $T_i$ and $T_f$ mark the initial and final temperatures of the era when $\Delta Y_{B-L}$ is a constant. Reproducing the observed baryon asymmetry, $Y_B^{\rm obs} = 8.7 \times 10^{-11}$~\cite{Aghanim:2018eyx}, requires a saxion~mass
\begin{align}
\label{eq:logbenchmark}
    m_S \simeq 135 \TeV \ N_{\rm DW}^{-1}  \left(\frac{g_*}{g_{\rm MSSM}} \right)^{ \scalebox{1.01}{$\frac{3}{2}$} } \left( \frac{0.01 \times 0.005~{\rm eV}^2}{\sum C_j m_{\nu_{j}}^2} \right) \left( \frac{7}{\ln \left( \frac{T_i}{T_f} \right)} \right) .
\end{align}

For $T > T_R$ (or for $T < T_{\rm RM}$), the universe is not radiation-dominated, and production becomes IR (or UV)-dominated. Again, this is summarized in Table~\ref{tab:YB} and illustrated in the left panel of Fig.~\ref{fig:schematic_YB}.

 If the reheat temperature is lower than the temperature where the saxion settles to its minimum, i.e., $T_R < T_S$, then Eq.~(\ref{eq:YB_RD}) does not hold because $\dot\theta$ is never constant during the radiation domination era, instead $\dot{\theta} \propto T^{3}$. In this case, $B-L$ production peaks at $T_S$, which is illustrated in the right panel of Fig.~\ref{fig:schematic_YB}. Then, the asymmetry may be obtained by first computing the redshift-invariant quantity $\dot{n}_{B-L}/( H \rho_{\rm inf})$, with the inflaton energy density denoted by $\rho_{\rm inf}$.  This quantity is readily evaluated at $T_S$, see Eq.~\eqref{eq:nB_dot}, recalling that $\left| \dot{\theta} \right| \simeq N_{\rm DW} m_{S}$ at this time. Then we can normalize the quantity to $n_{B-L}/s$ at $T_R$:
\begin{equation}
\label{eq:YB_MD}
    Y_{B-L} = \left. \frac{{\dot {n}}_{B-L} }{H \rho_{\rm inf} }\right|_{T = T_S} \hspace{-0.5cm} \times \left. \frac{\rho_{\rm inf}}{s} \right|_{T = T_R} = \left(\frac{90}{\pi^2 g_*}\right)^{\frac{1}{2}} \left(\frac{45}{2 \pi^2 g_*}\right)
    \sum C_i m_{\nu_{i}}^2 
    \frac{\MPl N_{\rm DW} m_S}{v_{H_u}^4} \left( \frac{T_R}{T_S} \right)^7 ,
\end{equation}
where we have assumed inflationary reheating by perturbative decays of the inflaton so $H(T) = H(T_R) \times (T/T_R)^4$ for $T > T_R$. 

The result depends on the choice of the neutrino spectrum.  
We will show results for a normal hierarchy (NH), or inverted hierarchy (IH), assuming the lowest mass eigenvalue is negligible, so the overall mass scale is given by the mass differences determined by oscillations. 
Even if we saturate the upper bound $\sum m_{\nu}< 0.12$ eV from the Cosmic Microwave Background along with data from Baryon Acoustic Oscillations~\cite{Aghanim:2018eyx}, the predictions for this case are not so different from those of the inverted hierarchy case. Precisely speaking, the values of $\sum C_i m_{\nu_i}^2$ with the upper bound saturated are 8\% (normal hierarchy) and 16\% (inverted hierarchy) larger than that for the inverted hierarchy with a negligible lightest neutrino mass. 

\subsection{KSVZ}
\label{sec:KSVZ}
The KSVZ model includes a coupling
\begin{equation}
    W_{\rm KSVZ}=\lambda_{\Psi} P \bar{\Psi} \Psi
\end{equation}
with $\Psi$ a new colored quark charged under the PQ symmetry such that charges $PQ_{\Psi}+PQ_{\bar{\Psi}}+PQ_P=0$.  This coupling is the origin of the mixed PQ-QCD anomaly which allows the axion to solve the strong CP problem.   The $\lambda_{\Psi}$ coupling plays an important role in the thermalization of the rotation.

The KSVZ model was carefully examined in Refs.~\cite{Co:2020jtv,Kawamura:2021xpu}.
We refer readers to these references for details, including the thermalization of the rotating PQ field. Here we focus on the implications of a factor of six enhancement in the baryon asymmetry production efficiency compared to Ref.~\cite{Co:2020jtv}.  This factor of six is the result of supersymmetrizing the Weinberg operator in Eq.~(\ref{eq:Wnu}), allowing lepton asymmetry production from scattering involving superpartners.  This factor is independent of the UV completion of the axion and applies to the DFSZ case as well.  The existence of superpartners in the bath also changes the efficiency of baryon asymmetry production by affecting the equilibrium Boltzmann equations and conserved quantities given in Appendix~\ref{app:cB}.

As a benchmark, the observed baryon asymmetry is reproduced for a saxion mass
\begin{align}
    m_S \simeq 190 \TeV \ \left(\frac{1}{N_{\rm DW}} \right)  \left(\frac{g_*}{g_{\rm MSSM}} \right)^{ \scalebox{1.01}{$\frac{3}{2}$} } \left( \frac{0.0106 \times 0.005~{\rm eV}^2}{\sum C_j m_{\nu_{j}}^2} \right) \left( \frac{5.3}{\ln \left( \frac{T_i}{T_f} \right)} \right) \quad {\rm (KSVZ)},
\end{align}
where we use $C_i = 0.0106$ based on Table~\ref{tab:KSVZ_C_i}. In the determination of  $C_{i}$, we have gone beyond the one-generation approximation of Ref.~\cite{Co:2020jtv}. This value of $C_{i} = 0.0106$ applies when the anomaly coefficients for the weak and strong interaction are identical $c_{W}=c_{g}(=1)$ and when all Yukawa couplings are in equilibrium. To get the benchmark value 5.3, we take $T_i = T_R = 2 \times 10^9 \GeV$ and $T_f = T_{\rm RM} = 10^7 \GeV$.

\subsection{DFSZ}
\label{sec:DFSZ}

In the DFSZ case, the effective $\mu$-term depends upon the value of the scalar field.  This effective $\mu$-terms arises from the superpotential coupling
\begin{equation}
\label{eq:Wmu}
    W_{\mu} = \lambda \frac{P^n H_{u} H_{d}}{M^{n-1}}.
\end{equation}
The idea of relating the $\mu$-term to the scale of Peccei-Quinn symmetry is sometimes known as the Kim-Nilles mechanism, which was originally explored for the $n=2$ case in~\cite{Kim:1983dt}.  Because the value of $P$ changes during the universe's history, so too will the masses of the Higgs fields. As discussed below, this can impact the way in which the lepton asymmetry is transferred to the bath via Eq.~\eqref{eq:Wnu}.

The superpotential of Eq.~\eqref{eq:Wmu} gives a temperature dependent $\mu(T) = \lambda P^{n}/M^{n-1}$.  At temperatures before $P$ settles to its minimum, this scales as $R^{-3n/2}$, which is proportional to $T^{3n/2}$ during radiation domination. We define a temperature $T_\mu$ at which the temperature and the effective $\mu(T)$ are equal,
\begin{align}
\label{eq:T_mu}
    T_\mu = \left( \frac{T_S^{\frac{3n}{2}}}{\mu} \right)^{ \frac{1}{\frac{3n}{2}-1}} =
    \begin{cases}
    10^9 \GeV \left( \frac{T_S}{100 \TeV} \right)^3
    \left( \frac{\rm TeV}{\mu} \right)^2 & {\rm for} \,\, n = 1 \\
    10^6 \GeV \left( \frac{T_S}{100 \TeV} \right)^{ \scalebox{0.9}{$\frac{3}{2}$} }
    \left( \frac{\rm TeV}{\mu} \right)^{ \scalebox{0.9}{$\frac{1}{2}$} } & {\rm for} \,\, n = 2
    \end{cases}
    ,
\end{align}
where $\mu$ is the present-day value to which $\mu(T)$ settles for temperatures below $T_S$. For temperatures $T>T_{\mu}$, scattering via the Weinberg operator is ineffective as the lepton-number violation is limited to even higher dimension operators generated by integrating out the Higgs superfields.  

So, the earliest temperature at which the chiral asymmetry may be effectively transferred to $B-L$ is  $T_{\mu}$.  However, the reheat temperature $T_R$ is sometimes limited by BBN constraints~\cite{Kawasaki:2008qe,Kawasaki:2017bqm} to values that are smaller than $T_{\mu}$. In this case, the earliest temperature relevant for transfer to $B-L$ is $T_R$. Based on Eq.~(\ref{eq:logbenchmark}) and $N_{\rm DW} = 3 n$, a benchmark prediction of the saxion mass is
\begin{align}
    m_S \simeq 39 \TeV \times \left( \frac{1}{n} \right)  \left(\frac{g_*}{g_{\rm MSSM}} \right)^{ \scalebox{1.01}{$\frac{3}{2}$} } \left( \frac{0.0153 \times 0.005~{\rm eV}^2}{\sum C_j m_{\nu_{j}}^2} \right) \left( \frac{5.3}{\ln \left( \frac{T_i}{T_f} \right)} \right) \quad {\rm (DFSZ)}.
\end{align}
To get the benchmark value of 5.3 in the parentheses, we have taken $T_i = T_R = 2\times 10^9 \GeV$ and $T_f = T_S = 10^7 \GeV$. The benchmark value of $C_{i}$ corresponds to the case where all Yukawa interactions and the gaugino mass are in equilibrium; see Table \ref{tab:DFSZ_C_i} in the Appendix.

The saxion mass, which we assume to be of the same order as the soft scalar masses of the MSSM, may be $\mathcal{O}(10)$ TeV; this is consistent with the observed Higgs boson mass if the ratio of the Higgs field vacuum expectation values $\tan{\beta} \gg 1$. Larger $m_{S}$ is also possible, which could reproduce the Higgs boson mass for more modest values of $\tan \beta$.

\section{Detailed analysis of the DFSZ model}
\label{sec:details_DFSZ}

We now analyze the DFSZ model in detail. We discuss the thermalization of $P$ via the coupling with the Higgs superfields in Eq.~(\ref{eq:Wmu}). We then show the allowed parameter space, determining both the minimum values of $m_{S}$ consistent with the generation of the baryon asymmetry and also the values of $m_{S}$ predicted by the production of both the baryon asymmetry and the dark matter abundance. We analyze the cases where the asymmetry is generated during reheating or the subsequent radiation-dominated era and the case where the saxion eventually dominates the energy. We discuss complications that may arise from the possible fragmentation of the rotation into Q-balls and explain how they can be avoided.
\subsection{Thermalization}
\label{sec:thermalization}

If the saxion does not thermalize sufficiently early, it will come to dominate the energy density of the universe.  In this case, when it ultimately decays, it will produce entropy which can dilute the baryon asymmetry.

We assume that the dominant interactions of the saxion are  via the coupling in the superpotential that gives the effective $\mu$-term, Eq.~(\ref{eq:Wmu}). Then the saxion can be thermalized via its interaction with the Higgsino at a rate given by
\begin{equation}
\label{eq:Gamma_S_higgsino}
\Gamma_{S\widetilde{H}\widetilde{H}} \simeq 0.1 \frac{\mu^2 (T+m_S)}{S^2} \left(\frac{S}{N_{\rm DW} f_a}\right)^{2n} ,
\end{equation}
where the term with $T$ or $m_S$ corresponds to the scattering or the decay rate, respectively.  

For $n = 1$, the rate is independent of the evolution of the saxion field value $S$.  The thermalization temperature is found by setting this rate equal to $3H$, and is given by
\begin{equation}
\label{eq:Tth}
T_{\rm th} \simeq \left(\frac{90}{\pi^2 g_*}\right)^{\frac{1}{2}} \frac{\mu^2 M_{\rm Pl}}{30 N_{\rm DW}^2 f_a^2} 
\simeq 200 \TeV \left(\frac{\mu}{\rm TeV}\right)^2
\left(\frac{10^8 \GeV}{f_a}\right)^2 \left(\frac{3}{N_{\rm DW}}\right)^2 \left( \frac{g_{\rm MSSM}}{g_*(T_{\rm th})} \right)^{ \scalebox{1.01}{$\frac{1}{2}$} },
\end{equation}
which is valid for $T_{\rm th} \gg m_S$, often the case for parameters of interest. 
The above expression assumes radiation domination. If the reheat temperature is below this $T_{\rm th}$, thermalization  instead occurs during the period of  inflationary reheating, and the actual thermalization temperature becomes lower than that in Eq.~(\ref{eq:Tth}) (but above $T_R$) due to an enhanced Hubble rate with respect to that for radiation domination. However, thermalization of the saxion during inflationary reheating will not create more entropy than already created by the inflaton.  So, the precise value $T_{\rm th}$ will  be irrelevant; instead, the value of $T_{R}$ will be important for analysis of the baryon asymmetry.

For $n > 1$ and $S > N_{\rm DW} f_a$, the thermalization rate depends on $S$, so we need the scaling of $S$ with temperature.  Conservation of $S$ number implies that $S$ scales as $R^{-3/2}$. During radiation domination, $R \propto T^{-1}$, so $\Gamma_{S\widetilde{H}\widetilde{H}} \propto (T+m_S)S^{2n-2} \propto (T+m_S)T^{3n-3}$ increases with increasing $T$ faster than a radiation-dominated $H \propto T^2$ does for any $n > 1$. Therefore, for $n>1$ the saxion may thermalize at a high temperature but then decouple from the thermal bath when $\Gamma_{S\widetilde{H}\widetilde{H}}$ drops below the Hubble expansion rate. However, there is a maximum  temperature at which the saxion can thermalize via Higgsino scattering, namely  $T \sim T_\mu$. Above this temperature, Higgsinos are out of equilibrium because their mass exceeds the temperature. To test whether thermalization occurs at this point, we equate $\left. \Gamma_{S \tilde{H} \tilde{H}} \right|_{T= T_{\mu}}= 3 H(T_{\mu})$. Using Eq.~(\ref{eq:T_mu}),  for $n=2$ we find that $T_{S}$ drops out from this relation, and the following constraint on $f_a$ may be derived
\begin{equation}
\label{eq:faMaxSoTth_eq_Tmu}
    f_a \lesssim 2 \times 10^9 \GeV 
    \left( \frac{\mu}{10 \TeV} \right)^{ \scalebox{1.01}{$\frac{1}{2}$} }
    \left( \frac{g_{\rm MSSM}}{g_*(T_\mu)} \right)^{ \scalebox{1.01}{$\frac{1}{4}$} }
    \left(\frac{6}{N_{\rm DW}}\right) \hspace{1 cm} \quad (n = 2).
\end{equation}
For $f_a$ larger than this critical value, the coupling of saxion is too weak to thermalize at $T_\mu$.  Instead, thermalization waits until after $T_S$ and occurs at the lower $T_{\rm th}$ given in Eq.~(\ref{eq:Tth}). For low inflationary reheat temperatures (or the case where the saxion comes to dominate), this critical value Eq.~(\ref{eq:faMaxSoTth_eq_Tmu}) is modified due to an enhanced Hubble expansion rate and a different scaling of $S$ with respect to temperature during a matter-dominated era. The inflaton- and saxion-dominated cases will be discussed in Sec.~\ref{sec:param_n2} and Sec.~\ref{sec:param_n2_SaxDom}, respectively.

Another possible thermalization channel is via the saxion scattering with the $W$ gauge boson. This occurs with a rate given by
\begin{equation}
\label{eq:Gamma_SWW}
    \Gamma_{SWW} = n^2 \times b \frac{T^3}{S^2},
\end{equation}
where $b \simeq 10^{-5}$~\cite{Bodeker:2006ij, Laine:2010cq, Mukaida:2012qn}. Even when the saxion-$W$ scattering does not completely thermalize the saxion, such scattering can play an important role in generating the thermal bath necessary for the Higgsinos to thermalize the saxion. This will be discussed in Sec.~\ref{sec:param_n2_SaxDom}.

\subsection{No saxion domination}
\label{sec:RD}

The analysis of baryon asymmetry and dark matter production from axion rotations depends on whether the saxion comes to dominate the energy density and creates entropy upon its thermalization. In this section, we focus on the case where the saxion is thermalized sufficiently early so this does not occur. Then, for much of the parameter space, the baryon asymmetry production dominantly occurs during the radiation domination era following inflationary reheating. We also consider the possibility that the dominant production occurs during inflationary reheating, which can happen for low reheat temperatures. The case with saxion domination is analyzed in Sec.~\ref{sec:SD}.

\subsubsection{$n=1$}
\label{sec:param_n1}

In this section we give results for the $n=1$ case, where the $\mu$-term arises through a renormalizable coupling defined in Eq.~(\ref{eq:Wmu}).   First, we discuss whether both the baryon asymmetry and dark matter may be generated by the dynamics of the axion field. In this $n=1$ case, consistency with bounds on the axion decay constant from red giant cooling significantly constrains the ability to simultaneously realize the baryon asymmetry and axion dark matter.  The generation of  dark matter is discussed in  \ref{discussion:No_KMM} below.  Then, independent of the origin of dark matter, we focus on the determination of the lowest scale of supersymmetry breaking consistent with the successful generation of the baryon asymmetry.  In \ref{discussion:min_mS} we outline how to find this minimum scale.  In \ref{discussion:results_min_mS} we present results, including the dependence on $\tan{\beta}$ and the reheat temperature $T_{R}$. Finally, drawing from the knowledge from \ref{discussion:results_min_mS}, we present and discuss the parameter space for achieving both the baryon asymmetry and axion dark matter in \ref{discussion:results_KMM_DM}, including the effects of the reheat temperature.

\begin{figure}[t]
\includegraphics[width=0.495\linewidth]{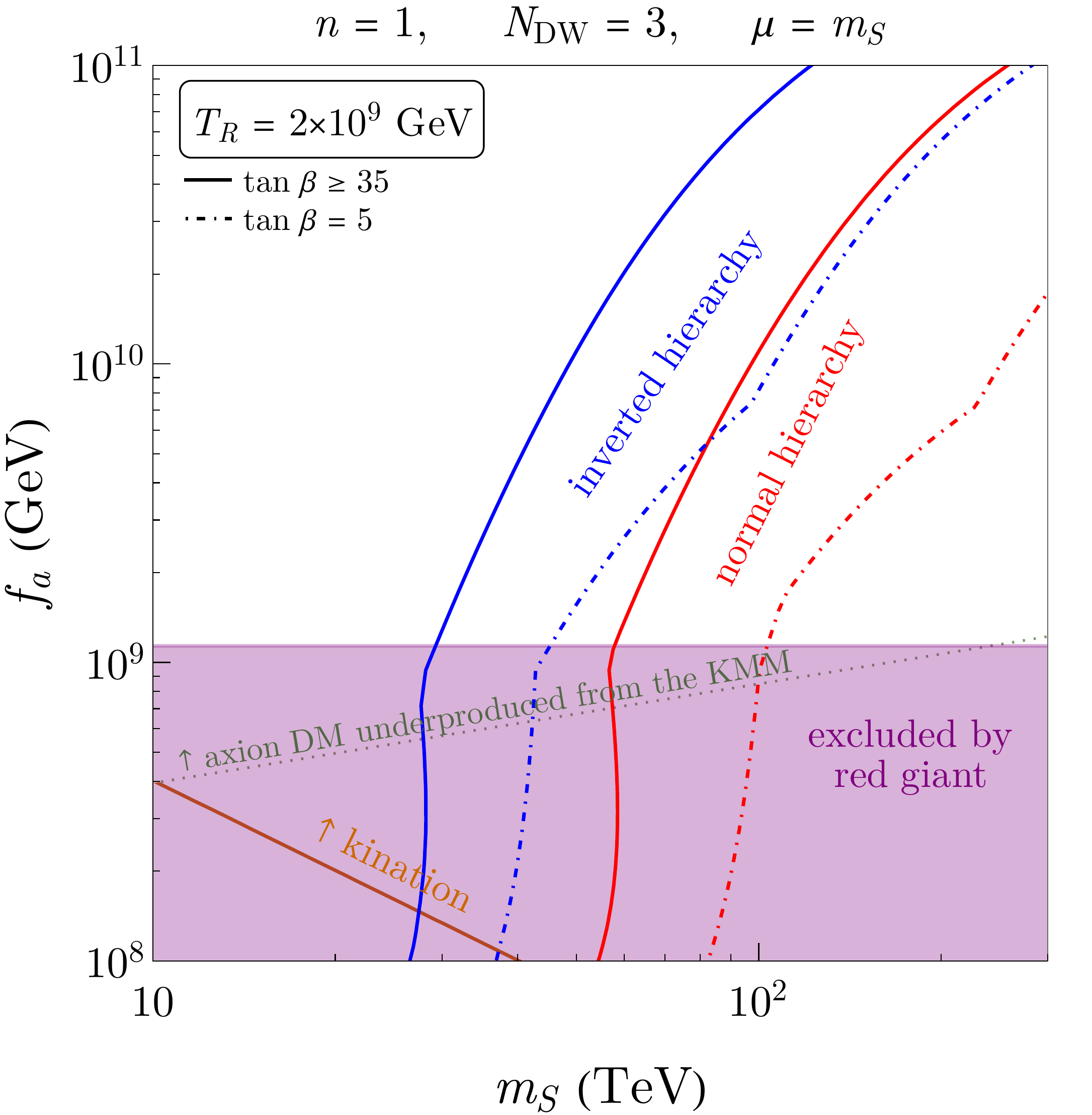}
\includegraphics[width=0.495\linewidth]{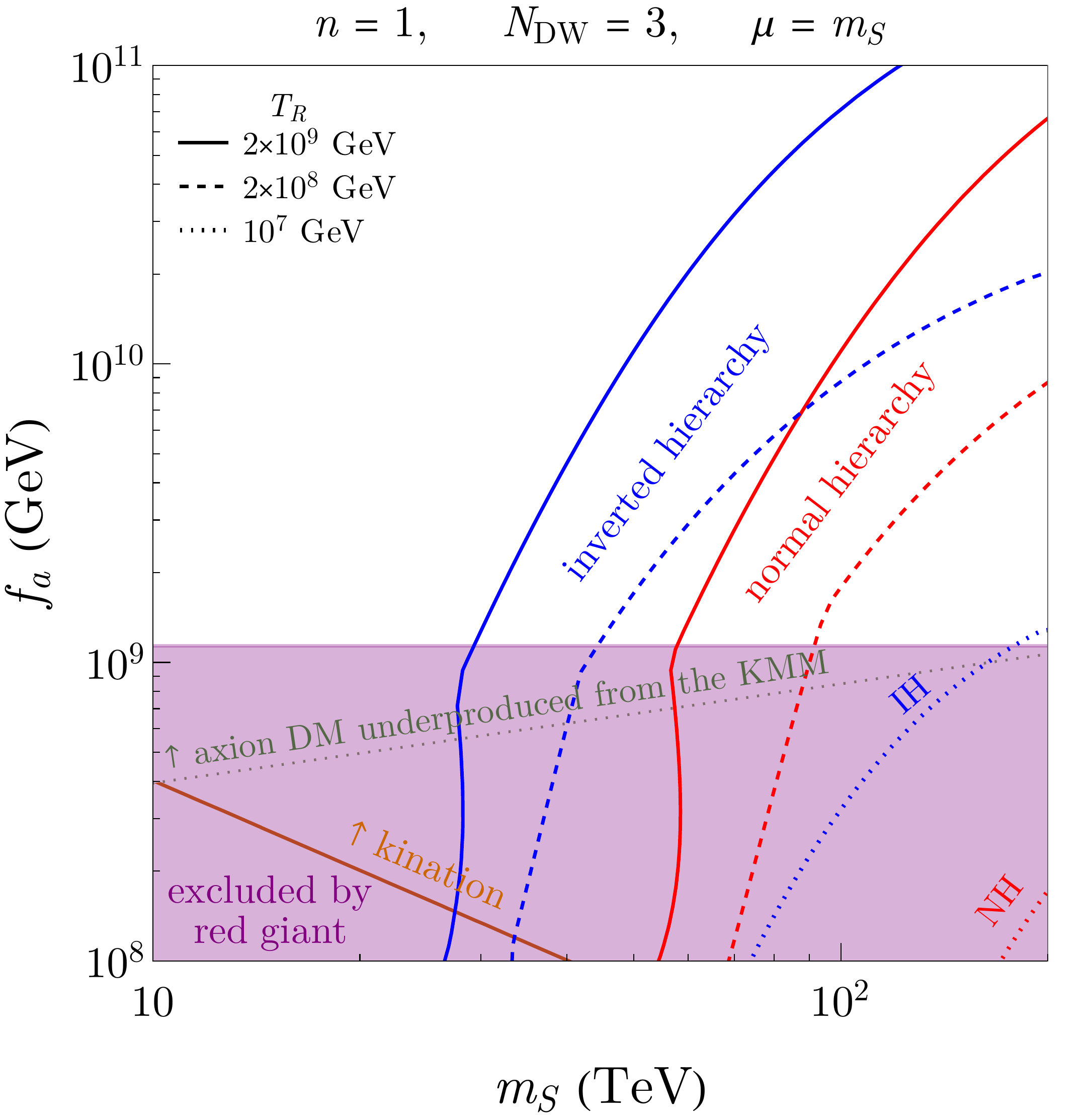}
\caption{Minimum $m_S$ for  $n=1$, domain wall number $N_{\rm DW} = 3$, and $\mu = m_S$. The baryon asymmetry can be correctly reproduced on and to the right of (blue/red) lines with the associated cosmology described in \ref{discussion:min_mS} and \ref{discussion:results_min_mS}. Different colors distinguish the assumed neutrino mass spectra. In the left panel with $T_R = 2 \times 10^9 \GeV$, solid curves are valid for all $\tan \beta \ge 35$, while dot-dashed curves correspond to $\tan \beta =5$. In the right panel, the solid, dashed, and dotted line styles indicate reheat temperatures $T_R = 2 \times 10^9 \GeV, 2 \times 10^8 \GeV, 10^7 \GeV$. The effects of $\tan\beta$ and $T_R$ are described in \ref{discussion:results_min_mS}. Above the green dotted line, as discussed in \ref{discussion:No_KMM}, kinetic misalignment underproduces axion dark matter.  The possibility of generating sufficient dark matter using a larger $\mu$ is discussed in \ref{discussion:results_KMM_DM} with results shown in Fig.~\ref{fig:n1NDW3_DM}. The purple region is excluded by observations of red giants~\cite{Capozzi:2020cbu, Straniero:2020iyi}.}
\label{fig:n1NDW3}
\end{figure}

\begin{enumerate}[label=\textbf{(I) Axion dark matter?},ref=(\underline{I}), wide, labelindent=0pt]
\item  \label{discussion:No_KMM}
Below the green dotted line in Fig.~\ref{fig:n1NDW3}, the dark matter abundance would be successfully explained by the kinetic misalignment mechanism.  
Above the green dotted line, axion dark matter is necessarily underproduced.  This is because even the maximum possible charge yield,  achieved when the saxion dominates, $Y_\theta = 3 r T_{\rm th} / 4 N_{\rm DW} m_S$ with $T_{\rm th}$ given in Eq.~(\ref{eq:Tth}), is too low to provide axion dark matter. Low values of $f_a$ in the purple shaded region of Fig.~\ref{fig:n1NDW3} are excluded by red giant brightness observations that bound axion-electron couplings~\cite{Capozzi:2020cbu, Straniero:2020iyi}.  The incompatibility of these regions shows that generation of all of the dark matter is not possible with the parameters shown. 
Here we have assumed $\mu = m_S$. Higher values of $\mu$ relative to $m_{S}$ shift this green dotted line upward, eventually allowing compatibility with the bound. We discuss this possibility further in \ref{discussion:results_KMM_DM}.

Moreover, if additional thermalization channels beyond those described in Sec.~\ref{sec:thermalization} are present, then it would be possible to increase $T_{\rm th}$ and hence the maximal yield.  This could allow the KMM to reproduce the observed DM abundance for larger $f_a$; see Ref.~\cite{Co:2019jts}.

Here, we do not include such channels. So, above this green line, an additional source of dark matter would be required.  We assume that whatever produces the balance of the dark matter budget does not disturb the prediction of the baryon asymmetry.  This would be the case, for example, if the dark matter were produced by thermal freeze-out of an LSP.\footnote{Given the superpartner scales considered here, this might require a hierarchy between gaugino/Higgsino masses and scalar masses.  Note, a thermally produced wino LSP would require a cored DM profile to avoid indirect detection bounds, see, e.g.~\cite{Cohen:2013ama,Fan:2013faa}.} 
\end{enumerate}

\begin{enumerate}[label=\textbf{(II) Finding the minimum $m_S$:},ref=(\underline{II}), wide, labelindent=0pt]
\item  \label{discussion:min_mS}
Even in cases where it is impossible to reproduce the full DM abundance, it is nonetheless of interest to understand what sets the minimum superpartner scale $m_{S}$ consistent with the production of the baryon asymmetry.  Since the size of the baryon asymmetry is proportional to $\dot\theta$ and hence $m_S$, this minimum $m_S$ scale can be found by maximizing the baryon asymmetry production efficiency.

The optimal cosmological evolution to obtain the smallest $m_S$ can be obtained as follows.  It is best to minimize $T_f = \max(T_S, T_{\rm RM})$ so the logarithmic enhancement in Eq.~(\ref{eq:YB_RD}) is maximized, but this should be done while  avoiding entropy production that would dilute the asymmetry. Thus, the maximum baryon generation efficiency is achieved if neither the saxion nor the rotation comes to dominate the total energy density.  This is accomplished if $T_{\rm RM} = \min(T_{\rm th},T_S)$. 
This ensures saxion thermalization (which occurs at $T_{\rm th}$) happens early enough to avoid entropy production.  It also ensures that the rapid redshift of the energy of rotation (which begins at $T_S$) occurs early enough that the rotation does not come to dominate; this will make the radiation-dominated era---and hence the period of logarithmically enhanced baryon production---as long as possible.
 This procedure also minimizes $T_f$.

\begin{figure}[t!]
\includegraphics[width=0.495\linewidth]{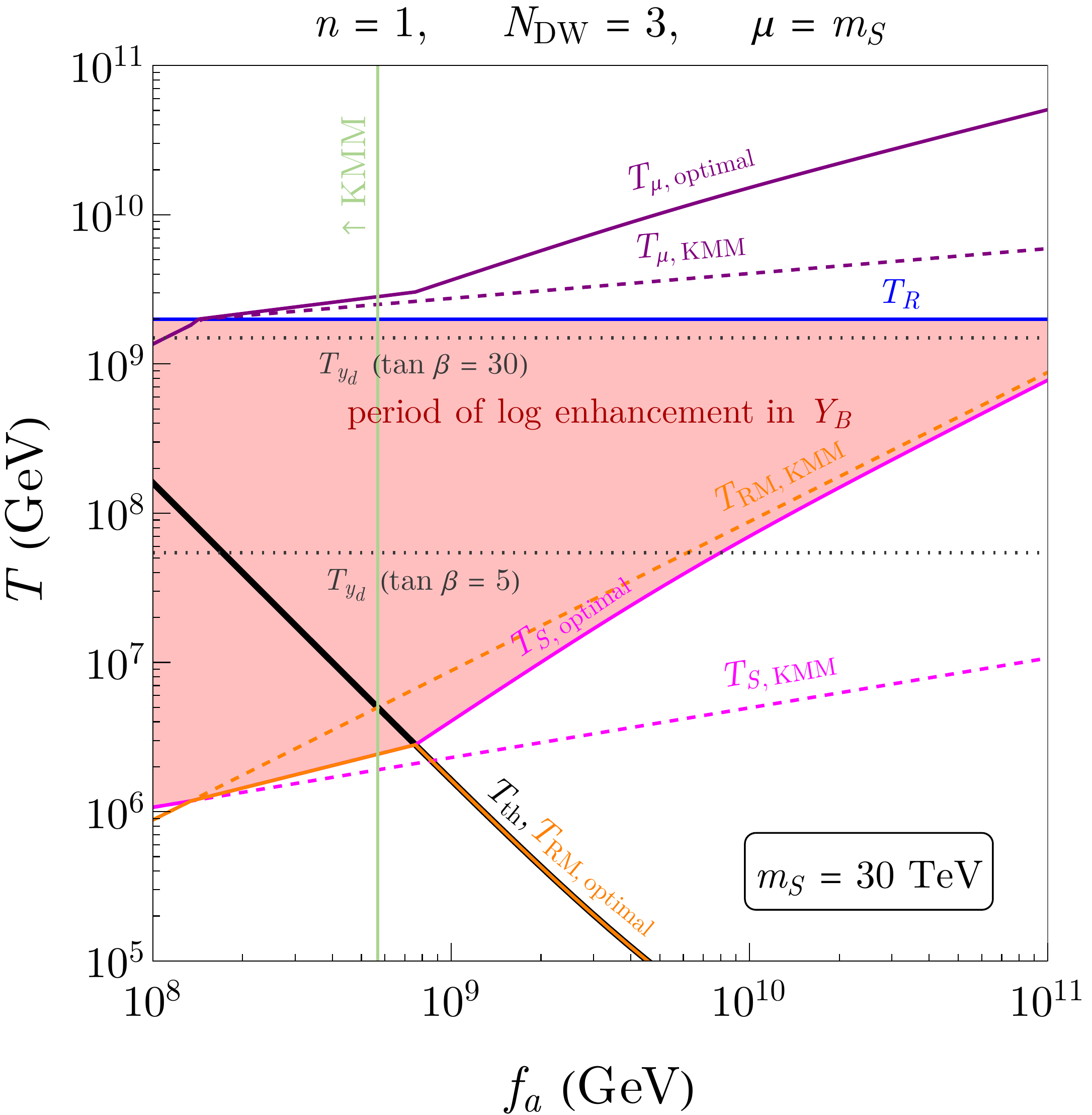}
\includegraphics[width=0.495\linewidth]{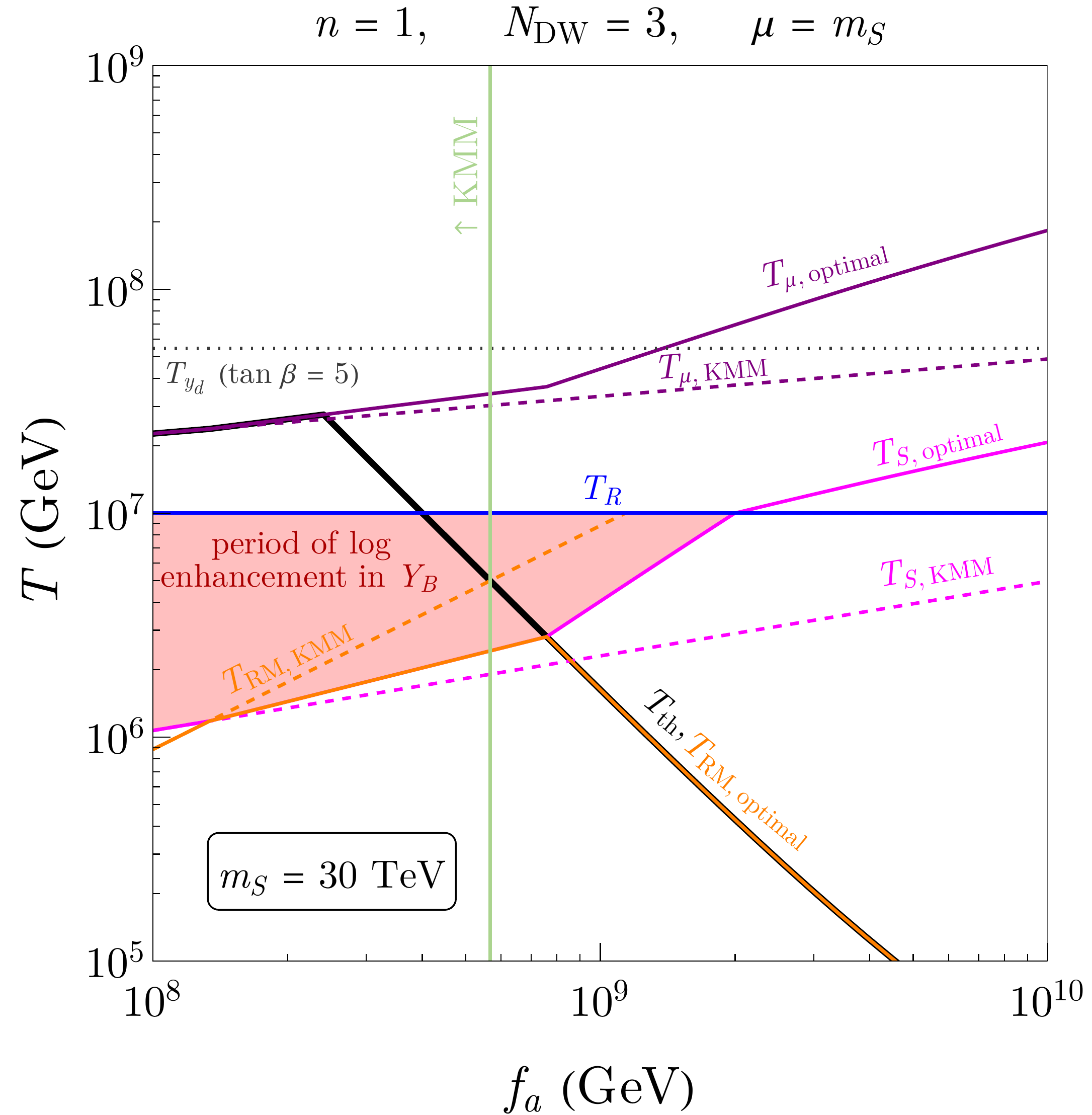}
\caption{Cosmologically relevant temperatures as functions of $f_a$ for fixed $m_S = 30 \TeV$ and $T_R = 2 \times 10^9 \GeV$ (left panel) and $T_R = 10^7 \GeV$ (right panel). The red shaded region between the reheat temperature $T_R$ and $\max(T_S, T_{\rm RM})$ indicates the range of temperatures where the dominant baryon asymmetry is produced. During this epoch, the baryon asymmetry produced per Hubble time is constant so the total $Y_{B-L}$ receives a logarithmic enhancement, see Eq.~(\ref{eq:YB_RD}). The subscript ``KMM'' refers to values that ensure axion dark matter is produced by the kinetic misalignment mechanism; see Eqs.~(\ref{eq:TS_KMM}) and (\ref{eq:TRM_KMM}) for example. This can be satisfied only to the left of the vertical green line.  This corresponds to the region below the green line in Fig.~\ref{fig:n1NDW3}. The subscript ``optimal'' indicates the cosmological scenario where $Y_B$ is most efficiently produced, leading to the smallest required $m_S$, corresponding to the curve segments above the green dotted line in Fig.~\ref{fig:n1NDW3}. The optimal scenario is achieved by choosing the charge yield $Y_\theta$ such that $T_{\rm RM,\,optimal} = \min(T_{S,\,{\rm optimal}}, T_{\rm th})$ so as to avoid rotation or saxion domination.  For $f_a > 1.5~(7.5) \times 10^8 \GeV$, $T_{\rm RM,\,optimal}$ follows $T_S$ ($T_{\rm th}$), while $T_{\mu,\,{\rm optimal}}$ and $T_{S,\,{\rm optimal}}$ change accordingly. Temperatures marked $T_{y_{d}}$ indicate when interactions involving the down-Yukawa coupling come into equilibrium for different choices of $\tan \beta$; different $C_{i}$ will apply above and below these lines, see~\ref{discussion:results_min_mS}.  Temperatures denoted $T_{\mu}$ show when Higgsinos come into thermal equilibrium.}
\label{fig:n1NDW3_temp}
\end{figure}

To clarify this cosmological history that produces the maximum asymmetry, we have shown relevant temperatures as functions of $f_a$ for $m_S = 30 \TeV$ in Fig.~\ref{fig:n1NDW3_temp}. The left (right) panel is for $T_R = 2 \times 10^9 \GeV$ ($10^7 \GeV$).  As discussed above, the optimal scenario for efficient baryon asymmetry production is to choose the PQ charge $Y_\theta$ (and hence the energy density in the saxion and rotation) so that  $T_{\rm RM} = \min(T_S, T_{\rm th})$. For $f_a > 7.5 \times 10^8 \GeV$, this imposes  $T_{\rm RM} = T_{\rm th}$, and we refer to this $T_{\rm RM}$ as $T_{\rm RM,\,optimal}$, according to Eq.~(\ref{eq:TRM_r_Tth}). Once we have fixed $T_{\rm RM}$ in this way, $T_S$ and $T_\mu$ may be determined. We have shown these as the $T_{S,\,{\rm optimal}}$ and $T_{\mu,\,{\rm optimal}}$ curves, which deviate from the values required by the KMM shown in dashed lines.

There is one final minor complication.  The region above the orange line in Fig.~\ref{fig:n1NDW3} would lead to a period of matter domination followed by kination domination had we assumed axion dark matter from the KMM, i.e., $T_{\rm RM,\,KMM} > T_{S,\,{\rm KMM}}$ using Eqs.~(\ref{eq:TS_KMM}) and (\ref{eq:TRM_KMM}). However, the goal for Fig.~\ref{fig:n1NDW3} is to derive the minimum $m_S$ rather than requiring the KMM. The optimal choice for $T_{\rm RM}$ to find this minimal $m_{S}$ is not $T_{\rm RM,\,KMM}$, but rather  $T_{\rm RM,\,optimal} = T_{S,\,{\rm optimal}}$, and this choice is applied in the region between the green dotted and the orange lines in Fig.~\ref{fig:n1NDW3} with $m_S \gtrsim 10 \TeV$. This is the case between $f_a = (1.5\mathchar`-7.5) \times 10^8 \GeV$ in both panels of Fig.~\ref{fig:n1NDW3_temp}, where $T_{\rm RM,\,optimal} = T_{S,\,{\rm optimal}}$. This optimal cosmology corresponds to Fig.~\ref{fig:schematic_rho} but with the blue curve shifted downwards and to the left so that the radiation energy density is equal to that of rotation at the kink in the rotation energy density. 
\end{enumerate}

\begin{enumerate}[label=\textbf{(III) Results on minimum $m_S$:},ref=(\underline{III}), wide, labelindent=0pt]
\item  \label{discussion:results_min_mS}

The blue and red curves in Fig.~\ref{fig:n1NDW3} show the minimum values of $m_{S}$ for which the baryon asymmetry may be achieved, with different colors corresponding to the choice of the neutrino mass spectrum.
The solid curves in both panels of Fig.~\ref{fig:n1NDW3} are identical and assume $T_R = 2 \times 10^9 \GeV$. Although the dot-dashed curves in the left panel also assume $T_R = 2 \times 10^9 \GeV$, they assume a different value of 
$\tan\beta$, whose effect is to be discussed below. 
For high $T_R$ such as this, $Y_B$ primarily depends on $m_S$, and the dependence on $T_R$ is logarithmic because of its role in setting $T_i$ in Eq.~(\ref{eq:YB_RD}). For these curves in the left panel of Fig.~\ref{fig:n1NDW3}, the dependence on $f_a$ is also only logarithmic and enters via its impact on $T_f = T_S$.  This explains the nearly vertical segments of the curves starting at low values of $f_a$. Indeed, starting at the bottom of these curves, we have the baryon asymmetry generated during a radiation-dominated era with a logarithmic enhancement. Moving to larger $f_a$, it eventually becomes impossible to reproduce the dark matter abundance above the green dotted line as explained in \ref{discussion:No_KMM}. Above this point, the most efficient generation of the asymmetry may be found by ensuring that the saxion does not come to dominate the energy density (and thus generate entropy) as described in \ref{discussion:min_mS}. Consequently, a kink in the curve develops here because the PQ charge must be such that $T_{\rm RM}$ only occurs at thermalization temperature $T_{\rm th}$ so as to avoid this dilution. For the curve segments below the green dotted line and above the orange line, the PQ yield needs to be chosen in a way such that the rotation does not dominate the energy density either; see discussion in \ref{discussion:min_mS}.

\noindent
{\boldmath \bf Effects of $\tan\beta$:}
The dot-dashed curves in the left panel assume a lower value for $\tan\beta = 5$ than the solid curves. The value of $\tan \beta$ can impact the baryon asymmetry via its effect on the down- and electron-Yukawa couplings.  When interactions involving the down- or electron-Yukawa coupling are out of equilibrium, this may change the chemical potentials and hence $C_i(T)$ in the baryon asymmetry of Eq.~(\ref{eq:DeltaY_B-L}). In constructing Fig.~\ref{fig:n1NDW3}, we have used the relevant $C_{i}$ for each temperature range; see Table~\ref{tab:DFSZ_C_i} and Appendix~\ref{app:cB} for details.  See Fig.~\ref{fig:conservation_laws} for the temperatures at which the Yukawa interactions come into equilibrium. We find that the values of $C_i$ depend most sensitively on whether the  down-Yukawa interaction is in equilibrium, and they are relatively insensitive to whether the electron-Yukawa interaction is. In the left panel of Fig.~\ref{fig:n1NDW3}, dot-dashed lines assume $\tan \beta = 5$, while solid lines assume down-Yukawa interactions are in equilibrium. The solid lines apply for all $\tan\beta \ge 35$ because the interactions come into equilibrium at temperatures higher than the reheat temperature $T_R = 2 \times 10^9 \GeV$ assumed in this panel.  The dot-dashed lines with lower $\tan\beta$ shift to higher $m_S$ compared to the solid lines with higher $\tan\beta$ because out-of-equilibrium down-Yukawa couplings reduce the coefficients $C_i$ and thus the efficiency in producing $Y_B$ and a larger $\dot\theta$ is needed to compensate.
All lines in the right panel of Fig.~\ref{fig:n1NDW3} assume that down-Yukawa interactions are in equilibrium, which is valid for $\tan\beta > 10$ (2) in the case of the dashed (dotted) lines with $T_R = 2 \times 10^8 \GeV$ ($10^7 \GeV$). 

\noindent
{\boldmath \bf Impact of reheat temperature:} 
In Fig.~\ref{fig:n1NDW3}, the solid and dot-dashed curves in the left panel and the solid lines in the right panel assume $T_R = 2 \times 10^9 \GeV$, whereas the dashed (dotted) lines in the right panel are for $T_R = 2 \times 10^8 \GeV$ ($10^7 \GeV$). The predictions are affected because the logarithmic enhancement of Eq.~(\ref{eq:YB_RD}), if present, starts at $T_i = T_R$.  It is also possible that for sufficiently low $T_{R}$, the baryon asymmetry is dominantly produced during the period of inflationary reheating. We will discuss such effects below after commenting on the constraints from BBN.

A reheat temperature $T_R = 2 \times 10^9 \GeV$ with $m_S\sim \TeV$ requires either i) $R$-parity violation, ii) a small gravitino mass $m_{3/2} \sim 100 \GeV$, or iii) a sneutrino as the next-to-LSP that is nearly degenerate with the gravitino LSP. For $m_{3/2} \gtrsim 7 \TeV$ the bound on $T_R$ rapidly weakens, so we expect that the dashed and dotted curves are valid without any additional assumptions. See Appendix~\ref{app:relics} for more details on BBN constraints.

As $T_R$ decreases, the generation of the baryon asymmetry is less efficient, and higher values of $m_S$ are  needed to reproduce the observed baryon abundance. At minimum, this is because lowering $T_R$ reduces $T_i$, the onset of the radiation-dominated era that is responsible for the logarithmic enhancement in the generation of the asymmetry in Eq.~(\ref{eq:YB_RD}). This explains why the dashed and dotted curves are shifted to higher $m_S$ than the solid curves at low values of $f_a$. For higher $f_a$, the curve bends further and becomes a straight line because, although $T_{\rm RM}$ still follows $T_{\rm th}$, eventually the resultant $T_{S,\,{\rm optimal}}$ exceeds $T_R$.  When this occurs,  $Y_B$ is no longer dominantly produced during an radiation-dominated era but rather during inflationary reheating. The logarithmic enhancement disappears, and the baryon asymmetry is diluted by entropy produced from the reheating, as in Eq.~(\ref{eq:YB_MD}).  The result is that the baryon asymmetry is sensitive to $T_S$ and therefore $f_a$. This cosmological evolution may be clarified by examining the right panel of Fig.~\ref{fig:n1NDW3_temp}.
There, $T_{S,\,{\rm optimal}}$ can be seen to deviate from $T_{S,\,{\rm KMM}}$ at $f_a \simeq 1.5 \times 10^8 \GeV$ (when $T_{\rm RM,\,optimal}$ starts to track $T_{S,\,{\rm optimal}}$), change its slope at $f_a \simeq 7.5 \times 10^8 \GeV$ (when $T_{\rm RM,\,optimal}$ starts to track $T_{\rm th}$), and then eventually exceed $T_R = 10^7 \GeV$ for $f_a \gtrsim 2 \times 10^9 \GeV$.

\end{enumerate}

\begin{figure}[t!]
\includegraphics[width=0.495\linewidth]{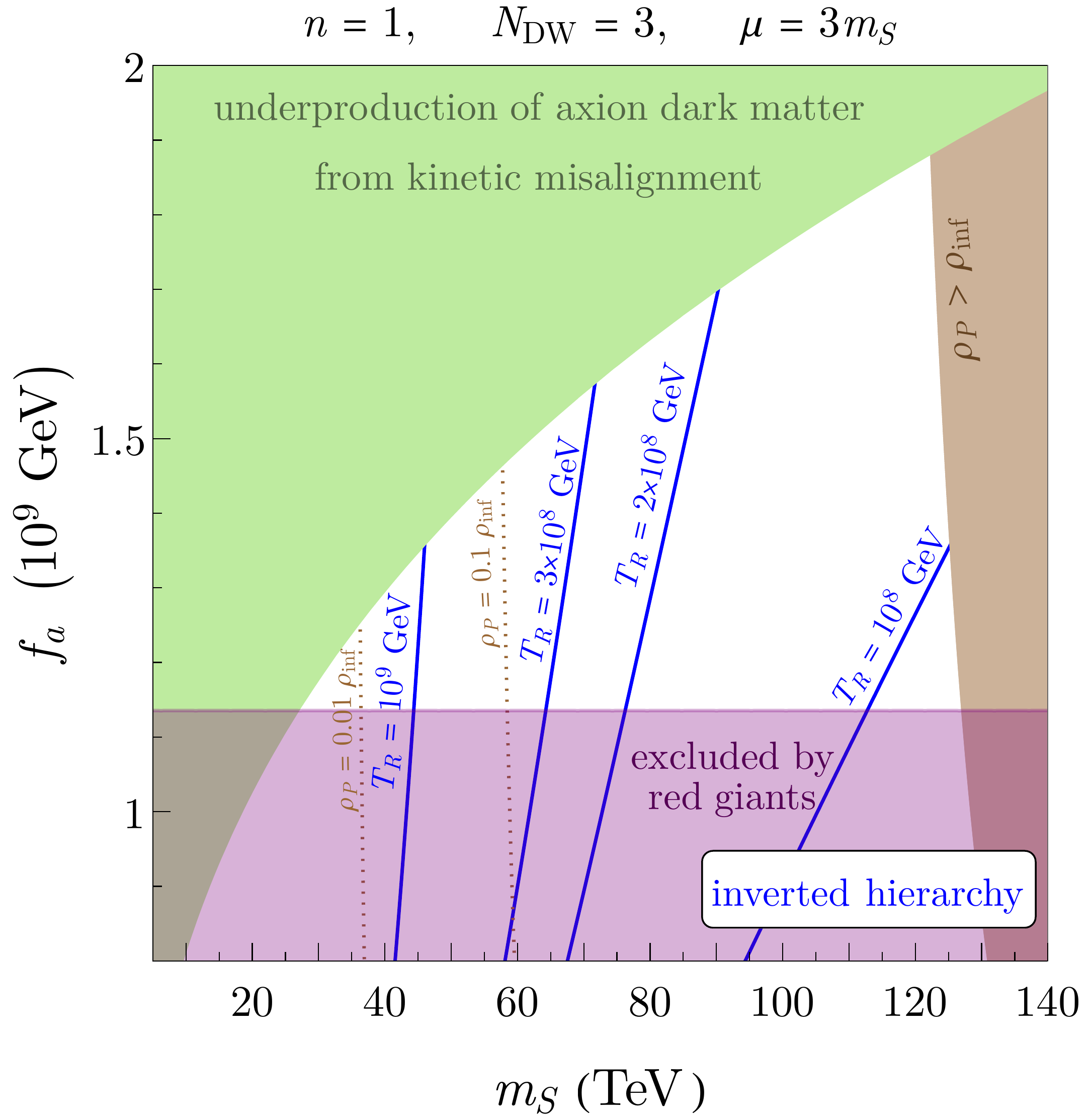}
\includegraphics[width=0.495\linewidth]{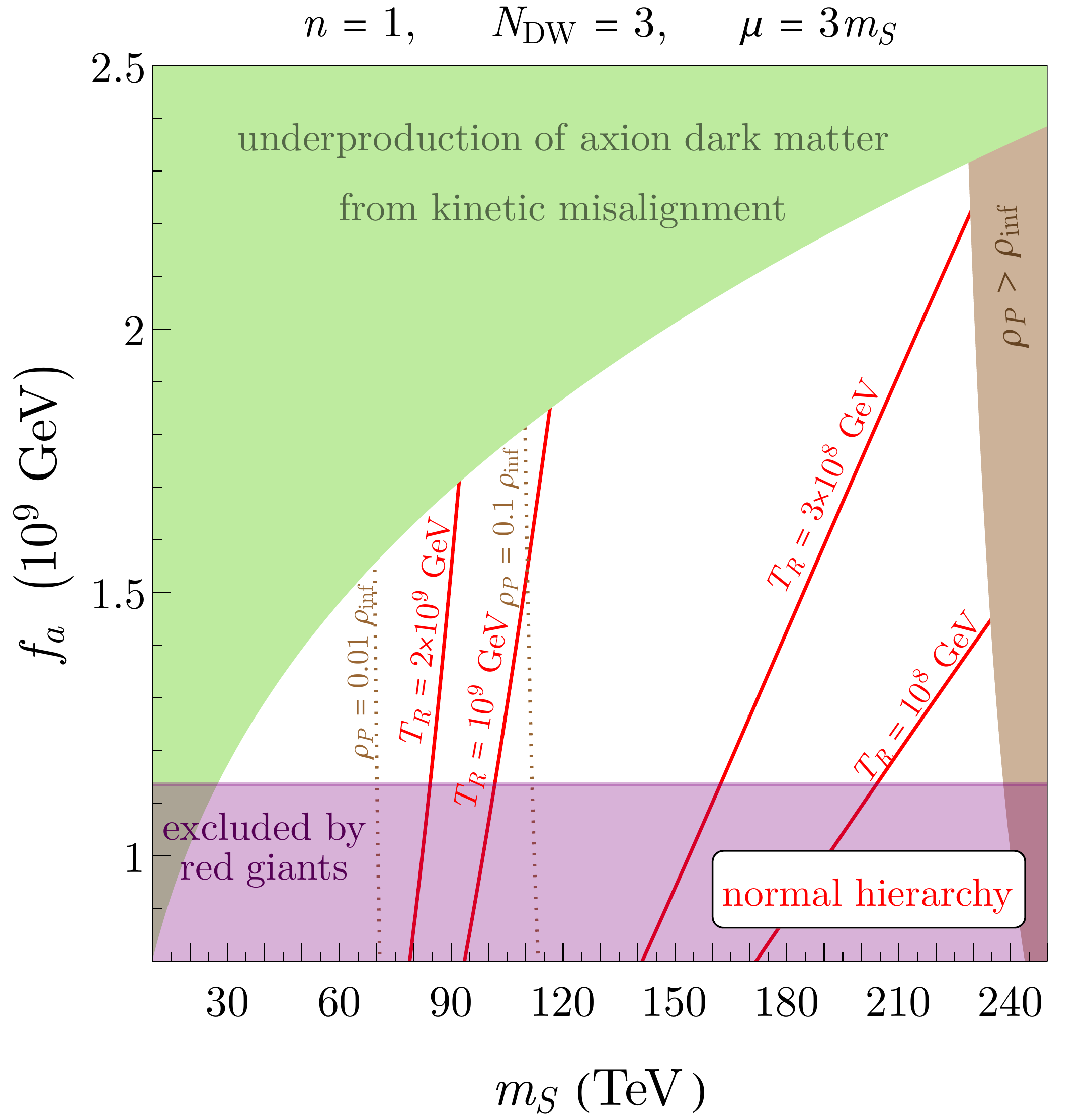}
\caption{Parameter space reproducing both the observed dark matter abundance and the baryon asymmetry. The left (right) panel assumes the neutrino mass spectrum with an inverted (normal) hierarchy. The blue and red contours show the reheat temperatures required to reproduce both the baryon asymmetry from lepto-axiogenesis and dark matter from kinetic misalignment. The kinetic misalignment mechanism predicts a period of matter domination followed by kination domination in the entire parameter space shown here. The green region leads to underproduction of axion dark matter from kinetic misalignment because of entropy production from saxion domination, i.e., $T_{\rm RM} > T_{\rm th}$ using Eqs.~(\ref{eq:TRM_general}) and (\ref{eq:Tth}). The purple region is excluded by the red giant brightness observations. The brown region is excluded because the required PQ charge leads to an energy density of the PQ field $\rho_P$ exceeding that of the inflaton, while the brown dotted contours show lower values of $\rho_P/\rho_{\rm inf}$.
}
\label{fig:n1NDW3_DM}
\end{figure}

\begin{enumerate}[label=\textbf{(IV) Results on dark matter:},ref=(\underline{IV}), wide, labelindent=0pt]
\item  \label{discussion:results_KMM_DM}
We now focus on the region where axion dark matter can be accounted for by kinetic misalignment, i.e., below the dotted green line in Fig.~\ref{fig:n1NDW3}. As can be seen in that figure and explained in \ref{discussion:No_KMM}, if $\mu = m_S$, this possibility is in tension with bounds from observations of red giants. 
However, if this strict relation between $\mu$ and $m_{S}$ is modified, we find that it is possible to produce dark matter in this way.  For larger $\mu$, the saxion thermalization rate in Eq.~(\ref{eq:Gamma_S_higgsino}) is enhanced and therefore the maximum yield $Y_{\theta}= 3 r T_{\rm th}/4 N_{\rm DW} m_{S}$ increases, so the green dotted line in Fig.~\ref{fig:n1NDW3} shifts upward. And for $\mu=3m_{S}$, the green line is above the purple boundary for $m_S \gtrsim 30 \TeV$, and axion dark matter from kinetic misalignment becomes viable. This benchmark case is shown in Fig.~\ref{fig:n1NDW3_DM}.
Given the narrow range in $m_S$ of interest there, we improve the precision of the prediction by going beyond the analytic evaluation of $Y_B$ that relies on estimating the production of asymmetry per Hubble time $\Delta Y_B$.  We instead numerically solve the coupled Boltzmann equations of the inflaton and radiation, while adding an energy component from the axion rotation on top of this background evolution. We numerically integrate $\dot n_{B-L} R^3$ using Eq.~(\ref{eq:nB_dot}) to obtain the baryon asymmetry. We find the predictions of $m_S$ are modified (increased) by up to a factors of two for a fixed $T_R$ using this more sophisticated treatment. In the left (right) panel of Fig.~\ref{fig:n1NDW3_DM}, an inverted (normal) neutrino mass hierarchy is assumed, and the predictions are shown by the blue (red) contours. We include contours of $T_R$ to show how the reheat temperature affects the prediction. 
The brown region is excluded because the required energy density in the complex field $P$, comprised of contributions from rotation and the saxion $\rho_P \equiv \rho_S + \rho_\theta  ~(\simeq 2 \rho_\theta$ for $A \simeq m_S$ according to Eq.~(\ref{eq:r})), exceeds that of the inflaton. The origin of this region may be understood by noting that  larger values of $m_S$ require less efficient production of $Y_B$, which may be achieved by a smaller logarithmic enhancement during radiation domination by decreasing the ratio between $T_R$ and $T_{\rm RM}$.  For fixed $T_{R}$, this requires a larger $T_{\rm RM}$ in Eq.~(\ref{eq:TRM_general}).  However, eventually $T_{\rm RM}$ becomes as large as $T_R/2$, at which point $\rho_P \simeq 2 \rho_\theta = \rho_{\rm inf}$, and an inconsistency arises because $P$ would instead drive an epoch of inflation.  Two brown dotted curves are shown for $\rho_P/\rho_{\rm inf} = 0.1$ and $0.01$, which are perhaps more realistic energy densities for $P$. In summary, the saxion mass is now predicted to have a strict upper limit  of 125 (240) TeV for inverted (normal) hierarchy with lower $m_S$ preferred for a realistic $\rho_P$. This mass range is intriguingly consistent with the supersymmetry-breaking scale determined from the observed value of the Higgs boson~mass. 
\end{enumerate}

\subsubsection{$n=2$}
\label{sec:param_n2}

\begin{figure}
\includegraphics[width=0.495\linewidth]{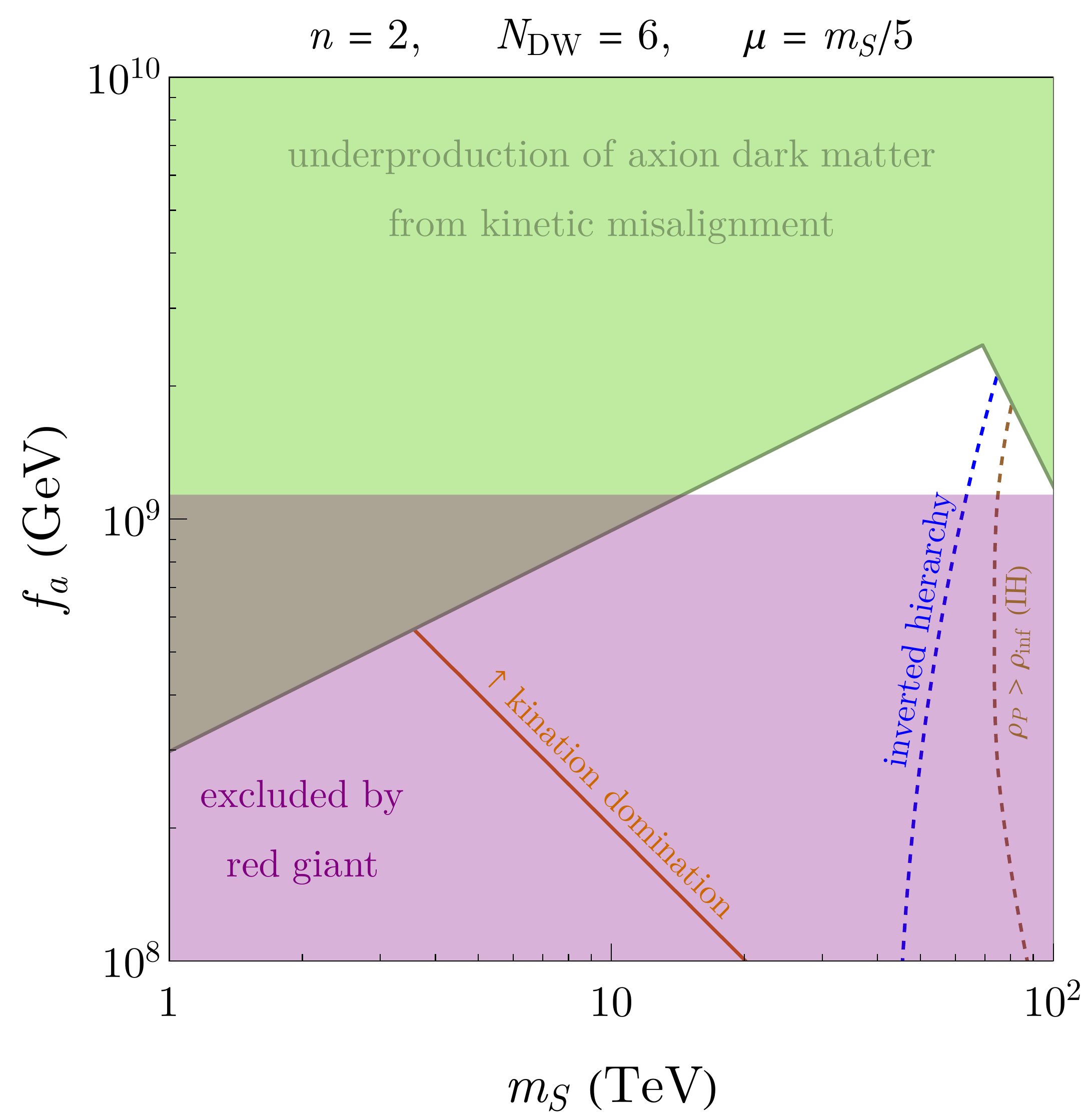}
\includegraphics[width=0.495\linewidth]{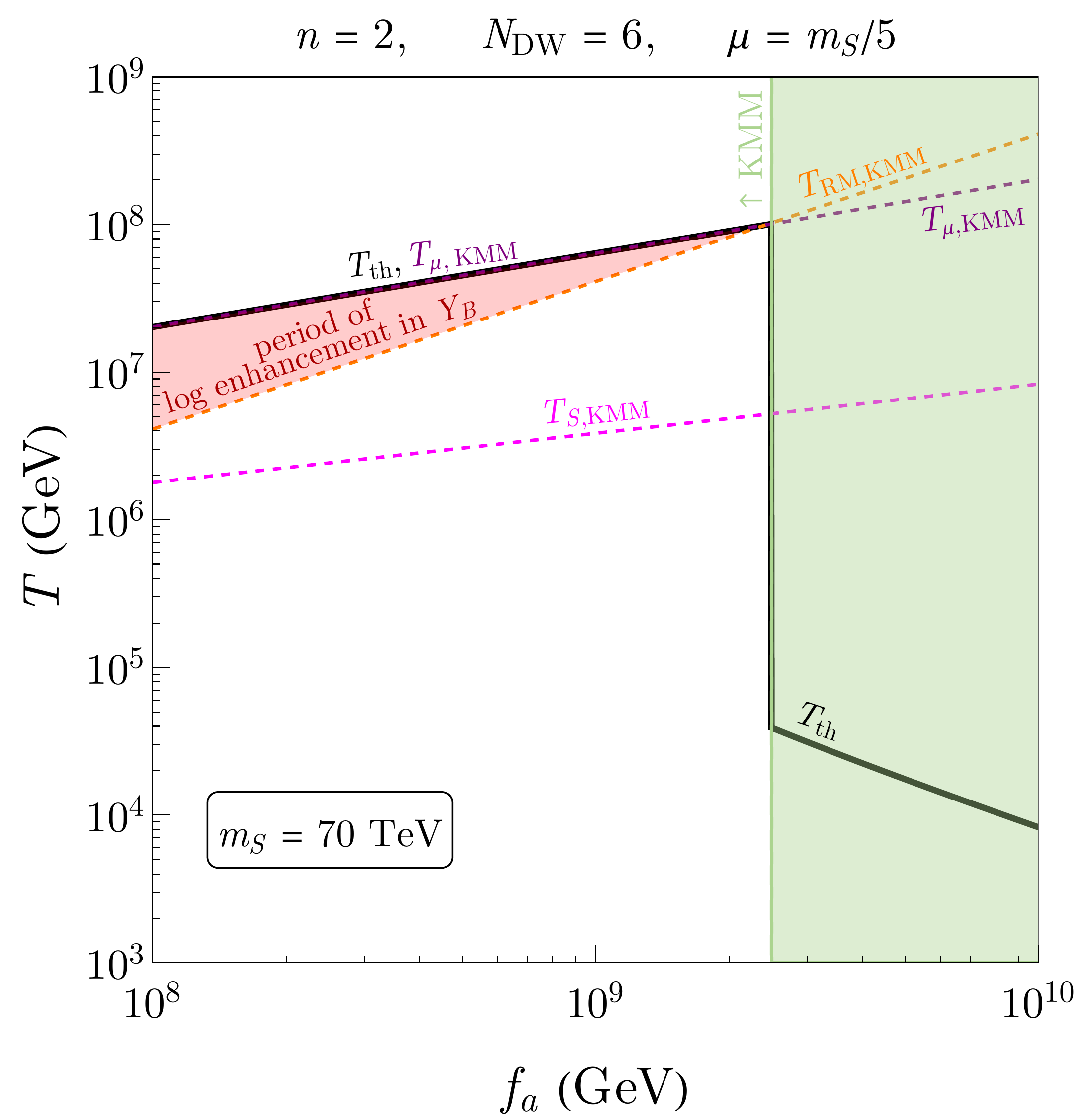}
\caption{Analysis of the $n=2$, $N_{\rm DW}=6$ case. {\bf Left:}  The baryon asymmetry and the dark matter abundance are correctly reproduced along the blue dashed line for the case of an inverted hierarchy neutrino spectrum, for sufficiently high reheat temperatures. Lower reheat temperatures lead to higher $m_S$, up to the brown dashed line where the $P$ field starts to drive inflation. In contrast, the normal hierarchy case leads to overproduction of the baryon asymmetry. The purple region is excluded by observations of red giants, while the green region underproduces dark matter. {\bf Right:} Temperatures in this combined dark matter/baryon asymmetry scenario for fixed $m_{S}$ = 70 TeV.  The temperature $T_{S,\,{\rm KMM}}$ (magenta dashed) indicates where the saxion reaches its minimum.  $T_{\mu,\,{\rm KMM}}$  indicates the temperature below which the Higgsinos are in thermal equilibrium.  $T_{\rm RM}$ (yellow dashed) indicates the temperature where the rotational energy would come to dominate.  For low $f_{a}$, $T_{S,\,{\rm KMM}}$ is reached first, and no era of rotational energy domination occurs. The green shaded region corresponds to the green shaded region in the left panel where the KMM is unable to fully reproduce the dark matter density.}
\label{fig:n2NDW6_mu}
\end{figure}

We now move to the case of $n=2$ and $N_{\rm DW} = 6$.  Our focus in this case will be on the parameter space where both the baryon asymmetry and dark matter abundance result from the axion rotation.  This case is presented in the left panel of Fig.~\ref{fig:n2NDW6_mu} for $\mu = m_S/5$.  The blue curve shows the minimum values of $m_S$ compatible with axion dark matter and the baryon asymmetry both arising from axion rotations. This minimum $m_S$ is achieved when $T_R$ is sufficiently high, as discussed below. The blue curve assumes the inverted hierarchy neutrino mass spectra as labeled, while the normal hierarchy case overproduces the baryon asymmetry in this parameter space.  We will see that the requirement of the successful generation of both the baryon asymmetry and dark matter prefers a relatively small region of $m_S$ ranging from around 60 TeV to nearly 100 TeV and $f_a = (1\mathchar`-2) \times 10^9 \GeV$.

Above the orange line in the left panel of Fig.~\ref{fig:n2NDW6_mu}, a period of matter domination followed by kination domination exists because $T_{\rm RM} > T_{\rm KR}$. 
In this case, the era of logarithmically enhanced baryon production ends at $T_{\rm RM}$, when the matter domination begins.  The region above the orange line also gives a potential signal in the modification of primordial gravitational waves~\cite{Co:2021lkc, Gouttenoire:2021wzu, Gouttenoire:2021jhk}. 
This is discussed around Eq.~(\ref{eq:YB_RD}) and illustrated by the red shaded region in the right panel of Fig.~\ref{fig:n2NDW6_mu}. To accurately obtain the final $Y_B$, rather than using the analytic estimate given in Eq.~(\ref{eq:YB_RD}), we numerically integrate $\dot n_{B-L} R^3$ from $T_\mu$ to a temperature much lower than $T_{\rm RM}$; this improves the precision of the prediction on $m_S$ and changes the prediction by a factor as large as 2. 

In the green shaded region, axions from kinetic misalignment cannot account for all of dark matter because the necessary axion yield $Y_\theta = 3 r T_{\rm th} / 4 N_{\rm DW} m_S$ requires a $T_{\rm th}$ value that is higher than can be achieved from saxion-Higgsino scattering given in Eq.~(\ref{eq:Tth}). In particular, above (below) the positively-sloped boundary of the green region, thermalization occurs below (at) $T_\mu$; see Eq.~(\ref{eq:faMaxSoTth_eq_Tmu}). This thermalization condition is also the origin of the vertical green line labeled with KMM in the right panel, which shows various temperatures as functions of $f_a$ for the benchmark point $m_S = 70 \TeV$. On the other hand, below the negatively-sloped boundary of the green region, we have $T_{\rm th} = T_\mu$ and $T_{\rm th} > T_{\rm RM}$ so that the saxion does not create entropy upon thermalization. Lastly, as can be seen in the right panel, the era that dominates the production of the baryon asymmetry begins at $T_\mu$, and therefore the result is independent of $T_R$ as long as $T_R > T_\mu$. Using $T_\mu$ from Eq.~(\ref{eq:T_mu}) and $T_S$ from Eq.~(\ref{eq:TS_KMM}), one finds $T_\mu \simeq 6 \times 10^7 \GeV~(f_a / 10^9 \GeV)^{1/2} (m_S / 5\mu)^{1/2}$. The calculated asymmetry will be valid for all $T_R$ larger than this value.

In other words, by lowering $T_R$, one can explain the baryon asymmetry and dark matter in the region to the right of the magenta line. However, there is a limit on how low $T_R$ can be: in the low $T_R$ and high $m_S$ limits, $T_R$ and $T_{\rm RM}$ approach each other, and when $T_R = 2 T_{\rm RM}$, the energy density of the complex field $\rho_P \equiv \rho_S + \rho_\theta$ is equal to that of the inflaton $\rho_{\rm inf}$ if $A \simeq m_S$, i.e., $\rho_S \simeq \rho_\theta$ based on Eq.~(\ref{eq:r}). The resulting upper bound on $m_S$ is shown by the brown line; see a related discussion in Sec.~\ref{sec:param_n1}. 
To the right of this brown curve, $\rho_P > \rho_{\rm inf}$, which is inconsistent because $P$ would drive inflation. 
The constraints involving $T_R$ are obtained by calculating $Y_B$ numerically, also including the coupled Boltzmann equations for inflationary reheating. 

As the sum of the neutrino masses decreases, both the blue and brown curves will move to the right, so it is possible to reproduce both the baryon asymmetry and the dark matter abundance for all of the white region to the right of the magenta curve.  For small enough neutrino mass, the blue curve will reach the intersection of the purple and green regions at the right of the figure, at which point, the window closes.  

In summary, simultaneous production of dark matter and the baryon asymmetry is possible between $m_S$ of 60-100 TeV depending on the sum of the neutrino masses, and $f_a$ should lie in the window $(1\mathchar`-2) \times 10^9 \GeV$.  The NH case (with vanishing lowest eigenvalues) is excluded by observations of red giants. 

Which neutrino spectra are allowed, however, depends on $\mu$.
We have assumed $\mu=m_S/5$ in Fig.~\ref{fig:n2NDW6_mu}. 
Smaller $\mu$ would decrease the thermalization rate $\Gamma_{S\widetilde{H}\widetilde{H}}$. This would make the negatively-sloped boundary of the green region, set by $\Gamma_{S\widetilde{H}\widetilde{H}} = 3H$ at $T_\mu$, shift downward. The positively-sloped boundary, set by $T_\mu = T_{\rm RM}$, would shift upward because $T_\mu \propto \mu^{-1/2}$ and $T_{\rm RM} \propto f_a$. Finally, a small $\mu$ increases $T_\mu$ and therefore the logarithmic enhancement in $Y_{B-L}$, which in turn requires a smaller $\dot\theta \propto m_S$ to compensate for the increased efficiency in producing $Y_{B-L}$. This shifts the prediction curves to the left. Numerically, $\mu < m_S/10$ makes the NH case with a vanishing lowest eigenvalue viable for a small range of saxion masses.

We do not analyze the baryon asymmetry in the green shaded region, where axion dark matter is underproduced by kinetic misalignment, because we find that in some of the parameter space the onset of the $P$ rotation can be initiated by the saxion thermal mass. This is because the thermal mass is proportional to $\mu$, which is in turn enhanced at high temperatures by $S^{n - 1}$.
 Rotations initiated by the thermal mass complicate the determination of the optimal cosmological evolution for the most efficient baryon asymmetry production. The thermal mass also leads to potential formation of Q-balls whose presence makes the baryon asymmetry calculation uncertain; see Sec.~\ref{sec:Qballs}. 

\subsection{Saxion domination}
\label{sec:SD}

In this subsection, we discuss a different cosmology where both dark matter and the baryon asymmetry may still be produced by axion rotations.  We do not optimize the production of the baryon asymmetry nor utilize $T_R$ to explore the parameter space as in Sec.~\ref{sec:RD}, but rather study the case where the saxion dominates the energy density before it is thermalized.  In this case, the saxion creates entropy that dilutes the baryon asymmetry produced immediately after inflationary reheating.  Consequently, the final asymmetry is dominantly produced after saxion thermalization and is therefore independent of the inflationary reheat temperature $T_R$. The predicted $m_S$ is generically larger than the optimal cases studied in Sec.~\ref{sec:RD} giving the most efficient baryon asymmetry production.

We require both $Y_B$ and dark matter from axion rotations. Then, this scenario makes a prediction for $(m_S, f_a)$ as a function of $r$, defined in Eq.~(\ref{eq:r}) as the ratio of the axion rotation to the saxion oscillation energy densities. The reason for the unique prediction is as follows.  For a given $\mu$, the relation $T_{\rm RM} = r T_{\rm th}$ from Eq.~(\ref{eq:TRM_r_Tth}) is satisfied along a contour in the $(m_S, f_a)$ plane because $T_{\rm RM}$ and $T_{\rm th}$ are independently determined by $(m_S, f_a)$.
In particular, $T_{\rm RM}$ is given by Eq.~(\ref{eq:TRM_general}) with $Y_{\theta}$ from kinetic misalignment using Eq.~(\ref{eq:Ytheta_KMM}), and see Sec.~\ref{sec:thermalization} for discussions of $T_{\rm th}$ for different values of $n$. Finally, a successful production of $Y_B$ picks out a single point along this contour once the neutrino mass spectrum has been specified.   We find that, for $\mu = m_S$, the predicted values of $f_a$ are in tension with red giant bounds except for the normal hierarchy case with $n = 1$ and $r \simeq 1$. 

We now comment on the effect of changing $\mu$.  If the value of $\mu$ is increased, this makes thermalization more efficient and would increase $T_{\rm th}$. This breaks the relation $T_{\rm RM} =r T_{th}$.   However, this relation can be restored by going to higher $f_{a}$  since $T_{\rm RM} \propto f_a$ and $T_{\rm th}$ decreases with increasing $f_a$. The predictions for $\mu = 3 m_S$ will be shown and discussed.

\subsubsection{$n=1$}
\label{sec:param_n1_SaxDom}

\begin{figure}
\includegraphics[width=0.495\linewidth]{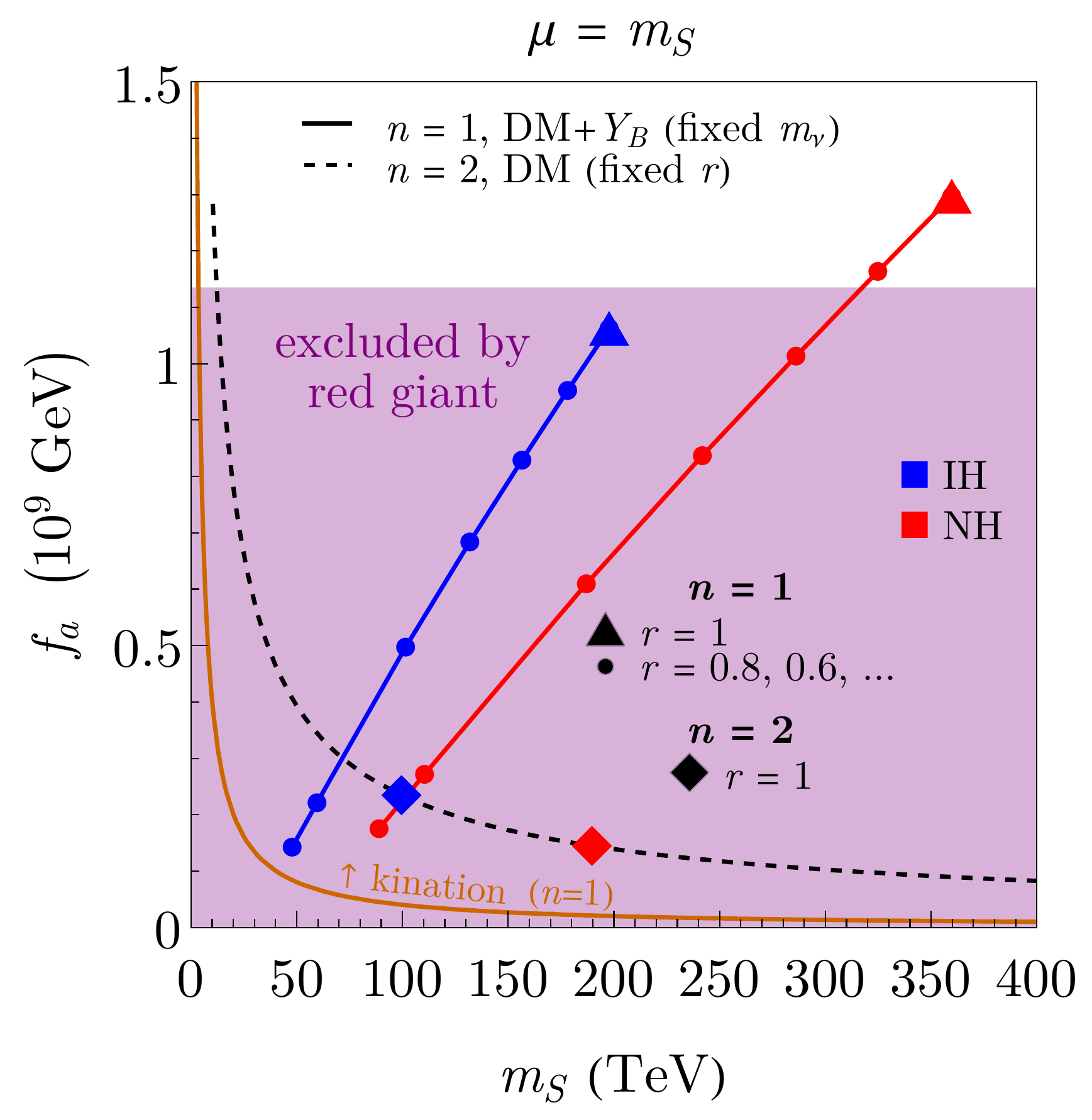}
\includegraphics[width=0.495\linewidth]{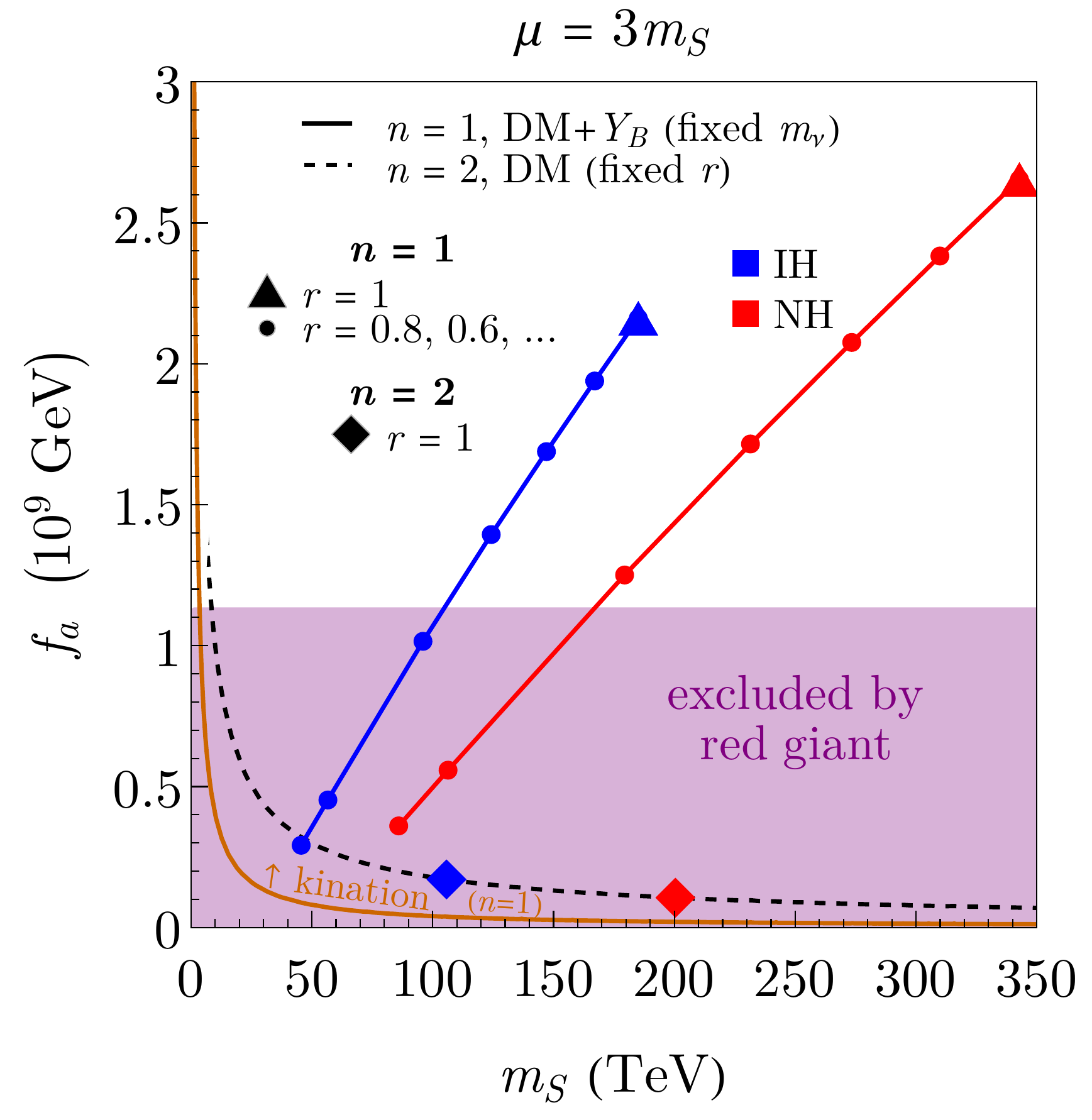}
\caption{Predictions for $m_S$ and $f_a$ from the baryon asymmetry via lepto-axiogenesis and axion dark matter from kinetic misalignment in the scenario where the saxion dominates. The left (right) panel is for $\mu = m_S$ ($\mu = 3 m_S$). The symbols connected by the solid (dashed) lines are for $n = 1$ ($n = 2$). The triangle and diamond symbols assume the saxion has the same energy as the axion rotation, i.e., $r = 1$ as defined in Eq.~(\ref{eq:r}). The circles below triangles denote lower values of $r$ in steps of $0.2$ from $r = 0.8$ until $r = 0.02$, after which it is $0.03$ and $0.01$. The colors refer to the chosen neutrino mass spectrum as labeled. The colored lines  connect the predictions of both kinetic misalignment and lepto-axiogenesis for $n = 1$ 
with various values of $r$, whereas the dashed black line is the prediction of kinetic misalignment alone for $n = 2$ assuming $r = 1$ (predictions for $n=2$, $r < 1$ are not included due to complications involving thermalization, see text). In the regions above the orange curves, axion dark matter from kinetic misalignment predicts eras with matter and kination domination, which may leave imprints in primordial gravitational waves~\cite{Co:2021lkc, Gouttenoire:2021wzu, Gouttenoire:2021jhk}.}
\label{fig:sax_dom}
\end{figure}

The thermalization temperature for $n=1$ is given in Eq.~(\ref{eq:Tth}). In the saxion domination case,
$T_{\rm RM} = r T_{\rm th}$ with $T_{\rm RM}$ determined by requiring dark matter from kinetic misalignment, and this gives a relation between $m_S$ and $f_a$ for a given $r$. Furthermore, to accurately derive $Y_B$, we numerically solve the coupled Boltzmann equations of the saxion and radiation with the thermalization rate given in Eq.~(\ref{eq:Gamma_S_higgsino}) and then integrate $\dot n_{B-L} R^3$. This then makes a single point prediction $(m_S, f_a)$ when given $\mu$, $r$, and a neutrino spectrum. The final predictions are shown by the symbols connected by the solid lines in Fig.~\ref{fig:sax_dom}. (The diamonds connected by the dashed black curve are for $n = 2$ and will be discussed below.) The triangles at the top are the predictions assuming $r = 1$, while the circles below them show the predictions for smaller $r$, decreasing in steps of $0.2$ until $r = 0.2$, below which the circles are for $r = 0.03$ and $r = 0.01$. The two colors indicate different neutrino mass spectra. The left (right) panel of Fig.~\ref{fig:sax_dom} shows the predictions for $\mu = m_S$ ($\mu = 3 m_S$). The predictions are in a small region with $f_a = (1\mathchar`-3) \times 10^9 \GeV$.  The required values of $m_S$ range from 100-360 TeV depending upon the choice of neutrino spectrum and are in an interesting range considering the observed Higgs boson mass.

\subsubsection{$n=2$}
\label{sec:param_n2_SaxDom}

We continue to analyze the case where the saxion comes to dominate the energy density of the universe before being thermalized, i.e., $T_{\rm RM} > T_{\rm th}$, but now for $n=2$. The results are given in Fig.~\ref{fig:sax_dom}. The diamonds show the points predicted by requiring both the baryon asymmetry and dark matter abundance, which are in tension with the red-giant observations. We nevertheless analyze this $n=2$ scenario to obtain the prediction from the dark matter abundance (black dashed curves) but with an underproduced baryon asymmetry. The prediction is sharp and points to $m_S \simeq 10 \TeV$ and $f_a \simeq (1.2\mathchar`-1.3) \times 10^9 \GeV$ as shown by the black dashed segment above the purple region. The truncation of the black dashed curves at low $m_S$ is due to thermalization constraint discussed below.

The thermalization analysis for $n = 2$ is more involved than for $n = 1$ since $\Gamma_{S\widetilde{H}\widetilde{H}}$ increases with $(T+m_S) S^{2n-2}$, and thermalization can potentially occur at temperatures higher than $T_S$ when the non-trivial scaling of $S$ may matter.  For the saxion to thermalize via scattering with Higgsinos, a thermal bath must be present with a temperature larger than the Higgsino mass parameter $\mu(T) = \mu \times (S(T) / N_{\rm DW} f_a)^2$. This bath can in principle originate from inflationary reheating or from the saxion scattering with the $W$ gauge boson. In what follows, we assume the high $T_R$ and/or large initial $S$ limit so the inflationary reheating contribution to the bath is negligible around thermalization. (For instance, for the initial $S$ close to the Planck scale, $T_R > \mathcal{O}(10^7) \GeV$ is sufficient for this assumption to hold.) The origin of the bath must be from the saxion-$W$ scattering. The temperature of the bath that originates in this way   increases~\cite{Co:2020xaf} during saxion reheating because the temperature dependence of the rate given in
Eq.~(\ref{eq:Gamma_SWW}). 

As the temperature increases, it may eventually become equal to the effective $\mu(T)$ at a temperature we call $T_{{\rm th}, i}$, at which point thermalization via Higgsinos is initiated. Thermalization may then suddenly complete via saxion-Higgsino scattering, and the temperature increases abruptly to $T_{\rm th}$ as the saxion energy is suddenly converted to the bath.  This occurs as long as $\Gamma_{S\widetilde{H}\widetilde{H}} > H$. We assume the saxion field value does not change significantly after thermalization, $S_{\rm th} \simeq S_{{\rm th}, i}$, which is the case if the initial rotation is nearly circular ($r \simeq 1$). The temperature right after thermalization $T_{\rm th}$ can be computed by requiring 1) conservation of energy $\rho_S = m_S^2 S_{{\rm th},i}^2 = \frac{\pi^2}{30} g_* T_{\rm th}^4$, 2) initial radiation created by $W$ scattering $\rho_{S}   \frac{\Gamma_{SWW}}{H} = m_S^2 S_{{\rm th}, i}^2 \frac{\Gamma_{SWW}}{H} = \frac{\pi^2}{30} g_* T_{{\rm th}, i}^4$ with the subscript ``${{\rm th}, i}$'' indicating evaluation right before the abrupt thermalization, and 3) the condition for Higgsinos to just come into thermal equilibrium $T_{{\rm th}, i} = \mu(T_{{\rm th}, i}) = \mu \times (S_{{\rm th}, i} / N_{\rm DW} f_a)^2$. 
We obtain
\begin{equation}
    T_{\rm th} = 7 \times 10^6 \GeV 
    \left( \frac{N_{\rm DW}}{6} \right)^{ \scalebox{1.01}{$\frac{1}{3}$} }
    \left( \frac{m_S}{20 \TeV} \right)^{ \scalebox{1.01}{$\frac{1}{2}$} }
    \left( \frac{m_S}{\mu} \right)^{ \scalebox{1.01}{$\frac{1}{6}$} }
    \left( \frac{f_a}{6 \times 10^8 \GeV} \right)^{ \scalebox{1.01}{$\frac{1}{3}$} } .
\end{equation}
Using this expression, we find that along the black dashed line of Fig.~\ref{fig:sax_dom}, the PQ charge yield $Y_\theta = 3 r T_{\rm th} / 4 N_{\rm DW} m_S$ matches the value required by the observed dark matter abundance via kinetic misalignment, i.e., Eq.~(\ref{eq:Ytheta_KMM}).

The black dashed line is truncated at low $m_S$ because $\Gamma_{S\widetilde{H}\widetilde{H}} < H$ when $T = T_{{\rm th}, i} = \mu(T_{{\rm th}, i})$.  That is to say, even though a bath has been created via saxion-W scattering that allows Higgsinos to come into equilibrium, the interaction rate between Higgsinos and the saxion is still too small to complete thermalization at this time.   In this case, only a small fraction, $\Gamma_{S\widetilde{H}\widetilde{H}}/H$, of the saxion energy density is transferred into the bath at this time. And since $\Gamma_{S\widetilde{H}\widetilde{H}}$ decreases faster than $H$ when $S$ is still away from the minimum at $N_{\rm DW} f_a$, thermalization is only possible after $S$ settles to the minimum so that $\Gamma_{S\widetilde{H}\widetilde{H}} \propto (T+m_S)$ can eventually overtake $H$. However, in this regime, we find that axion dark matter is underproduced because the low $T_{\rm th}$ gives an insufficient PQ charge $Y_\theta = 3 r T_{\rm th} / 4 N_{\rm DW} m_S$.

In deriving this black dashed line, we have assumed $r = 1$. One may be tempted to naively extend the calculation to derive the prediction for lower values of $r$ because $r$ seemingly appears to affect only $S_{\rm th} / S_{{\rm th}, i}$. For $r < 1$, $\Gamma_{S\widetilde{H}\widetilde{H}} \propto S^2$ may first be larger than $H$ when $T$ reaches $T_{{\rm th}, i}$ but become smaller than $H$ before reaching complete thermalization at $T_{\rm th}$. The condition $\Gamma_{S\widetilde{H}\widetilde{H}} > H$ should instead be evaluated at $T_{\rm th}$ with $S_{\rm th}$ rather than at $T_{{\rm th}, i}$. However, we note that thermalization via Higgsino scattering may be further complicated by the fact that the value of $S$ can get close to the origin in some portion of the cycle when the rotations are very elliptical, $r \ll 1$. During this portion, the saxion-Higgsino scattering may proceed because $T < \mu(S)$ near the origin, while $T > \mu(S)$ when $P$ is far away from the origin. We do not pursue this possibility further.

Finally, we show diamonds along the black dashed line to indicate the prediction from lepto-axiogenesis for the different hierarchical neutrino mass spectra. In deriving these predictions, we again numerically solve the coupled Boltzmann equations for the saxion and radiation with a non-trivial thermalization rate scaling and then integrate $\dot n_{B-L} R^3$ to obtain the final $Y_B$. As can be seen in the figure, lower values of $f_a$ are preferred by the predictions, but such low $f_a$ is in tension with the red giant constraints. In fact, even the degenerate limit of the neutrino spectrum leads to underproduction of the baryon asymmetry. Lowering $\mu$ may increase the predicted values of $f_a$, moving towards compatibility with the red-giant bound, but the black dashed line is truncated at lower $f_a$ . As a result, we do not find viable parameter space for a sufficient baryon asymmetry after marginalizing over $\mu$.

\subsection{Interpretation of results for one-field model}
\label{sec:one-field}

\begin{figure}
\includegraphics[width=\linewidth]{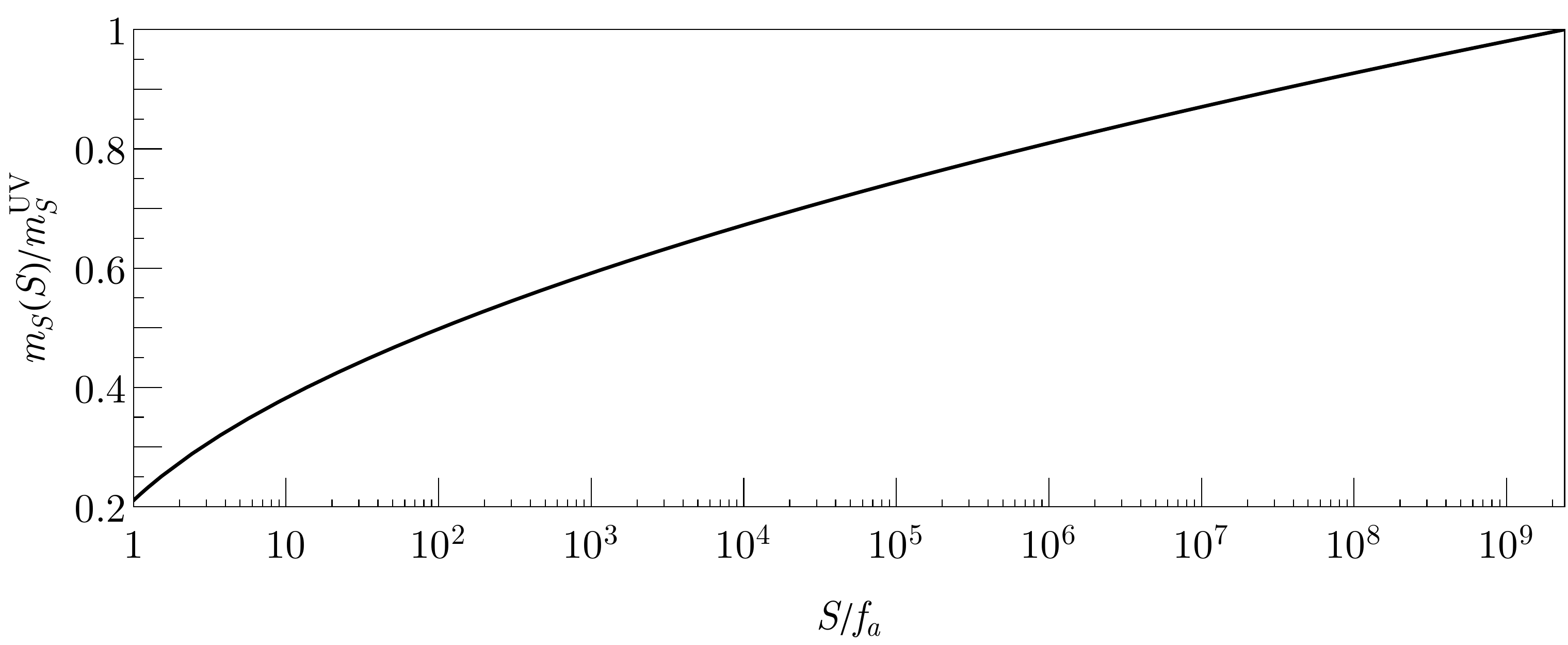}
\caption{Curvature of the PQ-field potential normalized to the UV value in the one-field model as a function of the ratio of the field value $S$ to the axion decay constant $f_a$.}
\label{fig:mS_log}
\end{figure}

In this subsection, we re-interpret the results presented in Figs.~\ref{fig:n1NDW3}, \ref{fig:n1NDW3_DM}, \ref{fig:n2NDW6_mu}, and \ref{fig:sax_dom} for the one-field model defined in Eq.~(\ref{eq:one-field}). This model requires special treatment because, unlike the two-field model of Eq.~(\ref{eq:two-field}), the curvature of the potential in the radial direction is logarithmically enhanced at large field values for $S$.

In what follows, we define $m_S^{\rm UV}$ as the curvature at $S = M_{\rm Pl}$.  We expect this value to be comparable to other scalar masses in the UV. We will discuss how our earlier results are modified under the understanding that the $x$-axes of the figures will now refer to $m_S^{\rm UV}$.  We denote $m_S(z)$ as the curvature at lower energy scales; $z$ may refer to the $S$ field value or a temperature to indicate the corresponding field value $S(T)$ at $T$, i.e., $m_S(T) = m_S(S(T))$. The field dependence of $m_S(S)$ is shown in Fig.~\ref{fig:mS_log}. Because the change of $m_S(z)$ is only logarithmic, the overall effect on the results is modest. In what follows, we will discuss this effect in detail. Our strategy will be to fix $f_{a}$ and find the value of $m_{S}^{\rm UV}$ that would reproduce the physics of a field-independent $m_{S}$ in each case.

We begin by discussing the effects of the evolution of $m_S(z)$ on saxion thermalization. The green dotted line in Fig.~\ref{fig:n1NDW3}, the green boundary of Fig.~\ref{fig:n1NDW3_DM}, and the positively-sloped green boundary of Fig.~\ref{fig:n2NDW6_mu} are all determined by thermalization requirements, and they are set such that $Y_\theta = 3 r T_{\rm th} / 4 N_{\rm DW} m_S(T_{\rm th})$ reproduces the required PQ charge yield $Y_{\theta,\,{\rm KMM}}$ in Eq.~(\ref{eq:Ytheta_KMM}).
In Figs.~\ref{fig:n1NDW3} and \ref{fig:n1NDW3_DM}, $T_{\rm th}$ is determined via Eq.~(\ref{eq:Tth}) and scales as the low-energy value of the Higgsino mass squared, $\mu^2$, which we expect to be $\left(m_S^{\rm UV}\right)^2$.
Thus, $Y_\theta \propto \left(m_S^{\rm UV}\right)^2 / m_S(T_{\rm th})$. For a fixed $f_a$, we can find the correct value of $m_S^{\rm UV}$ by ensuring $\left(m_S^{\rm UV}\right)^2 / m_S(T_{\rm th})$ is equal to the constant $m_S$ of our previous analysis.   

Thermalization occurs when the saxion is at (close to) the minimum, for low  (high) $m_S$, as can be seen in Fig.~\ref{fig:n1NDW3_temp}. Therefore, $m_S(T_{\rm th}) \simeq (0.2\mathchar`-0.5) m_S^{\rm UV}$ according to Fig.~\ref{fig:mS_log}, and thus $\left(m_S^{\rm UV}\right)^2 / m_S(T_{\rm th}) = (2\mathchar`-5)m_S^{\rm UV}$. The green line/region will then shift to the left by a factor of 2-5 to compensate for this. On the other hand, for Fig.~\ref{fig:n2NDW6_mu}, the positively-sloped boundaries are in fact unaffected because the condition given in Eq.~(\ref{eq:faMaxSoTth_eq_Tmu}) depends on only $\mu \approx m_S^{\rm UV}$. The negatively-sloped boundaries are set by $T_{\rm th} = T_{\rm RM}$, where $T_{\rm th} = T_\mu$. This condition translates to $m_S(T_{\rm RM})/(m_S(T_S) / \mu)^{1/2}$ being equal to the $m_S$ derived for the fixed curvature case;
this condition has an accidental cancellation numerically so the boundaries do not move appreciably.

We now discuss how the predictions of $m_S$ that reproduce the baryon asymmetry are affected by $m_S(z)$. During the epoch where the $\Delta Y_{B-L}$ is a constant, the dependence of the total $Y_{B-L}$ on $m_S(T)$ is through a now slightly temperature dependent $\left|\dot\theta(T) \right|= N_{\rm DW} m_{S}(T)$ for $T_i > T > T_f$. For Figs.~\ref{fig:n1NDW3} and \ref{fig:n1NDW3_DM} ($n=1$), $T_i$ is $T_R$ and $T_f$ is often $T_S$ (see Fig.~\ref{fig:n1NDW3_temp}). For Fig.~\ref{fig:n2NDW6_mu} ($n=2$), $T_i$ is $T_\mu$, which is not much above $T_S$, and $T_f$ is often $T_S$. Since $m_S$ is lower at these lower temperatures compared to $m_S^{\rm UV}$, the effect is to reduce the efficiency of $Y_{B-L}$ production.  To compensate for this, $m_S^{\rm UV}$ needs to increase by a factor of a few. This results in a shift of the prediction curves to the right. As a result of this shift and the left shift of the green dotted line in Fig.~\ref{fig:n1NDW3}, the hierarchical cases shown by blue and red curves are more easily compatible with the red giant bound and the green constraint for dark matter. In other words, a viable parameter space for dark matter would open up with a milder hierarchy between $\mu$ and $m_S$ than $\mu = 3 m_S$ assumed in Fig.~\ref{fig:n1NDW3_DM}. The shift would be more prominent in Fig.~\ref{fig:n2NDW6_mu} than Fig.~\ref{fig:n1NDW3} because the former case involves $m_S(T)$ only at temperatures close to $T_S$.

The brown regions/curves in Figs.~\ref{fig:n1NDW3_DM} and \ref{fig:n2NDW6_mu} are also affected by a changing $m_S$. As explained in Secs.~\ref{sec:param_n1} and~\ref{sec:param_n2}, these constraints occur because a successful production of $Y_B$ would require $T_{\rm RM} > T_R/2$, which would result in a period of inflation driven by the saxion. First, as explained above, an evolving $m_S(T)$ decreases the efficiency of production relative to the constant case, so to reproduce the baryon asymmetry, a larger $m_S^{\rm UV}$ is required. Second, because $m_{S}(T_{\rm RM})$ is smaller than $m_{S}^{\rm UV}$,  the saxion will take longer to come to dominate, and so $T_{\rm RM}$ is smaller in the case where the saxion mass evolves.  This means the constraint is relaxed, which also shifts the brown regions/curves to higher $m_S^{\rm UV}$.

Lastly, in Fig.~\ref{fig:sax_dom}, the dominant (logarithmically enhanced) era of asymmetry production is present between $T_{\rm th}$ and $T_{\rm RM} = r T_{\rm th}$. For $r = \mathcal{O}(1)$, $T_{\rm RM} \simeq T_{\rm th} \gg T_S$, so $m_S(T)$ during this era is $\mathcal{O}(0.5) m_S^{\rm UV}$, and the predicted points will shift to the right by a factor of 2 or so. The predicted points for $r \ll 1$ are excluded by red giants whether or not we account for the effect of $m_S(z)$.

On balance, for the one-field model, larger values of $m_{S}$ are preferred than in the two-field case, often by a factor of few.

We discuss the potential domain wall problem in the one-field model. After the initiation but before the thermalization, the rotation is generically not circular. For non-circular motion, fluctuations of the PQ breaking field can be produced by parametric resonance~\cite{Dolgov:1989us, Traschen:1990sw, Kofman:1994rk, Shtanov:1994ce, Kofman:1997yn,Co:2020dya,Co:2020jtv}. The PQ symmetry may be non-thermally restored by the fluctuations and broken again once the fluctuations are reduced by the expansion of the universe. If this actually occurs, a domain wall-string network is produced, which is stable if $N_{\rm DM}>1$ and causes a domain wall problem. Unlike the case without angular momentum~\cite{Kasuya:1996ns,Kasuya:1998td}, it is not clear if the restoration actually occurs, since the non-zero angular momentum provides an effective potential that strongly disfavors the origin of the field space. We leave the investigation of the dynamics via numerical lattice computation to future work, and only note that the one-field model may require $N_{\rm DW}=1$ or explicit PQ breaking that can destroy the domain walls~$N_{\rm DW}>1$~\cite{Sikivie:1982qv}.

\subsection{Q-balls}
\label{sec:Qballs}
If the potential of the $S$ field is nearly quadratic, a small correction may make the potential flatter than a quadratic one, for which a non-topological soliton called a Q-ball may be formed~\cite{Coleman:1985ki,Kusenko:1997zq,Kusenko:1997si,Kasuya:1999wu,Dine:2003ax}.  Q-ball formation can complicate the thermal history.  If formed, Q-balls will localize the PQ charge inside them. It is unclear as to what the spatial distribution of the $\dot{\theta}$ will be as the universe evolves in the presence of these Q-balls.  This uncertainty would confuse the evaluation of the baryon asymmetry.

Most discussions of Q-balls have taken place in the context of potentials with minima near the origin in field space. It is possible that the symmetry-breaking potential of $P$ allows the Q-balls to decay or even prevents its initial formation.  While understanding the dynamics of the Q-balls associated with a symmetry-breaking potential such as the one needed for the axion is of interest, we leave it for future work.  For now, we assume that the properties of the Q-balls in the present setup are identical to the more familiar ones associated with potentials that have minima at the origin.  We then comment on which cases Q-ball formation might confuse the calculation of the baryon asymmetry, while keeping in mind that future investigations might mitigate these concerns.  Histories that include Q-ball formation may actually ultimately prove viable.

For $n=1$, the thermal potential, given in the second terms of Eqs.~(\ref{eq:Vth_high_mu}) and (\ref{eq:Vth_low_mu}) below, is flatter than a quadratic one for both $\mu(S)>T$ and $\mu(S)<T$, so once Q-balls are formed, they would remain stable until $T \ll m_S$ when the quantum correction to the soft mass of $S$ from interactions with the Higgs fields dominates over the thermal potential. In this case, the estimation of the baryon asymmetry would potentially be rendered invalid, because the Q-balls would be present during the epoch that is important for the generation of the asymmetry. We may avoid the era of a flat potential by coupling $P$ to additional fields, $W= y_{\psi} P \psi \bar{\psi}$. Because we will require a large $y_{\psi}$, the $\psi$ fields receive a large mass from the large $P$ field value and are not present in the thermal bath. Assuming that it is gauge-singlet, $\psi$ also does not introduce a coupling of $P$ to gauge bosons.  So, the effect of $\psi$ is to introduce a modification of the zero-temperature potential.  Assuming that the soft mass squared of $\psi \bar{\psi}$ is positive, quantum corrections to the soft mass of $P$ induced by this coupling steepens the zero-temperature potential and can destabilize Q-balls.
So, with an ${\mathcal O(1)}$ coupling $y_{\psi}$, for $\mu(S)>T$ the non-quadratic part of the potential of $S$ is
\begin{align}
\label{eq:Vth_high_mu}
V \supset \kappa m_S^2 S^2 {\rm ln}\frac{S}{\mu} + \alpha_2^2 T^4 {\rm ln} \frac{S}{T} ,
\end{align}
where $\kappa \sim 1/(16\pi^2)$.  Q-ball solutions exist if $V/S^2$ is minimized at non-zero $S$. The above potential has a minimum at $S^2\sim \alpha^2 T^4 / (\kappa m_S^2)$. Requiring self-consistency with the condition $\mu(S)>T$, we obtain
\begin{align}
\label{eq:Q_Tbound_large}
    T > \frac{1}{g_2^2} \left(\frac{\kappa}{1/16\pi^2}\right)^{1/2}N_{\rm DW} \frac{m_S}{\mu} f_a \quad \quad ({\rm Q\mathchar`-balls: \; } \mu(S)>T).
\end{align}
The Q-ball solution may also exist in the regime $\mu(S)<T$,  for which
the non-quadratic part of the potential of $S$ is
\begin{align}
\label{eq:Vth_low_mu}
    V \supset \kappa m_S^2 S^2 {\rm ln}\frac{S}{\mu} - c_T y^4 S^4,
\end{align}
where $y = \mu / (f_a N_{\rm DW})$ is the coupling between $P$ and $H_u H_d$, $ c_T \sim 1/(16\pi^2)$.
Note that the thermal trilinear term $-y^3 S^3 T$ is absent since the Higgs field obtains a large thermal mass $\sim g T$ and the IR singularity is removed. The minimum of $V/S^2$ is at $S^2\sim \kappa m_S^2/(c_Ty^4)$. For consistency, this should satisfy $y S < T$, so we obtain
\begin{align}
\label{eq:Q_Tbound_small}
    T > \left(\frac{\kappa}{c_T}\right)^{1/2}N_{\rm DW} \frac{m_S}{\mu} f_a  \quad \quad ({\rm Q\mathchar`-balls: \; } \mu(S)<T).
\end{align}
Comparing Eq.~\eqref{eq:Q_Tbound_large} with Eq.~\eqref{eq:Q_Tbound_small}, the latter gives a slightly stronger condition, so Q-balls disappear when Eq.~\eqref{eq:Q_Tbound_small} is violated.
Unless $T_R \gg f_a$, the production of $B-L$ asymmetry dominantly occurs after Q-balls disappear, so the estimation of $B-L$ asymmetry is not affected by the production of Q-balls.  Given current bounds on $f_a$ and constraints on $T_R$ from BBN, we do not expect $T_R \gg f_a$.

For $n=2$, the potential of $S$ is flatter than a quadratic one only for $\mu(S) >T$.  Therefore, even if Q-balls are formed, once the field value of $S$ inside the Q-balls is such that $\mu(S)<T$, Q-balls should disappear. However, this can occur only at a temperature below $T_\mu$, since the field value of $S$ inside the Q-balls is larger than the average field value. With Q-balls at temperatures below $T_\mu$,  the estimation of $B-L$ asymmetry may be affected. We may avoid this by a coupling $W= P \psi \bar{\psi}$ as in $n=1$.
So, for $n=2$, when the condition in Eq.~\eqref{eq:Q_Tbound_large} is violated, Q-balls disappear.

So, for both $n=1$ and $2$, even if Q-balls form at the early stage of the evolution of the axion rotation, they can disappear by the era when $B-L$ asymmetry is produced by lepto-axiogenesis if there exists a coupling to extra fields $\psi \bar{\psi}$. We stress again that this extra couplings may not be necessary because the symmetry breaking potential of $S$ may lead to additional effects that destabilize the Q-balls.

We note that the Q-ball formation may lead to production of domain walls. Indeed, Q-ball formation is a result of the growth of fluctuations. As in the parametric resonance during the oscillation of the PQ symmetry breaking field~\cite{Kasuya:1996ns,Kasuya:1998td}, the growth of fluctuations may non-thermally restore the PQ symmetry and produce domain walls. Since $N_{\rm DW} >1$ for the DFSZ model, domain walls are stable and will come to  dominate the universe. However, we expect that the symmetry restoration would not occur in the two-field model since the PQ symmetry-breaking fields are fixed on the moduli space where the PQ symmetry is broken. In the one-field model, on the other hand, the symmetry restoration might occur. Whether or not the domain wall production actually occurs should be investigated by numerical computation; it is possible that the non-zero angular momentum in field space tends to expel the field from the center and prevent the symmetry restoration.

In summary, it remains to be seen whether or not Q-balls, if formed, are ultimately problematic, and whether they disturb the calculation of the baryon asymmetry.  However, coupling  the PQ-field to other fields induces quantum corrections to the saxion potential that steepen it and can  avert Q-ball production.

\section{Discussion}
\label{sec:discussion}

In this work we have explored the possibility that the observed baryon asymmetry arises from the interplay of early-universe dynamics of the axion and the origin of neutrino masses.  Under this assumption, 
we could obtain information on the mass of the saxion, the radial mode of the complex field that contains the axion.  In models of gravity mediation, the mass of the saxion would be comparable to the masses 
of the MSSM particles. So, one can interpret the results as predictions for the masses of the superpartners. We have investigated the DFSZ model in detail including the successful thermalization of the saxion.

For a hierarchical neutrino mass spectrum, the scalar mass may be as low as $\mathcal{O}(10)$ TeV. The observed Higgs boson mass in this case may be explained by moderately large $\tan\beta$.
For the scalar mass of $\mathcal{O}(10)$ TeV, the gaugino masses given by the anomaly mediation~\cite{Randall:1998uk,Giudice:1998xp} is below $\mathcal{O}(100)$ GeV, so singlet SUSY-breaking fields must be present to give phenomenologically viable gaugino masses. This generically leads to the Polonyi problem~\cite{Coughlan:1983ci}, which can be avoided by a large coupling between the SUSY-breaking fields and the inflaton~\cite{Linde:1996cx,Takahashi:2011as,Nakayama:2012mf} or a coupling between the SUSY-breaking fields and a pseudo-flat direction~\cite{Harigaya:2013ns}.

Successful thermalization of the rotation typically requires $\mu$ different from $m_S$ by an $\mathcal{O}(1)$ factor. For $\mu > m_S$, electroweak symmetry breaking requires the soft masses of the Higgs fields to be also larger than $m_S$ by an $\mathcal{O}(1)$ factor.

If reheat temperatures are somewhat lower than the maximum value considered here, or if the saxion comes to dominate the energy density of the universe at some point in its history, then the scalar mass is required to be larger. Interestingly, after requiring the kinetic misalignment mechanism to explain the observed dark matter abundance, we find the scalar mass is at most 300 TeV. (One can check that the predicted scalar mass is still small enough that the tachyonic instability to create a helical magnetic field is ineffective, so the associated overproduction of the baryon asymmetry recently noted in Ref.~\cite{Domcke:2022uue} is avoided.) The scalar mass of 300 TeV is compatible with the scenario without singlet supersymmetry-breaking fields~\cite{Giudice:1998xp}, also known as mini-split SUSY, pure gravity mediation, spread SUSY, etc. In this scenario, the infamous Polonyi problem and the BBN gravitino problem are absent, the SUSY flavor/CP problem  mitigated, and the observed Higgs boson mass in this case can be explained with $\tan\beta$ of order unity~\cite{Wells:2003tf,ArkaniHamed:2004fb,Giudice:2004tc,Wells:2004di,Ibe:2006de,Acharya:2007rc,Hall:2011jd,Ibe:2011aa,Arvanitaki:2012ps,ArkaniHamed:2012gw}. The dominant contribution to the gaugino mass is given by anomaly mediation~\cite{Randall:1998uk,Giudice:1998xp}, and the gauginos may be searched for at the LHC.

As for the axions, we find a preferred region that simultaneously predicts the dark matter and the baryon asymmetry with $f_{a} \sim 10^{9}$ GeV, just above the current bound from observations of red giants.  This presents a target for experimental searches including the Broadband Reflector Experiment for Axion Detection (BREAD) \cite{BREAD:2021tpx}, the Axion Resonant InterAction Detection Experiment (ARIADNE) \cite{Arvanitaki:2012cn,ARIADNE:2017tdd}, or other future detectors \cite{Adams:2022pbo}.

Questions regarding the dynamics of the rotating axion field remain.  As discussed in Sec.~\ref{sec:Qballs}, Q-balls can form when the saxion potential is flatter than a quadratic one. The spatial distribution of the angular velocity of the axion field after Q-balls form, but prior to their decay, is of importance to accurately estimate the efficiency of axiogenesis scenarios.  In Sec.~\ref{sec:Qballs}, we introduced new couplings of the PQ-field to hasten the disappearance of the Q-balls, rendering them irrelevant. However, even in the absence of these additional couplings, we expect  Q-balls to eventually decay since the zero-temperature potential does not admit isolated Q-ball solutions. When the decay actually occurs requires additional investigation, perhaps with the help of a lattice computation.   Because the requirement of a large initial field value constrains the potential of the saxion to be nearly quadratic in axiogenesis,  the condition for an epoch of Q-ball formation should be  satisfied rather generically.  This makes answering the fate of axiogenesis in the presence of Q-balls a particularly interesting question.

\section*{Acknowledgements}
The work is  supported by the U.S. Department of Energy, Office of Science, under Award number
 DE-SC0011842 at the University of Minnesota (R.C.) and DE-SC0007859 at the University of Michigan (A.P.).

\appendix

\section{Computation of chemical potentials}
\label{app:cB}
In this appendix, we calculate the chemical potentials for Eq.~(\ref{eq:nDotB-L_General}), which in turn allow us to compute the $C_{i}(T)$ of Eq.~(\ref{eq:nB_dot}) that are necessary to evaluate the baryon asymmetry.
To calculate the chemical potentials we apply the principle of detailed balance to scattering processes in equilibrium~\cite{Harvey:1990qw}.  This sets the sum of the chemical potentials participating in a given reaction to zero.  If a certain scattering process is out of equilibrium, we replace the equilibrium condition with a corresponding conservation law.  Solving the resulting system of equations allows for the determination of the chemical potentials.

We will discuss the equilibrium condition for each scattering process and the corresponding conservation laws.  In the present case, the scattering  processes include Yukawa interactions, electroweak and strong sphaleron processes, gaugino masses, and the $\mu$-term.

\subsection{All interactions in equilibrium}

At low temperatures, all Yukawa couplings, sphaleron processes, and mass terms are in thermal equilibrium. Because of the explicit PQ breaking by the QCD anomaly, the rotation is slowly washed out, and it would vanish at the true thermal equilibrium. However, the washout rate is much smaller than the Hubble expansion rate, and the true thermal equilibrium is never reached~\cite{Co:2019wyp}. Instead, we should consider a quasi-equilibrium state where $\dot{\theta}$ is taken to be constant with its value determined by the potential of the saxion. The quasi-equilibrium can be found by taking the time derivatives of the MSSM particle number asymmetry in the Boltzmann equations to vanish. The solution to this system of equations depends on the magnitudes of coupling constants. However, because the  up-Yukawa coupling is small, it can be set to zero to a good approximation~\cite{Co:2021qgl}. And while the goal is to find the (quasi)-equilibrium values for the case where the chiral symmetry is completely broken, this procedure, wherein we take the parameter which breaks the chiral asymmetry the least (the up-Yukawa) to vanish, will reproduce the leading contribution to the asymmetry.
Then, with this prescription for the up-Yukawa coupling in place, taking the time-derivatives to be zero is equivalent to applying the principle of detailed balance to each scattering process. 

The equilibrium conditions for the remaining Yukawa interactions are
\begin{align}
    \mu_{\ell_i}+\mu_{\bar{e}_i}+\mu_{\widetilde{H}_d}+\mu_{\lambda} & = 0, \label{eq:E-type-Yukawa}\\
    \mu_{Q_2}+\mu_{\bar{u}_2}+\mu_{\widetilde{H}_u}+\mu_{\lambda} & = 0, \\
    \mu_{Q_3}+\mu_{\bar{u}_3}+\mu_{\widetilde{H}_u}+\mu_{\lambda} & = 0, \\
    \mu_{Q_i}+\mu_{\bar{d}_j}+\mu_{\widetilde{H}_d}+\mu_{\lambda} & = 0. \label{eq:D-type-Yukawa}
\end{align}
We have chosen to express equilibrium conditions in terms of the fermionic part of each chiral supermultiplet, and $\mu_\lambda$ is the chemical potential of gauginos. Since the doublet quarks and squarks couple to all gauginos, as long as the gauge interaction is in thermal equilibrium, all gauginos have the same chemical potential.  The scalar and fermionic chemical potentials---owing to in equilibrium interactions with gauginos---are related by
\begin{equation}
    \mu_{\lambda}+\mu_{\psi}-\mu_{\phi} = 0, \label{eq:scalarFermionRelation}
\end{equation}
where $\phi$ and $\psi$ represent the scalar and fermion part of a chiral supermultiplet, respectively.
While the charged lepton and up-quark Yukawa interactions may be taken to be flavor diagonal, in general, there will be off-diagonal components for the down-quarks, see Eq.~(\ref{eq:D-type-Yukawa}). Note that four of the nine equations in  Eq.~(\ref{eq:D-type-Yukawa}) are redundant.  Among the linearly dependent equilibrium conditions, it is convenient to use those which violate conservation laws with the largest rate. It is that rate which sets the temperature at which the conservation law is broken and the corresponding equilibrium condition is satisfied.  We choose the $(i,j) = (1,1)$, $(1,2)$, $(2,2)$, $(2,3)$, and $(3,3)$ parts of Eq.~(\ref{eq:D-type-Yukawa}). The equilibrium conditions for the electroweak and strong sphalerons are
\begin{align}
    \sum_{k=1}^{N_g}(3\mu_{Q_k}+\mu_{\ell_k})+\mu_{\widetilde{H}_u}+\mu_{\widetilde{H}_d}+4\mu_{\lambda}+c_W \mu_{\theta} & = 0,  \label{eq:WeakSphaleron}\\
    \sum_{k=1}^{N_g}(2\mu_{Q_k}+\mu_{\bar{u}_k}+\mu_{\bar{d}_k})+6\mu_{\lambda}+c_g \mu_{\theta} & = 0, \label{eq:StrongSphaleron}
\end{align}
where $c_W$ and $c_g$ are the weak and strong anomaly coefficients of the PQ symmetry.  These anomaly coefficients are set to zero in the DFSZ case, but not the KSVZ case. Because $\rho_\theta$ is given as $-\dot{\theta} n_{\theta}$, $\mu_{\theta}$ must be $-\dot{\theta}$.  Other interactions to consider are chiral-symmetry violation by the gaugino mass and either the standard MSSM $\mu$-term (for KSVZ) or the interaction in Eq.~(\ref{eq:Wmu}) (for DFSZ), which give
\begin{align}
    \mu_{\lambda} & = 0, \\
    \mu_{\widetilde{H}_u}+\mu_{\widetilde{H}_d}+\frac{n}{N_{\rm DW}}\mu_{\theta} & = 0. \label{eq:mufromWmu}
\end{align}
Setting $n=0$ in Eq.~(\ref{eq:mufromWmu}) corresponds to taking the standard MSSM $\mu$-term for KSVZ, while taking $n \neq 0$ corresponds to the DFSZ case.

In addition to the above detailed balance relations, we must also impose the conservation laws to determine the asymmetry. With all interactions in thermal equilibrium, the only conservation laws are those of weak hypercharge, $Y=0$, and $B/3-L_i$ for each generation~$i$.
$B/3-L_i$ is violated by the superpotential in Eq.~(\ref{eq:Wnu}), but this interaction is never close to equilibrium, and it is therefore a small perturbation that may be neglected for the computation of chemical potentials.  For $\mu_{i}, m_i \ll T$, the net fermion and boson densities are given by $n_{\psi}-n_{\psi^\dag} = \frac{g}{6}T^2\mu_{\psi}$ and $n_{\phi}-n_{\bar{\phi}} = \frac{g}{3}T^2\mu_{\phi}$, so the hypercharge and $B/3-L_i$ conservation conditions can be expressed in terms of chemical potentials as
\begin{align}
    \sum_{k=1}^{N_g}(\mu_{Q_k}-\mu_{\ell_k}-2\mu_{\bar{u}_k}+\mu_{\bar{d}_k}+\mu_{\bar{e}_k})+\mu_{\widetilde{H}_u}-\mu_{\widetilde{H}_d} & = 0, \\
    \sum_{k=1}^{N_g}(2\mu_{Q_k}-\mu_{\bar{u}_k}-\mu_{\bar{d}_k})-6\mu_{\ell_i}+3\mu_{\bar{e}_i}-2\mu_{\lambda} & = 0.
\end{align}

\subsection{Out of equilibrium Yukawa interactions}

The conservation law that is broken at the lowest temperature is $\bar{e}_1$ number conservation.  This is finally broken when the scattering rate involving the electron Yukawa coupling with rate $\alpha_2 y_e^2 T$ overtakes the Hubble expansion rate.  When $\bar{e}_1$ number is conserved, Eq.~(\ref{eq:E-type-Yukawa}) for $i=1$ should be replaced by
\begin{align}
\mu_{\bar{e}_1}=0.
\end{align}
This will result in a solution to the chemical potentials for leptons that depends on the generation. The 
conservation law that persists to the next lowest temperature
is $\bar{u}_1-\bar{d}_1$ number, which is broken by the down-Yukawa interaction with a rate $\alpha_3 |Y^d_{11}|^2 T$.  When $\bar{u}_1-\bar{d}_1$ number is conserved, the $(i,j) = (1,1)$ component of Eq.~(\ref{eq:D-type-Yukawa}) should be replaced~by
\begin{align}
    \mu_{\bar{u}_1}-\mu_{\bar{d}_1} = 0.
\end{align}
The last symmetry we consider is $3B_1-B$, which is broken by off-diagonal down-type Yukawa interactions of the first generation with the second and third generations. Because of the large charm and top Yukawa couplings, we take a quark basis where the down-type Yukawa matrix $Y^d = V_{\rm CKM} {\rm diag} (y_d,y_s,y_b)$ and the up-type Yukawa is diagonal.  In this basis, the dominant contributions to $3B_1-B$ breaking are from the interactions of $Q_1$ with $d_2$ and $d_3$, so the rate of symmetry breaking is $\alpha_3 (|Y^d_{12}|^2+|Y^d_{13}|^2) T$.  When $3B_1-B$ is conserved, the $(i,j) = (1,2)$ component of Eq.~(\ref{eq:D-type-Yukawa}) should be replaced by
\begin{equation}
2(2\mu_{Q_1}-\mu_{\bar{u}_1}-\mu_{\bar{d}_1})-(2\mu_{Q_2}-\mu_{\bar{u}_2}-\mu_{\bar{d}_2})-(2\mu_{Q_3}-\mu_{\bar{u}_3}-\mu_{\bar{d}_3}) = 0.
\end{equation}
Other Yukawa interactions could be  out of equilibrium, but this would occur at high enough temperatures that the $B-L$ production by lepto-axiogenesis is subdominant.

\begin{figure}
\includegraphics[width=0.6\linewidth]{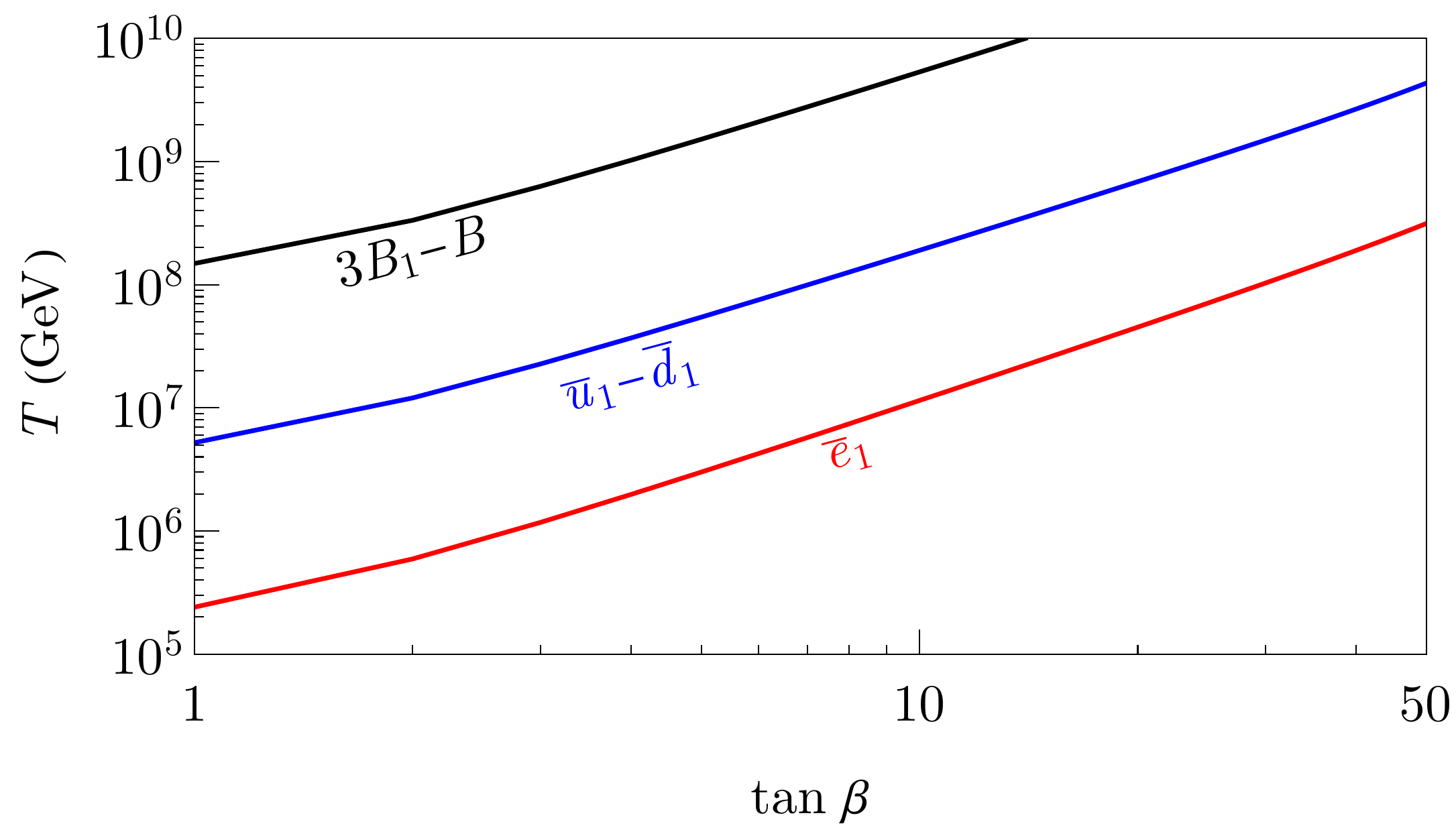}
\caption{Temperatures at which conservation laws are broken by Yukawa interactions coming into equilibrium.  These temperatures are functions of $\tan \beta$, the ratio of Higgs field vacuum expectation values.  The $\bar{e}_1$ number is broken by the electron-Yukawa interaction, $\bar{u}_1-\bar{d}_1$ by the down-quark-Yukawa interaction, and $3B_1-B$ by off-diagonal down-type-quark-Yukawa interactions.}
\label{fig:conservation_laws}
\end{figure}

The temperatures at which these different conservation laws are broken are shown in Fig.~\ref{fig:conservation_laws}.  These are functions of $\tan \beta$ because of the dependence of the MSSM Yukawa matrices on $\tan \beta$.  Ignoring threshold corrections from integrating out superpartners, $Y^u = Y^u_{\rm SM}/\sin \beta$, $Y^d = Y^d_{\rm SM}/\cos \beta$, and $Y^e = Y^e_{\rm SM}/\cos \beta$.  We take the gauge and SM Yukawa couplings defined at a scale of 10 TeV from Ref.~\cite{Antusch:2013jca} and then run them using the 1-loop RGEs~\cite{Martin:1993zk} of the MSSM.  In Fig.~\ref{fig:conservation_laws}, it is assumed that the Hubble parameter is that of a radiation-dominated universe with $g_*=g_{\rm MSSM}=228.75$.  A universe with fewer relativistic degrees of freedom would break the symmetries at a higher temperature, while a universe not dominated by radiation would break them at a lower temperature. 
  
Whether or not the high temperatures where these new conservation laws apply are compatible with constraints from supersymmetric relics, see Appendix~\ref{app:relics}, depends on the details of the spectrum.

\subsection{Out of equilibrium gaugino masses and $\mu$-term}

At sufficiently high temperatures, scattering due to the gaugino mass or the $\mu$-term may be ineffective.   

First, we discuss the gaugino mass.  The rate of chiral-symmetry violation by the gaugino mass is $\Gamma \sim m_{\lambda}^2/T$.  Equating this rate with the Hubble expansion rate during a radiation-dominated era, we find this interaction goes out of equilibrium for temperatures above 
\begin{equation}
\label{eq:Tlambda}
T_{\lambda} \simeq  (m_{\lambda}^2 M_{\rm Pl})^{1/3} =10^{8} {\rm \; GeV} \left( \frac{m_{\lambda}}{1 {\rm \; TeV}} \right)^{2/3} .
\end{equation}
 
Above these temperatures, the chemical potential associated with the gauginos $\mu_\lambda$ will no longer vanish, and it will enter into the equations that result from the Yukawa interactions and weak and strong sphalerons, see Eqs.~\eqref{eq:E-type-Yukawa}-\eqref{eq:D-type-Yukawa}, \eqref{eq:WeakSphaleron}, and \eqref{eq:StrongSphaleron}.
Another chiral symmetry, $R$-symmetry, is present, and we must impose an additional conservation law in our system of equations. 

To find the new conservation law, we must identify the relevant $R$-symmetry.  It  
should be non-anomalous with respect to $SU(3)$ and $SU(2)$.  One set of $R$-charge assignments for this symmetry for the MSSM superfields is given in Table~\ref{tab:charges}.  Because this symmetry is an $R$-symmetry, the charge of the fermions is $Q_{\psi} = Q_R-1$.   Gauginos have $Q_{\lambda}=1$.  Contributions to would-be anomalies are $2N_{c}$ from gauginos, and $Q_R \times N$ from the chiral superfields, where $N$ counts the multiplicity.  This allows us to verify that the would-be $R\mathchar`-SU(2)\mathchar`-SU(2)$ anomaly cancels between winos and the leptons ($4-(4/3)N_{g}$=0).  The would-be $R\mathchar`-SU(3)\mathchar`-SU(3)$ anomaly  cancels between gluinos and the right-handed quarks $(6+2N_{g}(-1)=0)$, where we have combined the contributions from the up-type and down-type quarks.  In terms of chemical potentials, the conservation condition for this $R$-symmetry~is
\begin{multline}
    12\mu_{\lambda}+\sum_{k=1}^{N_g} \left[-3(\mu_{\bar{u}_k}+\mu_{\bar{d}_k})+\frac{1}{3}\mu_{\bar{e}_k}-\frac{8}{3}\mu_{\ell_k}+12(\mu_{Q_k}+\mu_{\lambda}) \right. \\ \left. +4(\mu_{\widetilde{H}_u}+\mu_{\widetilde{H}_d}+2\mu_{\lambda})+\frac{8}{3}(\mu_{\bar{e}_k}+\mu_{\lambda})-\frac{4}{3}(\mu_{\ell_k}+\mu_{\lambda}) \right] = 0.
\end{multline}
If the axino-gluino-gluon or axino-Higgs-Higgsino coupling  were in thermal equilibrium, the axino would contribute $\mu_{\lambda}$ or  $\mu_{\widetilde{H}_u}+\mu_{\widetilde{H}_d}+ \mu_\lambda$ to this expression for the conserved charge.  This affects the values of $C_i$ by at most a few percent, so we ignore these possibilities.
\begingroup
\setlength{\tabcolsep}{4pt} 
\renewcommand{\arraystretch}{1.2}
\begin{table}
    \centering
    \begin{tabular}{c|c|c|c|c|c|c|c|c|}
     &$Q$ & $U_{1}^{c}$ & $U_{2,3}^{c}$& $D^{c}$& $E^{c}$ &$L$ & $H_{u}$& $H_{d}$  \\ \hline \hline
    $Q_{R}$ & 1  & 0 & 0  & 0 &$\frac{4}{3}$& $-\frac{1}{3}$&1 & 1 \\ \hline
    $WW'$& $-1$ & $6$& $0$& $0$& $-\frac{10}{3}$& $\frac{7}{3}$& 1 & 1  \\
    \end{tabular}
    \caption{Charge assignments for the superfields for the additional symmetries present at high temperatures.   The $R$-symmetry $Q_{R}$ is present for large temperatures when the gaugino mass is ineffective. The $WW'$ symmetry (not an $R$-symmetry) is present when the $\mu$-term is ineffective. The symmetries are chosen to be non-anomalous with respect to $SU(2)$ and $SU(3)$, see text. }
    \label{tab:charges}
\end{table}
\endgroup

In the case of the KSVZ axion, the $\mu$-term is ineffective at temperature higher than Eq.~(\ref{eq:Tlambda}) with $m_\lambda$ replaced with $\mu$.
We do not expect that there is such a temperature regime in the  DFSZ case because the $\mu$-term itself will increase with temperature, keeping it in equilibrium. If such a temperature regime does exist, this results in yet another conserved symmetry, which can be taken to be a linear combination of the Weinberg-Wilczek Peccei-Quinn (WW) symmetry wherein $Q_{H_{u}}=Q_{H_{d}}=1$, and $Q_{Q}=Q_{L}=-1$ and two additional symmetries: $B+L$, and the symmetry under which only the right handed up quark is charged $Q_{u1}$, $(Q(U^c) =1)$.  The resulting charges for a non-anomalous symmetry are given by $WW'=WW+ \frac{5}{3}(B+L)+6 Q_{u1} -\frac{5}{3}(B-L)$, where we have added a multiple of the non-anomalous $B-L$ symmetry to give the more convenient charge assignments shown in Table \ref{tab:charges}.
In terms of chemical potentials, the $WW'$ conservation condition is
\begin{multline}    18(3\mu_{\bar{u}_1}+2\mu_{\lambda})+2(3\mu_{H_u}+3\mu_{H_d}+4\mu_{\lambda}) \\ + \sum_{k=1}^{N_g}\left[-6(3\mu_{Q_k}+2\mu_{\lambda})-\frac{10}{3}(3\mu_{\bar{e}_k}+2\mu_{\lambda})+\frac{14}{3}(3\mu_{\ell_k}+2\mu_{\lambda})\right]=0.
\end{multline}

\subsection{Results for  $C_{i}$}
\label{sec:C_i}
Solving the relevant system of equations for the $\mu_{i}$ allows determination of $C_{i}$, see Eqs.~(\ref{eq:nDotB-L_General}) and~(\ref{eq:nB_dot}).

\paragraph*{{\bf DFSZ:}}
In Table~\ref{tab:DFSZ_C_i}, we show $C_i$ for the various cases in the DFSZ model. When all Yukawa interactions and the gaugino mass term are in equilibrium, $C_1=C_2=C_3=0.0459 \frac{n}{N_{\rm DW}}$, independent of the PMNS mixing angles.  When the electron Yukawa is out of equilibrium, the $C_i$ coefficients are slightly different from each other and depend on the PMNS mixing angles.  But using PMNS mixing angles $\theta_{12}=34^{\circ}$, $\theta_{23}=48^{\circ}$, $\theta_{23}=8.5^{\circ}$, the $C_i$ coefficients are still all $0.046 \frac{n}{N_{\rm DW}}$ to two significant digits.  Whether the down-Yukawa interaction is out of equilibrium has a more significant impact; in this case the coefficients become $C_1 = 0.0229 \frac{n}{N_{\rm DW}}$, $C_2 = 0.0203 \frac{n}{N_{\rm DW}}$, and $C_3=0.0182 \frac{n}{N_{\rm DW}}$, and the resulting asymmetry can be affected by more than a factor of 2. Whether or not the off-diagonal Yukawa interactions with the down quark are in equilibrium has no effect on $C_i$.

Whether the gaugino mass term is in equilibrium has a small effect.  For the cases when the down Yukawa is in equilibrium, the difference is roughly $3\%$, pushing $C_i$ to $0.0446 \frac{n}{N_{\rm DW}}$. The effect on the cases where the down is out of equilibrium is similarly small.

\begin{table}[t]
\begin{tabular}{c|c|c|c|}
\cline{2-4}
                                      & \multicolumn{3}{c|}{DFSZ}                                  \\ \hline
\multicolumn{1}{|c|}{$m_{\lambda}$ Efficient} & \multicolumn{1}{c|}{All Yukawas Efficient} & $y_e$ Inefficient & $y_e$ and $y_d$ Inefficient                                                                                                    \\ \Xhline{1.2pt}  
\multicolumn{1}{|c|}{$C_1$}           & \multicolumn{1}{c|}{$0.0459 \frac{n}{N_{\rm DW}}$} & $0.0455 \frac{n}{N_{\rm DW}}$ & $0.0229 \frac{n}{N_{\rm DW}}$\\ \hline
\multicolumn{1}{|c|}{$C_2$}           & \multicolumn{1}{c|}{$0.0459 \frac{n}{N_{\rm DW}}$} & $0.0458 \frac{n}{N_{\rm DW}}$ & $0.0203 \frac{n}{N_{\rm DW}}$ \\ \hline
\multicolumn{1}{|c|}{$C_3$}           & \multicolumn{1}{c|}{$0.0459 \frac{n}{N_{\rm DW}}$} & $0.0461 \frac{n}{N_{\rm DW}}$ & $0.0182 \frac{n}{N_{\rm DW}}$                                                  \\ \hline \hline

\multicolumn{1}{|c|}{$m_{\lambda}$ Inefficient} & \multicolumn{1}{c|}{All Yukawas Efficient} & $y_e$ Inefficient & $y_e$ and $y_d$ Inefficient                                                                                                    \\ \Xhline{1.2pt}
\multicolumn{1}{|c|}{$C_1$}           & \multicolumn{1}{c|}{$0.0446 \frac{n}{N_{\rm DW}}$} & $0.0439 \frac{n}{N_{\rm DW}}$ & $0.0213 \frac{n}{N_{\rm DW}}$ \\ \hline
\multicolumn{1}{|c|}{$C_2$}           & \multicolumn{1}{c|}{$0.0446 \frac{n}{N_{\rm DW}}$} & $0.0444 \frac{n}{N_{\rm DW}}$ & $0.0189 \frac{n}{N_{\rm DW}}$ \\ \hline
\multicolumn{1}{|c|}{$C_3$}           & \multicolumn{1}{c|}{$0.0446 \frac{n}{N_{\rm DW}}$} & $0.0449 \frac{n}{N_{\rm DW}}$ & $0.0170 \frac{n}{N_{\rm DW}}$    
                                         \\ \hline

\end{tabular}
\caption{$C_i$ coefficients in the DFSZ model when different reactions are in equilibrium.  The first group of rows corresponds to the case when scattering through the gaugino mass is in equilibrium, and the second group corresponds to the case where it is not.  The first column of numbers corresponds to the low-temperature case when all Yukawa interactions are in equilibrium.  The second corresponds to the case when only interactions via the electron-Yukawa are out of equilibrium, and the third also has down-Yukawa interactions out of equilibrium.}
\label{tab:DFSZ_C_i}
\vspace{0.5cm}
\begin{tabular}{c|c|c|c|}
\cline{2-4}
                                      & \multicolumn{3}{c|}{KSVZ}                                  \\ \hline
\multicolumn{1}{|c|}{$m_{\lambda}$ and $\mu$ Efficient} & \multicolumn{1}{c|}{All Yukawas Efficient} & $y_e$ Inefficient & $y_e$ and $y_d$ Inefficient                                                                                                    \\ \Xhline{1.2pt}  
\multicolumn{1}{|c|}{$C_1$}           & \multicolumn{1}{c|}{$0.0037 c_g + 0.0069 c_W$} & $0.0016 c_g + 0.0082 c_W$ & $-0.0063 c_g + 0.0083 c_W$\\ \hline
\multicolumn{1}{|c|}{$C_2$}           & \multicolumn{1}{c|}{$0.0037 c_g + 0.0069 c_W$} & $0.0033 c_g + 0.0072 c_W$ & $-0.0055 c_g + 0.0074 c_W$ \\ \hline
\multicolumn{1}{|c|}{$C_3$}           & \multicolumn{1}{c|}{$0.0037 c_g + 0.0069 c_W$} & $0.0047 c_g + 0.0064 c_W$ & $-0.0050 c_g + 0.0066 c_W$                                                  \\ \hline \hline

\multicolumn{1}{|c|}{$m_{\lambda}$ Inefficient} & \multicolumn{1}{c|}{All Yukawas Efficient} & $y_e$ Inefficient & $y_e$ and $y_d$ Inefficient                                                                                                    \\ \Xhline{1.2pt}
\multicolumn{1}{|c|}{$C_1$}           & \multicolumn{1}{c|}{$0.0037 c_g + 0.0089 c_W$} & $0.0016 c_g + 0.0098 c_W$ & $-0.0063 c_g + 0.0083 c_W$ \\ \hline
\multicolumn{1}{|c|}{$C_2$}           & \multicolumn{1}{c|}{$0.0037 c_g + 0.0089 c_W$} & $0.0033 c_g + 0.0091 c_W$ & $-0.0055 c_g + 0.0074 c_W$ \\ \hline
\multicolumn{1}{|c|}{$C_3$}           & \multicolumn{1}{c|}{$0.0037 c_g + 0.0089 c_W$} & $0.0047 c_g + 0.0085 c_W$ & $-0.0050 c_g + 0.0067 c_W$    
                                         \\ \hline \hline

\multicolumn{1}{|c|}{$m_{\lambda}$ and $\mu$ Inefficient} & \multicolumn{1}{c|}{All Yukawas Efficient} & $y_e$ Inefficient & $y_e$ and $y_d$ Inefficient                                                                                                    \\ \Xhline{1.2pt}
\multicolumn{1}{|c|}{$C_1$}           & \multicolumn{1}{c|}{$-0.0107 c_g + 0.0063 c_W$} & $-0.0126 c_g + 0.0072 c_W$ & $-0.0127 c_g + 0.0071 c_W$ \\ \hline
\multicolumn{1}{|c|}{$C_2$}           & \multicolumn{1}{c|}{$-0.0107 c_g + 0.0063 c_W$} & $-0.0111 c_g + 0.0064 c_W$ & $-0.0112 c_g + 0.0064 c_W$ \\ \hline
\multicolumn{1}{|c|}{$C_3$}           & \multicolumn{1}{c|}{$-0.0107 c_g + 0.0063 c_W$} & $-0.0098 c_g + 0.0059 c_W$ & $-0.0101 c_g + 0.0058 c_W$  
                                       \\ \hline
\end{tabular}
\caption{$C_i$ coefficients in the KSVZ model when different reactions are in equilibrium.  The first group of rows corresponds to the case when scattering via the gaugino mass and $\mu$-term are in equilibrium.  The second group corresponds to the case where the $\mu$-term is in equilibrium but the gaugino mass is not.  The third group corresponds to the case where both the gaugino mass and $\mu$-term are out of equilibrium.  The first column corresponds to the low-temperature case when all Yukawa interactions are in equilibrium.  The second gives results when only the interactions via the electron-Yukawa is out of equilibrium, and the third also has down-Yukawa interactions out of equilibrium. In the standard normalization of the axion-gluon coupling, $c_g = 1$.}
\label{tab:KSVZ_C_i}
\end{table}

\paragraph*{{\bf KSVZ:}}
In Table~\ref{tab:KSVZ_C_i}, we show $C_i$ for the various cases in the KSVZ model.  The case when the off-diagonal Yukawa interactions with the down quark are out of equilibrium is not shown because it has a small impact on the result.     Whether or not these interactions are in equilibrium has no effect when scattering via the $\mu$-term is efficient, and an effect only on the level of several percent when the $\mu$-term is inefficient.

\section{Scaling of baryon asymmetry production}
\label{app:YB_scaling}

To find the  baryon asymmetry, it is important to identify which cosmological epoch dominates production. 
The yield of the $B-L$ asymmetry produced per Hubble time $\Delta Y_{B-L}$ is given in Eq.~(\ref{eq:DeltaY_B-L}).  This quantity is redshift invariant after production if no entropy is subsequently produced.  In this case, the dominant epoch can be identified by examining the scaling of $\dot{\theta}$ and the Hubble rate $H$. On the other hand, if entropy is produced from inflationary reheating/saxion thermalization, it is more convenient to examine
\begin{equation}
    \frac{\Delta n_{B-L}}{\rho_{\rm matter}} \equiv \frac{\dot n_{B-L}}{\rho_{\rm matter} H} = \sum C_i(T) m_{\nu_{i}}^2 \frac{\dot\theta T^5}{\rho_{\rm matter} H v_{H_u}^4} ,
\end{equation}
which is redshift invariant following the production of the asymmetry   because both previously produced $n_{B-L}$ and the matter (inflaton or saxion) energy density $\rho_{\rm matter}$ scale as $R^{-3}$. In Table~\ref{tab:YB}, we summarize how these relevant quantities scale. If the final scaling of $\Delta n_{B-L} / s$ or $\Delta n_{B-L} / \rho_{\rm matter}$ features an increasing (decreasing) function of $R$, then the production is IR (UV)-dominated during the corresponding epoch. 

For example, if $T_S > T_R$, the table shows that production peaks at $T_S$ during inflationary reheating, labeled as the inflaton non-adiabatic, matter-dominated era MD$_{\rm NA}^{\rm inf}$. This is because $\Delta n_{B-L} / \rho_{\rm inf}$ is IR-dominated (UV-dominated) before (after) $T_S$ during MD$_{\rm NA}^{\rm inf}$, while $\Delta n_{B-L}/s$ stays UV-dominated in all subsequent eras with $T < T_S$. This result is illustrated in the right panel of Fig.~\ref{fig:schematic_YB}.

On the other hand, if $T_S < T_R$, the baryon asymmetry is produced in equal amount in each Hubble time, $\Delta n_{B-L} / s \propto R^0$, during a radiation-dominated era labeled by RD until $T_{f} = \max(T_S, T_{\rm RM})$. We first discuss the case without saxion domination, i.e., with early thermalization. At this $T_{f}$, the production subsequently becomes UV-dominated because, if $T < T_S$ during radiation domination, $\Delta n_{B-L} / s \propto R^{-3}$ or if $T < T_{\rm RM}$ (but $T > T_S$) there is a matter-dominated era by the rotation energy density MD$_{\rm A}^{\rm rot}$ and $\Delta n_{B-L} / s \propto R^{-1/2}$ in this era. This (adiabatic) matter-dominated era MD$_{\rm A}^{\rm rot}$ does not result in any entropy production as the energy density ultimately becomes subdominant to radiation due to the era where it scales as kination. 
This is the case where continuous production leads to the logarithmic enhancement discussed around Eq.~(\ref{eq:YB_RD}). This case is illustrated in the left panel of Fig.~\ref{fig:schematic_YB}.

\begingroup
\setlength{\tabcolsep}{4pt} 
\renewcommand{\arraystretch}{1.56} 
\begin{table}[t]
\begin{tabular}{|cc|c|c|c|c|c||c|c|}
\hline
\multicolumn{2}{|c|}{Epoch}                                              & $H$                                 & $T$                                 & $\Gamma_L$                          & $\rho_{\rm matter}$     & $\dot\theta$ & $\frac{\Delta n_{B-L}}{s}$    & $\frac{\Delta n_{B-L}}{\rho_{\rm matter}}$ \\ \Xhline{1.2pt}  
\multicolumn{1}{|c|}{\multirow{2}{*}{MD$_{\rm NA}^{\rm inf}$}} & $T>T_S$ & \multirow{2}{*}{$R^{-\frac{3}{2}}$} & \multirow{2}{*}{$R^{-\frac{3}{8}}$} & \multirow{2}{*}{$R^{-\frac{9}{8}}$} & \multirow{2}{*}{$R^{-3}$} & $R^0$        & --                 & $R^{\frac{21}{8}}$                \\ \cline{2-2} \cline{7-9} 
\multicolumn{1}{|l|}{}                                         & $T<T_S$ &                                     &                                     &                                     &                           & $R^{-3}$     & --                 & $R^{-\frac{3}{8}}$                \\ \Xhline{1.2pt}  
\multicolumn{1}{|c|}{\multirow{2}{*}{RD}}                      & $T>T_S$ & \multirow{2}{*}{$R^{-2}$}           & \multirow{2}{*}{$R^{-1}$}           & \multirow{2}{*}{$R^{-3}$}           & \multirow{2}{*}{--}       & $R^0$        & $R^0$              & --                                \\ \cline{2-2} \cline{7-9} 
\multicolumn{1}{|c|}{}                                         & $T<T_S$ &                                     &                                     &                                     &                           & $R^{-3}$     & $R^{-3}$           & --                                \\ \Xhline{1.2pt}  
\multicolumn{1}{|l|}{\multirow{2}{*}{MD$_{\rm NA}^{\rm osc}$ $\begin{dcases} {\Gamma_{S\widetilde{H}\widetilde{H}}} \vspace{0.3cm} \\   {\Gamma_{SWW}} \end{dcases}$ }} & $T>T_S$ & $R^{-\frac{3}{2}}$ & $R^{\frac{3}{2}}$ & $R^{\frac{9}{2}}$ & \multirow{2}{*}{$R^{-3}$} & $R^0$        & --                 & $R^{12}$                \\ \cline{2-5} \cline{7-9} 
\multicolumn{1}{|l|}{}                                         & $T>T_S$ &   $R^{-\frac{3}{2}}$     &      $R^{-\frac{1}{2}}$    &  $R^{-\frac{3}{2}}$   &                           & $R^0$     & --                 & $R^2$                \\ \Xhline{1.2pt} 
\multicolumn{1}{|c|}{MD$_{\rm A}^{\rm rot}$}                   & $T>T_S$ & $R^{-\frac{3}{2}}$                  & $R^{-1}$                            & $R^{-3}$                            & --                        & $R^0$        & $R^{-\frac{1}{2}}$ & --                                \\ \hline
\multicolumn{1}{|c|}{KD}                                       & $T<T_S$ & $R^{-3}$                            & $R^{-1}$                            & $R^{-3}$                            & --                        & $R^{-3}$     & $R^{-2}$           & --                                \\ \hline
\end{tabular}
\caption{Scaling of quantities relevant for the estimation of the $B-L$ asymmetry.  Positive (negative) exponents for $R$ in the final two columns indicate  IR (UV)-dominated production.  The case that scales as $R^{0}$ has equal contributions per Hubble time and so receives a logarithmic enhancement; see text for details. We note that $\rho_{\rm matter}$ represents either $\rho_{\rm inf}$ or $\rho_S$ depending on which one dominates and creates entropy.}
\label{tab:YB}
\end{table}
\endgroup

In the above discussion, we assumed that the saxion energy density was depleted by thermalization before dominating the total energy density. If instead the saxion comes to dominate and subsequently creates a large amount of entropy from its thermalization, any previously produced baryon asymmetry can be sufficiently diluted so that the production after saxion thermalization dominates. As discussed in Sec.~\ref{sec:SD}, during the non-adiabatic era before the end of thermalization, the relevant thermalization processes are saxion-Higgsino and saxion-$W$ scatterings for $n=1$ and $n=2$, respectively. Production of $n_{B-L}$ per Hubble time is listed in Table~\ref{tab:YB} for these two cases with the label MD$_{\rm NA}^{\rm osc}$, and one can see that production is IR-dominated for both cases. (We do not show the scaling for $T < T_S$ here;  it is never realized in our parameter space.) This verifies that the contribution produced subsequent to thermalization of the saxion dominates over that produced during thermalization. The production after thermalization is again logarithmically enhanced during a radiation-dominated era but now between $T_{\rm th}$ and $\max(T_{\rm RM}, T_S)$ with $T_{\rm RM}$ given by Eq.~(\ref{eq:TRM_r_Tth}). The results for the saxion domination scenario are presented in Sec.~\ref{sec:SD}.

\section{Constraints from supersymmetric relics}
\label{app:relics}

In this supersymmetric framework, there are a number of potentially long-lived relics.  These relics may provide  constraints on the theory. The constraints depend upon the identity of both the LSP, and if long-lived, the next-to-lightest supersymmetric particle (NLSP).  The predictions of Big Bang Nucleosynthesis (BBN) must not be disturbed, and, if stable, the LSP density may not exceed the dark matter density.

{\bf Non-gravitino/axino LSP: }
We first consider the case of the LSP being a superpartner of a Standard Model particle.   The constraint on the mass spectrum and/or the reheat temperature from BBN is discussed in
\cite{Kawasaki:2008qe}.
If the gravitino mass $m_{3/2} \sim$ TeV,  late gravitino decays will disturb BBN unless the reheat temperature $T_R \lsim 10^6$ GeV. The bound can be relaxed if the LSP is a slepton, but a charged LSP is strongly constrained by searches for heavy hydrogen~\cite{Burdin:2014xma} and a sneutrino LSP is excluded by direct detection experiments.

Because this value for $T_R$ is close to the typical $T_{\rm RM}$ (or even smaller)  this means that any logarithmic enhancement, see Eq.~(\ref{eq:YB_RD}), is necessarily absent, and the prediction for $m_{S}$ is somewhat modified (increased).  
If $m_{3/2} \sim$ 10 TeV, the upper bound is $T_R\lsim 10^8$ GeV.

It is conceivable that $m_{3/2}$ is quite large with mass $\gsim 100$ TeV, in which case the gravitino decays might be early enough to avoid conflicts with BBN and larger reheat temperatures might be allowed.   However, in this case, the scalar mass must be also $\mathcal{O}(100)$ TeV; otherwise for $m_{3/2}\gg m_S$, the $A$-term in Eq.~(\ref{eq:VP}) becomes much larger than $m_S$, and $P$ is trapped at a minimum with large $S$. Additionally, in order for the thermal freeze-out abundance of the LSP (say wino or Higgsino) not to be too large, a hierarchy of the type $m_{\rm LSP} \ll m_{S} \sim m_{3/2}$ is required.
Moreover, even if gravitino decays during BBN are avoided and the LSP thermal abundance is not too large, there is still a danger of non-thermal overproduction of the LSP
from gravitino decays.  For a gravitino mass of 100 TeV
and a LSP mass of a TeV, this constrains the reheat temperature $T_R < 2 \times 10^{9}$ GeV~\cite{Kawasaki:2008qe}.

The above upper bounds on the reheat temperature could be relaxed if $R$-parity is violated. In this case, we may assume a slepton LSP, thereby weakening the BBN constraints from gravitino decays, but without conflicting with heavy isotope searches nor direct detection. If there are no sparticles between the gravitino and the slepton(s), the upper bound becomes $T_R < 10^9~(10^{11}) $ GeV for $m_{3/2}\sim 1~(10)$ TeV. For $m_{3/2} > 100$ TeV, the LSP overproduction bound disappears and $T_R$ may be much above $10^9$ GeV.

{\bf Gravitino LSP: }
To avoid overproduction of a gravitino LSP from thermal processes requires a reheat temperature $T_R < 2\times 10^9 \GeV \times ({\rm TeV}/m_{3/2})$. In this case a logarithmic enhancement as in Eq.~\eqref{eq:YB_RD} can remain. However, avoiding disruption of BBN via decay of the (visible sector) NLSP puts strong constraints on the parameter space if $T_R$ is above the sparticle masses. The strongest constraints arise~\cite{Kawasaki:2008qe,Kawasaki:2017bqm} when where the NLSP has a large branching ratio to hadrons; constraints  are minimized for a sneutrino NLSP. This can be realized by taking the soft mass of ${\bf \bar{5}}$ to be smaller than that of ${\bf 10}$.
Then, the dominant constraint comes from a three-body decay involving a weak boson, whose branching ratio is ${\mathcal{O}(10^{-2})}$.
For the gravitino mass of TeV, the sneutrino NLSP lifetime is around $10^{5}$ s, and from the constraint on the decay into weak gauge bosons derived in~\cite{Kawasaki:2017bqm}, we obtain
\begin{equation}
m_{\tilde{\nu}}  Y_{\tilde{\nu}} \times {\rm Br}({\rm three\mathchar`-body}) < 10^{-14}   \Rightarrow m_{\tilde{\nu}} Y_{\tilde{\nu}} < 10^{-12}.
\end{equation}
The freeze-out abundance of a TeV-scale sneutrino would violate this bound.
To evade the bound requires $m_{\tilde{\nu}} > 10$ TeV, so that the lifetime of the sneutrino is shorter than 100 s. 
We may also avoid the bound by taking $m_{\tilde{\nu}}-m_{3/2} < m_{Z}$. In this case, the decay mode relevant for the BBN constraints becomes a four-body decay with a branching ratio $\sim 10^{-4}$, and the constraint is marginally satisfied.

If the cutoff scale is below the Planck scale, $m_{3/2} \ll m_{\rm NLSP}$ and the NLSP lifetime may be shorter. For example, with the cutoff scale around the string scale $\sim 10^{17}$ GeV, $m_{3/2}\sim 100$ GeV with $m_{\rm NLSP} \sim$ few TeV is possible. The lifetime is then shorter than 100~s, and $m_{\rm NLSP} Y_{\rm NLSP} < 10^{-10}$ and $10^{-7}$ is required for the NLSP with the leading hadronic decay mode and the sneutrino NLSP, respectively. This is satisfied for $m_{\rm NLSP}= {\mathcal O}(1)$ TeV.

The bound on the mass spectrum may be avoided if $R$-parity is violated, since the NLSP can decay much before BBN. The gravitino may still be long-lived enough to be dark matter.  In this case,  $R$-parity violating couplings could provide an additional source of asymmetry, see~\cite{Co:2021qgl}, but this is small if the couplings are not large.

{\bf Axino LSP: }
The axino should not be the LSP unless $R$-parity violation is introduced.  To see why this is so, recall the saxion is thermalized. Unless this thermalization occurs below the masses of the sparticles, the axino is also thermalized. Unless the axino mass $m_{\tilde{a}}$ is below $\mathcal{O}(100)$ eV, axino dark matter is overproduced. However, even a subdominant component of hot dark matter is constrained, so a stronger bound $m_{\tilde{a}} \lsim \mathcal{O}(10)$ eV applies~\cite{Xu:2021rwg}.

While in some of the parameter space saxion thermalization does occur at $T_{\rm th} \lsim m_{S}$, and it may be possible to avoid thermalization of the axino, it is nevertheless potentially produced in dangerous amounts via freeze-in at higher temperatures. Indeed, unless the $m_{\tilde{a}} \ll$ TeV---difficult in gravity mediation---the axino is still overproduced.  Indeed, the two-field model gives $m_{\tilde{a}} \simeq m_{3/2}$ because a non-zero vacuum expectation value for $X$ is induced by a supergravity tadpole.  In the one-field model, although $m_{\tilde{a}}$ vanishes at  tree-level, it is still generated by one-loop quantum corrections. The dominant contribution comes from the Yukawa coupling $y P \psi \bar{\psi}$ and the associated $A$-term, where this interaction is also responsible for the generation of the logarithmic potential.

The axino LSP can be viable if $R$-parity violation allows the axino to decay. For example, for $m_{\tilde{a}}$ above the electroweak scale, the axino can decay before BBN via the $L H_u$ operator without giving a too-large neutrino mass. The contribution to the baryon asymmetry from axiogenesis via $R$-parity violation~\cite{Co:2021qgl} is subdominant compared to the lepto-axiogenesis contribution. Such a large $m_{\tilde{a}}$ is readily obtained in the two-field model. In the one-field model, generating a loop-induced $m_{\tilde{a}}$ exceeding the electroweak scale places bounds on the supersymmetry-breaking scale. 
In gravity mediation with a singlet supersymmetry-breaking field, $A \sim m_{S}$, so  $m_{\tilde{a}}$ above the electroweak scale requires $m_S > 10$ TeV. In gravity mediation without singlets, $A \sim 0.01 m_S$, so $m_S > 10^6 \GeV$ would be required.

The upper bound on $T_{R}$ from BBN is not relaxed in comparison with other cases. Although the gravitino can have dominant decay $\tilde{G} \rightarrow \tilde{a} {a}$ if it is the NLSP, the axino anyway decays into SM particles, so the BBN constraint still applies.

\nocite{apsrev41Control}
\bibliographystyle{apsrev4-1}
\bibliography{DFSZ_Lepto-Axiogenesis}

\end{document}